\documentclass[english,10pt, onecolumn]{IEEEtran}
\usepackage[T1]{fontenc}
\usepackage[latin9]{inputenc}
\usepackage{float}
\usepackage{booktabs}
\usepackage{amsmath}
\usepackage{amsthm}
\usepackage{amssymb}
\usepackage{graphicx}
\usepackage{xcolor}
\makeatletter

\providecommand{\tabularnewline}{\\}
\floatstyle{ruled}
\newfloat{algorithm}{tbp}{loa}
\providecommand{\algorithmname}{Algorithm}
\floatname{algorithm}{\protect\algorithmname}

\theoremstyle{plain}
\newtheorem{thm}{\protect\theoremname}
\theoremstyle{plain}
\newtheorem{lem}[thm]{\protect\lemmaname}
\theoremstyle{plain}
\newtheorem{cor}[thm]{\protect\corollaryname}
\theoremstyle{plain}
\newtheorem{prop}[thm]{Proposition}

\IEEEoverridecommandlockouts

\newtheorem{lemma}{Lemma}
\newtheorem{definition}{Definition}

\usepackage{algorithmic,sidecap,cite}

\makeatother

\usepackage{babel}
\providecommand{\corollaryname}{Corollary}
\providecommand{\lemmaname}{Lemma}
\providecommand{\theoremname}{Theorem}

\begin{document}
\global\long\def\et#1{\tilde{\mathbf{e}}_{#1,h_{#1}}}%
\global\long\def\gtc{\tilde{\mathbf{G}}_{h_{X},h_{X|y}}}%
\global\long\def\gtl{\tilde{\mathbf{G}}_{h_{X}(l),h_{Y}(l)}}%
\global\long\def\gt{\tilde{\mathbf{G}}_{h_{X},h_{Y}}}%
\global\long\def\gh{\tilde{\mathbf{G}}_{h}}%
\global\long\def\ghat{\hat{\mathbf{G}}_{k_{1},k_{2}}}%
\global\long\def\gk{\hat{\mathbf{G}}_{k}}%
\global\long\def\ghatl{\hat{\mathbf{G}}_{k(l)}}%
\global\long\def\g#1{\tilde{\mathbf{G}}_{#1}}%
\global\long\def\cov{\text{Cov}}%

\global\long\def\ft#1{\tilde{\mathbf{f}}_{#1,h_{#1}}}%
\global\long\def\fth#1{\tilde{\mathbf{f}}_{#1,h}}%
\global\long\def\ftl#1{\tilde{\mathbf{f}}_{#1,h_{#1}(l)}}%
\global\long\def\et#1{\tilde{\mathbf{e}}_{#1,h_{#1}}}%
\global\long\def\eth#1{\tilde{\mathbf{e}}_{#1,h}}%
\global\long\def\ftlx#1#2{\tilde{\mathbf{f}}_{#1,h_{#1}\left(l_{#2}\right)}}%

\global\long\def\bE{\mathbb{E}}%
\global\long\def\var{\mathbb{V}}%
\global\long\def\bias{\mathbb{B}}%
\global\long\def\S{\mathcal{S}}%
\global\long\def\ez#1{\mathbb{E}_{\mathbf{#1}}}%

\global\long\def\dist#1{\mathbf{\rho}_{#1,k_{#1}}}%
\global\long\def\distp#1{\mathbf{\rho}_{#1,k_{#1}+1}}%

\global\long\def\fh#1{\hat{\mathbf{f}}_{#1,k_{#1}}}%
\global\long\def\fhl#1#2{\hat{\mathbf{f}}_{#1,k(#2)}}%
\global\long\def\fhp#1{\hat{\mathbf{f}}_{#1,k_{#1}+1}}%
\global\long\def\fb#1{\bar{\mathbf{f}}_{#1,k_{#1}}}%
\global\long\def\fbp#1{\bar{\mathbf{f}}_{#1,k_{#1}+1}}%

\global\long\def\ett#1#2{\tilde{\mathbf{e}}_{#1#2,h_{#1},h_{#2}}}%
\global\long\def\Y{\mathbf{Y}}%
\global\long\def\X{\mathbf{X}}%
\global\long\def\Z{\mathbf{Z}}%
\global\long\def\N{\mathbf{N}}%
\global\long\def\W{\mathbf{W}}%
\global\long\def\hx{\mathbf{h}_{X_{C}|x}}%
\global\long\def\hy{\mathbf{h}_{Y_{C}|y}}%

\global\long\def\gtay#1#2#3{\mathbf{#1}_{#2}^{(#3)}}%


\author{\IEEEauthorblockN{Kevin R. Moon\IEEEauthorrefmark{1}, Kumar Sricharan\IEEEauthorrefmark{2}, Alfred O. Hero III\IEEEauthorrefmark{3}}\\
\IEEEauthorblockA{\IEEEauthorrefmark{1}Dept. of Mathematics and Statistics, Utah State University,  kevin.moon@usu.edu} \\
\IEEEauthorblockA{\IEEEauthorrefmark{2}Intuit Inc., sricharan$\textunderscore$kumar@intuit.com} \\
\IEEEauthorblockA{\IEEEauthorrefmark{3}EECS Dept., University of Michigan, hero@eecs.umich.edu}  \thanks{This work was supported in part by the US Army Research Office under grants W911NF1910269 and W911NF1510479, and by the National Nuclear Security Administration in the US Department of Energy under grant DE-NA0003921. This paper appeared in part in the Proceedings of the 2017 IEEE Intl. Symposium on Information Theory (ISIT)~\cite{moon2017ensemble}.}}
\title{Ensemble Estimation of Generalized Mutual Information with Applications
to Genomics}
\maketitle
\begin{abstract}
Mutual information is a measure of the dependence between random variables
that has been used successfully in myriad applications in many fields.
Generalized mutual information measures that go beyond classical Shannon
mutual information have also received much interest in these applications.
We derive the mean squared error convergence rates of kernel density-based
plug-in estimators of general mutual information measures between
two multidimensional random variables $\X$ and $\Y$ for two cases:
1) $\X$ and $\Y$ are continuous; 2) $\X$ and $\Y$ may have a mixture
of discrete and continuous components. Using the derived rates, we
propose an ensemble estimator of these information measures called
GENIE by taking a weighted sum of the plug-in estimators with varied
bandwidths. The resulting ensemble estimators achieve the $1/N$ parametric
mean squared error convergence rate when the conditional densities
of the continuous variables are sufficiently smooth. To the best of
our knowledge, this is the first nonparametric mutual information
estimator known to achieve the parametric convergence rate for the
mixture case, which frequently arises in applications (e.g. variable
selection in classification). The estimator is simple to implement
and it uses the solution to an offline convex optimization problem
and simple plug-in estimators. A central limit theorem is also derived
for the ensemble estimators and minimax rates are derived for the
continuous case. We demonstrate the ensemble estimator for the mixed
case on simulated data and apply the proposed estimator to analyze
gene relationships in single cell data.
\end{abstract}


\section{Introduction}

\label{sec:intro}Mutual information (MI) is a measure of the amount
of shared information between a pair of random variables $\X$ and
$\Y$. MI estimation is related to the problem of estimating functionals
of probability distributions, which has received deserved attention
in recent years~\cite{gao2017estimating,zeng2018jackknife,jiao2015minimax,jiao2017maximum,wu2016minimax,krishnamurthy2014divergence,kandasamy2015nonparametric,singh2014exponential,singh2014renyi,sricharan2013ensemble,moon2014nips,moon2014isit,moon2018ensemble,noshad2017direct,wisler2018direct,berrett2019efficient}.
Many statistical problems rely in some form upon accurate estimation
of functionals of probability distributions including estimating the
decay rates of error probabilities~\cite{cover2012elements}, estimating
bounds on the Bayes error rate~\cite{avi1996bound,moon2015Bayes,chernoff1952measure,berisha2014bound},
and hypothesis testing~\cite{moon2015partI,unnikrishnan2011universal,naghshvar2012extrinsic}.
MI estimation, in particular, also has many applications in information
theory and machine learning including independent subspace analysis~\cite{pal2010estimation},
structure learning \cite{moon2017information}, fMRI data processing~\cite{chai2009fMRI},
forest density estimation~\cite{liu2012exponential}, clustering~\cite{lewi2006real},
neuron classification~\cite{schneidman2003information}, blind source
separation~\cite{hild2001blind}, intrinsically motivated reinforcement
learning \cite{mohamed2015variational,salge2014changing}, as well
as other data science applications such as sociology~\cite{reshef2011detecting},
computational biology~\cite{van2018recovering,krishnaswamy2014conditional,moon2019phate},
and improving neural network models~\cite{belghazi2018mine}. A particularly
common application is feature selection or extraction where features
are chosen to maximize the MI between the chosen features (represented
by $\mathbf{X}$) and the outcome variables (represented by $\mathbf{Y}$)
\cite{torkkola2003feature,vergara2014review,peng2005feature,kwak2002input,zhang2020label}. 

In many of these applications, the variables $\X$ and $\Y$ may have
any mixture of discrete and continuous components. In feature selection,
for example, the predictor labels may have discrete components (e.g.
classification labels) while the input variables may have a mixture
of discrete and continuous features. To the best of our knowledge,
there are currently no nonparametric MI estimators that are known
to achieve the parametric mean squared error (MSE) convergence rate
$1/N$ ($N$ is the number of samples) in this setting where $\X$
and/or $\Y$ contain a mixture of discrete and continuous components.
Instead, most existing estimators of MI focus on the cases where both
$\X$ and $\Y$ are either purely discrete or purely continuous. Also,
while many nonparametric estimators of MI exist, most have not been
generalized beyond Shannon or R\'{e}nyi information. Furthermore,
minimax convergence rates are currently unknown for the continuous
and the mixture cases.

In this paper, we provide a framework for nonparametric estimation
of a large class of MI measures where we only have available a finite
population of i.i.d. samples.\textbf{ }This framework can be applied
to accurately estimate general MI measures when either $\X$ and $\Y$
are purely continuous or the mixed case when $\X$ and $\Y$ may contain
a mixture of discrete and continuous components. We derive an MI estimator
for these cases that achieves the parametric MSE rate when the conditional
densities of the continuous variables are sufficiently smooth, thus
achieving the minimax rate (which we also derive) in this setting.
We call this estimator the \textbf{G}eneralized \textbf{EN}semble
\textbf{I}nformation \textbf{E}stimator (GENIE).

Our estimation method applies to other MI measures in addition to
Shannon information, which have been the focus of much recent interest.
An information measure based on a quadratic divergence was defined
in \cite{torkkola2003feature}. A density-resampled version of MI
was introduced in~\cite{krishnaswamy2014conditional} to better measure
gene relationships in single-cell data when sampling may not be uniform.
A MI measure based on the Pearson divergence was considered in \cite{sugiyama2012machine}.
Minimal spanning tree \cite{costa2004geodesic} and generalized nearest-neighbor
graph \cite{pal2010estimation} approaches have been developed for
estimating R\'{e}nyi information~\cite{csiszar1995generalized,verdu2015alpha,principe2010information},
which has been used in many applications (e.g.~\cite{tomamichel2018operational,dong2016gravity,hild2001blind,datta2009min,hayashi2017equivocations}).

\subsection{Related Work}

Many estimators for MI have been previously developed. Nearly all
of these estimators ignore the mixed case and focus on the case where
both $\X$ and $\Y$ are either purely continuous or purely discrete.
A popular $k$-nearest neighbor (nn)-based estimator was proposed
in \cite{kraskov2004estimating} which is a modification of the entropy
estimator derived in \cite{kozachenko1987sample}. However, these
estimators have only been shown to achieve the parametric convergence
rate when the dimension of each of the random variables is less than
3 \cite{gao2018demystifying}. Furthermore, these estimators focus
only on estimating the Shannon MI between purely continuous random
variables. Similarly, the estimators in \cite{pal2010estimation,ganguly2018nearest}
do not achieve the parametric rate and focus on the purely continuous
case. An adaptation of the Shannon MI estimator in~\cite{kraskov2004estimating}
was recently proposed to handle the discrete-continuous mixture case~\cite{gao2017estimating}.
While this estimator has been proven to be consistent, its convergence
rate is currently unknown. Central limit theorems have also been derived for several entropy estimators~\cite{berrett2019efficient,berrett2019twosample,berrett2019nonparametric}, which can then be applied to Shannon MI. However, it is not clear if these results can be extended to more general MI functionals.

A neural network-based estimator of Shannon MI was proposed in~\cite{belghazi2018mine}.
While this estimator is computationally efficient, its statistical
properties are largely unknown as the authors only prove convergence
in probability rates. It is also unclear how to extend this estimator
to other MI measures such as the R\'{e}nyi information. A jackknife
approach to estimating Shannon MI was also recently proposed~\cite{zeng2018jackknife}.
This approach provides an automatic selection of the kernel bandwidth
for a plug-in kernel density estimator (KDE) and does not require
boundary correction, which is generally a major difficulty in estimating
functionals of probability distributions. However, the MSE convergence
rate of this estimator is also unknown. 

Much work has focused on the problem of estimating the entropy of
purely discrete random variables~\cite{han2015minimax,jiao2015minimax,jiao2017maximum,wu2016minimax}.
Shannon MI can then be estimated by estimating the joint and marginal
entropies of $\X$ and $\Y$. However, it is not clear if discrete
methods can be extended successfully to the mixed-case. Quantizing
the continuous components of the data is one potential approach that
has been shown to be consistent for some quantization schemes in the
purely continuous case~\cite{darbellay1999MIest} but it is currently
unknown if similar approaches can be applied in the mixed-case. Also,
extending these estimators to general MI measures like R\'{e}nyi
information is not straightforward.

Recent work has focused on nonparametric divergence estimation for
continuous random variables. One approach \cite{krishnamurthy2014divergence,kandasamy2015nonparametric,singh2014exponential,singh2014renyi}
uses an optimal KDE to achieve the parametric convergence rate when
the densities are at least $d$ \cite{singh2014exponential,singh2014renyi}
or $d/2$ \cite{krishnamurthy2014divergence,kandasamy2015nonparametric}
times differentiable where $d$ is the dimension of the data. These
methods, like ours, assume that the densities are bounded away from
zero as this simplifies the analysis. However, this induces a boundary
on the densities' support set. For accurate estimation, the optimal
KDE approaches require knowledge of the density support boundary and
are difficult to construct near the boundary. Numerical integration
may also be required for estimating some divergence functionals under
this approach, which can be computationally expensive. In contrast,
our approach to MI estimation does not require numerical integration
and can be performed without knowledge of the support boundary.

Some methods for estimating distributional functionals have relaxed
the boundedness assumption on the densities \cite{singh2016finite,han2020optimal,berrett2019efficient}.
These approaches typically assume that the tails of the densities
decay at a sufficiently fast rate (e.g. sub-exponential or sub-Gaussian).
In \cite{han2020optimal,singh2016finite}, the authors only consider densities
with up to 2 derivatives as it is  difficult to exploit higher smoothness when the densities are not lower-bounded.

More closely related work \cite{sricharan2013ensemble,moon2014isit,moon2014nips,berrett2019efficient,moon2018ensemble,moon2017knn,noshad2017direct,wisler2018direct}
uses an ensemble approach to estimate entropy or divergence functionals
for continuous random variables. These works construct an ensemble
of simple plug-in estimators by varying the neighborhood size of density
estimators. They then take a weighted average of the estimators where
the weights are chosen to decrease the bias with only a small increase
in the variance. The parametric rate of convergence is achieved when
the densities are either $d$~\cite{sricharan2013ensemble,moon2014isit,moon2014nips,noshad2017direct}
or $d/2$ \cite{moon2018ensemble,moon2017knn,wisler2018direct} times
differentiable. These approaches are simple to implement as they only
require simple plug-in estimates and the solution of an offline convex
optimization problem. The ensemble approach also automatically corrects
for bias at the boundary of the densities' support set.

Finally, \cite{gao2015efficient} showed that $k$-nn or KDE based
approaches underestimate the MI when the MI is large. As MI increases,
the dependencies between random variables increase which results in
less smooth densities. Thus this isn't an issue when the densities
are smooth \cite{krishnamurthy2014divergence,kandasamy2015nonparametric,singh2014exponential,singh2014renyi,sricharan2013ensemble,moon2014isit,moon2014nips,moon2018ensemble}.

For the mixture setting, we focus on the important special case where
the components of each observation are assumed to decompose into discrete
and continuous dimensions. This enables the density to be factored:
$f_{X}(x)=f_{X_{D}}\left(x_{D}\right)f_{X_{C}|X_{D}}\left(x_{C}|x_{D}\right)$
where $x_{C}$ and $x_{D}$ are the continuous and discrete components
of $x$. We note that this excludes the more general case considered
by~\cite{gao2017estimating} where one or more components can have
discrete and continuous values simultaneously. However, our setting
is a common occurrence in many machine learning and statistical problems.
For example, a search within the UCI Machine Learning Repository~\cite{Dua:2019}
yields many datasets with such structure. Many statistical models
have also focused on similar settings~\cite{olkin1961multivariate,cox1999likelihood,train2008algorithms,little1985maximum}.
Thus we believe that this special case warrants its own treatment
and retain the more general case for future work. Despite the importance
of this mixed setting, no other MI estimators have been derived or
analyzed that achieve the parametric MSE convergence rate.

\subsection{Contributions}

In the context of this related work, we make the following novel contributions
in this paper: 
\begin{enumerate}
\item For purely continuous random variables, we derive the asymptotic bias
and variance of kernel density plug-in MI estimators for general MI
measures without boundary correction~\cite{karunamuni2005boundary}
(Section \ref{sec:MI_est}). 
\item We leverage the results for the purely continuous case to derive the
bias and variance of general kernel density plug-in MI estimators
when $\X$ and/or $\Y$ contain a mixture of discrete and continuous
components by reformulating the densities as a mixture of the conditional
density of the continuous variables given the discrete variables (Section
\ref{sec:mixed}). Note that this is a special case of the mixture
setting where discrete and continuous components are separated into
different dimensions.
\item We leverage this theory for the mixed case described above in conjunction
with the generalized theory of ensemble estimators \cite{moon2016arxiv,moon2016isit}
to derive GENIE. To the best of our knowledge, this is the first non-parametric
estimator of general MI measures that achieves a parametric rate of
MSE convergence of $O\left(1/N\right)$ when the densities are sufficiently smooth  for any mixed case (Section
\ref{sec:mixed_ensemble}) let alone the special case we consider,
where $N$ is the number of samples available from each distribution.
\item We prove a minimax lower bound for the convergence rate of MI estimators
in the purely continuous case (Section~\ref{subsec:Minimax}). This
unifies the minimax theory for estimating continuous entropy~\cite{birge1995estimation}
and divergence functionals~\cite{krishnamurthy2014divergence,kandasamy2015nonparametric}.
Neither of these approaches are directly extendable to the MI case
due to the dependence of the marginal distributions on the joint distribution
and the integral relationship between the joint and the marginals.
Therefore, we have tailored the proof to the MI estimation case. We
also show that the MI ensemble estimator achieves the minimax rate
when the densities are sufficiently smooth.
\item We derive a central limit theorem for the ensemble estimators (Section
\ref{subsec:clt}).
\item We apply the method to single-cell RNA-sequencing feature selection
problems (Section \ref{sec:experiments}).
\end{enumerate}
We note that KDE plug-in approaches to estimating functionals such
as entropy and MI are well-known and perhaps the simplest approach~\cite{levit1978asymptotically,goldstein1992optimal}.
Applying the generalized theory of ensemble estimation to the KDE
plug-in estimator does not raise the complexity of the estimators
substantially, either computationally or conceptually. Yet by employing
these simple methods, the resulting ensemble estimator is able to
achieve the minimax convergence rate for sufficiently smooth densities
without employing more complicated von-Mises expansions (as in~\cite{krishnamurthy2014divergence,kandasamy2015nonparametric})
or boundary correction (as in~\cite{krishnamurthy2014divergence,kandasamy2015nonparametric,singh2014exponential,singh2014renyi,sricharan2013ensemble})
to reduce the bias.

\section{Mutual Information Functionals}

We first define a family of MI functionals based on $f$-divergence
functionals which are defined as follows. Let $P$ and $Q$ be probability
measures on the Euclidean space $\mathcal{S}$. Let $g:(0,\infty)\rightarrow\mathbb{R}$ be the $f$-divergence shaping function.
The $f$-divergence functional associated with $g$ is~\cite{ali1966div,csiszar1967div}
\begin{equation}
D_{g}(P||Q):=\bE_{Q}\left[g\left(\frac{dP}{dQ}\right)\right],\label{eq:fdiv}
\end{equation}
where $\frac{dP}{dQ}$ is the Radon-Nikodym derivative and $\bE_{Q}$
indicates the expectation wrt to the measure $Q$. To obtain a true
divergence, we require $g$ to be convex and $g(1)=0.$ However, we
consider more general functionals and so we do not place these restrictions
on $g$.

A generalized MI functional can be derived from (\ref{eq:fdiv}).
Let $\X$ and $\Y$ be (potentially multivariate) random variables
with respective marginal probability measures $P_{X}$ and $P_{Y}$
and joint probability measure $P_{XY}$. Let $g$ be as before. Then
the MI functional associated with $g$ is 
\begin{equation}
I(\X;\Y):=D_{g}\left(\left.P_{X}P_{Y}\right\Vert P_{XY}\right).\label{eq:MI_general}
\end{equation}
Shannon MI can be obtained from (\ref{eq:MI_general}) by setting
$g(t)=-\log t$.

If $\X$ and $\Y$ are purely continuous random variables with respective
marginal probability densities $f_{X}$ and $f_{Y}$ and joint probability
density $f_{XY}$, then (\ref{eq:MI_general}) can be written as 
\begin{equation}
I(\X;\Y)=\int g\left(\frac{f_{X}\left(x\right)f_{Y}\left(y\right)}{f_{XY}\left(x,y\right)}\right)f_{XY}\left(x,y\right)dxdy.\label{eq:MI_cont}
\end{equation}

However, we are also interested in the case where $\X$ or $\Y$ may
have a mixture of discrete and continuous components. In this special
case, the distributions can be factored into a product of the conditional
density and the probability mass functions. The MI can then be expressed
as a sum of integrals which can then be individually estimated. To
do this, denote the continuous and discrete components of $\X$ as
$\X_{C}$ and $\X_{D}$, respectively. Denote $\Y_{C}$ and $\Y_{D}$
similarly. Let $\Z=\left(\mathbf{X},\mathbf{Y}\right)^{T}$ and let
$\Z_{C}$ and $\Z_{D}$ be the respective continuous and discrete
components of $\Z$. Consider the probability distributions $f_{XY}$,
$f_{X}$, $f_{Y}$ and the corresponding densities that are obtained
by conditioning on $\X_{D}$ and $\Y_{D}$, e.g. $f_{XY}\left(x_{C},x_{D},y_{C},y_{D}\right)=f_{X_{C}Y_{C}|X_{D}Y_{D}}\left(x_{C},y_{C}|x_{D},y_{D}\right)f_{X_{D}Y_{D}}\left(x_{D},y_{D}\right)$.
Then after factoring the distributions, (\ref{eq:MI_general}) can
be written as
\begin{align}
I(\X;\Y) & =\sum_{x_{D},y_{D}}\int g\left(\frac{f_{X}\left(x_{C},x_{D}\right)f_{Y}\left(y_{C},y_{D}\right)}{f_{XY}\left(x_{C},x_{D},y_{C},y_{D}\right)}\right)dF_{XY}\left(x_{C},x_{D},y_{C},y_{D}\right)\nonumber \\
 & =\sum_{z_{D}}f_{Z_{D}}\left(z_{D}\right)\int g\left(R_{1}\left(z_{C}\right)R_{2}\left(z_{D}\right)\right)f_{Z_{C}|Z_{D}}\left(z_{C}|z_{D}\right)dz_{C},\label{eq:MI_mixed}
\end{align}
where 
\begin{align*}
R_{1}(z_{C}) & =\frac{f_{X_{C}|X_{D}}\left(x_{C}|x_{D}\right)f_{Y_{C}|Y_{D}}\left(y_{C}|y_{D}\right)}{f_{X_{C}Y_{C}|X_{D}Y_{D}}\left(x_{C},y_{C}|x_{D},y_{D}\right)},\\
R_{2}\left(z_{D}\right) & =\frac{f_{X_{D}}\left(x_{D}\right)f_{Y_{D}}\left(y_{D}\right)}{f_{X_{D}Y_{D}}\left(x_{D},y_{D}\right)}.
\end{align*}
The expression $R_{1}$ is the ratio of the product of the conditional
densities $f_{X_{C}|X_{D}}$ and $f_{Y_{C}|Y_{D}}$ to the conditional
density $f_{X_{C}Y_{C}|X_{D}Y_{D}}$. It is a continuous function
of $z_{C}$. Similarly, the expression $R_{2}$ is the ratio of the
product of the probability mass functions (pmf) $f_{X_{D}}$ and $f_{Y_{D}}$
to the pmf $f_{X_{D}Y_{D}}$ and is a discrete function of $z_{D}$. 

In the following sections, we will obtain MSE convergence rates of
KDE plug-in estimators of the general MI measures described above.
We first focus on the case when $\X$ and $\Y$ are purely continuous
(Equation (\ref{eq:MI_cont})). We then generalize to the case where
$\X$ and $\Y$ may have any mixture of continuous and discrete components
(Equation (\ref{eq:MI_mixed})). The derived convergence rates can
then be used to derive ensemble estimators that achieve the parametric
MSE rate.

\section{Continuous Random Variables}

\label{sec:MI_est}For this section, we define KDE plug-in estimators
of general MI measures under the assumption that $\X$ and $\Y$ are
purely continuous. Thus $\X_{C}=\X$ and $\Y_{C}=\Y$ and we can write 

\begin{equation}
I(\X;\Y)=\int g\left(\frac{f_{X}\left(x\right)f_{Y}\left(y\right)}{f_{XY}\left(x,y\right)}\right)f_{XY}\left(x,y\right)dxdy.\label{eq:MI_cont2}
\end{equation}
We then derive the MSE convergence rate of the KDE plug-in estimator.
We also present a minimax lower bound for MI estimation in this continuous
setting.

To more easily generalize our results to the mixture case, we consider
a modified version of (\ref{eq:MI_cont2}) where the densities are
weighted as follows. Let $\nu$ be a 3-dimensional vector with $0<\nu_{i}\leq1$
for each $i\in\{1,2,3\}$. We can then write 
\begin{equation}
I_{\nu}(\X;\Y)=\int g\left(\frac{f_{X}\left(x\right)f_{Y}\left(y\right)\nu_{1}\nu_{2}}{f_{XY}\left(x,y\right)\nu_{3}}\right)f_{XY}\left(x,y\right)dxdy.\label{eq:MI_weight}
\end{equation}
The expression in (\ref{eq:MI_weight}) reduces to that in (\ref{eq:MI_cont2})
when $\nu_{i}=1$ for each $i\in\{1,2,3\}$. When we generalize to
the mixture case, the pmf estimators will be substituted into $\nu$.

\subsection{The KDE Plug-in Estimator}

Let $f_{X}(x)$, $f_{Y}(y)$, and $f_{XY}(x,y)$ be $d_{X}$, $d_{Y}$,
and $d_{X}+d_{Y}=d$-dimensional probability densities. Since we are
assuming for now that $\mathbf{X}$ and $\mathbf{Y}$ are continuous
with marginal densities $f_{X}$ and $f_{Y}$, the MI functional $I_{v}(\X;\Y)$
can be estimated using KDEs. Assume that $N$ i.i.d. samples $\left\{ \mathbf{Z}_{1},\dots,\mathbf{Z}_{N}\right\} $
are available from the joint density $f_{XY}$ with $\mathbf{Z}_{i}=\left(\mathbf{X}_{i},\mathbf{Y}_{i}\right)^{T}$.
Let $h_{X}$, $h_{Y}$ be kernel bandwidths. Let $K_{X}(\cdot)$ and
$K_{Y}(\cdot)$ be symmetric kernel functions with $\int K_{X}(x)dx=\int K_{Y}(y)dy=1$,
$||K_{X}||_{\infty},\,||K_{Y}||_{\infty}<\infty$ where $||K||_{\infty}=\sup_{x}|K(x)|$.
The KDEs for $f_{X}$, $f_{Y}$, and $f_{XY}=f_{Z}$, respectively,
are 
\begin{eqnarray}
\ft X(x) & = & \frac{1}{Nh_{X}^{d_{X}}}\sum_{i=1}^{N}K_{X}\left(\frac{x-\mathbf{X}_{i}}{h_{X}}\right),\label{eq:fx}\\
\ft Y(y) & = & \frac{1}{Nh_{Y}^{d_{Y}}}\sum_{i=1}^{N}K_{Y}\left(\frac{y-\mathbf{Y}_{i}}{h_{Y}}\right),\label{eq:fy}\\
\ft Z(x,y) & = & \frac{1}{Nh_{X}^{d_{X}}h_{Y}^{d_{Y}}}\sum_{i=1}^{N}K_{X}\left(\frac{x-\mathbf{X}_{i}}{h_{X}}\right)K_{Y}\left(\frac{y-\mathbf{Y}_{i}}{h_{Y}}\right),\label{eq:fz}
\end{eqnarray}
where $h_{Z}=(h_{X},h_{Y})$. Then $I_{\nu}(\X;\Y)$ can be estimated
with a KDE plug-in estimator: 
\begin{equation}
\gt=\frac{1}{N}\sum_{i=1}^{N}g\left(\frac{\ft X(\mathbf{X}_{i})\ft Y(\mathbf{Y}_{i})\nu_{1}\nu_{2}}{\ft Z(\mathbf{X}_{i},\mathbf{Y}_{i})\nu_{3}}\right).\label{eq:Gest}
\end{equation}
Note that in this estimator we evaluate the KDEs at each of the data
points. In practice, this is done using a leave-one-out KDE. This
enables us to avoid evaluating a high-dimensional integral and instead
estimate the integral with the empirical average in eq. (\ref{eq:Gest}).

\subsection{MSE Convergence Rate of the Continuous Plug-in Estimator}

We are interested in the MSE convergence rate of the KDE plug-in estimator
in eq. (\ref{eq:Gest}). The MSE of an estimator can be expressed
as the sum of the squared bias and the variance of the estimator.
We first focus on the bias of the estimator $\gt$. The bias of nonparametric
estimators typically depends on the smoothness of the functions that
are being estimated. In our case, we have multiple functions including
the joint and marginal densities and the function $g$. We quantify
the smoothness of the densities using the H\"{o}lder class $\Sigma(s,H)$:

\begin{definition}[H\"{o}lder Class] \label{def:holder}\emph{Let
$\mathcal{X}\subset\mathbb{R}^{d}$ be a compact space. For $q=(q_{1},\dots,q_{d}),$
$q_{i}\in\mathbb{N},$ define $|q|=\sum_{i=1}^{d}q_{i}$ and $D^{q}=\frac{\partial^{|q|}}{\partial x_{1}^{q_{1}}\dots\partial x_{d}^{q_{d}}}$.
The H\"{o}lder class $\Sigma(s,H)$ of functions on $L_{2}(\mathcal{X})$
consists of the functions $f$ that satisfy 
\[
\left|D^{q}f(x)-D^{q}f(y)\right|\leq H\left\Vert x-y\right\Vert ^{s-|q|},
\]
for all $x,\,y\in\mathcal{X}$ and for all $q$ s.t. $|q|\leq\left\lfloor s\right\rfloor $.
}\end{definition}

A key fact that comes from Definition~\ref{def:holder} is that if
a function $f$ belongs to $\Sigma(s,H),$ then it is $r=\left\lfloor s\right\rfloor $
times differentiable. Given this definition, the full assumptions
we make to derive bias convergence rates are:
\begin{itemize}
\item $(\mathcal{A}.0)$: The kernels $K_{X}$ and $K_{Y}$ are symmetric
product kernels with bounded support. 
\item $(\mathcal{A}.1)$: There exist constants $\epsilon_{0}$, $\epsilon_{\infty}$
such that $0<\epsilon_{0}\leq f_{X}(x)\leq\epsilon_{\infty}<\infty$
$\forall x\in\mathcal{S}_{X},$ $\epsilon_{0}\leq f_{Y}(y)\leq\epsilon_{\infty}$
$\forall y\in\mathcal{S}_{Y}$, and $\epsilon_{0}\leq f_{XY}(x,y)\leq\epsilon_{\infty}$
$\forall(x,y)\in\mathcal{S}_{X}\times\mathcal{S}_{Y}$.
\item $(\mathcal{A}.2)$: Each of the densities belong to $\Sigma(s,H)$
in the interior of their support sets with $s\geq2$.
\item $(\mathcal{A}.3)$: $g(t_{1}/t_{2})$ has an infinite number of mixed
derivatives wrt $t_{1}$ and $t_{2}$. 
\item $(\mathcal{A}.4)$: $\left|\frac{\partial^{k+l}g(t_{1}/t_{2})}{\partial t_{1}^{k}\partial t_{2}^{l}}\right|/(k!l!)$,
$k,l=0,1,\dots$ are strictly upper bounded for $\epsilon_{0}\leq t_{1},t_{2}\leq\epsilon_{\infty}.$
\item $(\mathcal{A}.5)$: Let $K$ be either $K_{X}$ or $K_{Y}$, $\mathcal{S}$
either $\mathcal{S}_{X}$ or $\mathcal{S}_{Y}$, $h$ either $h_{X}$
or $h_{Y}$, $f$ either $f_{X}$ or $f_{Y}$,  and $d$ either $d_{X}$
or $d_{Y}$. Let $q=(q_1,\dots,q_d)$ with $q_i\in\mathbb{N}$ and $|q|\leq r=\left\lfloor s\right\rfloor $. Then we assume for any positive integers $t$ and $l$ that 
\begin{equation}
\int_{x\in\mathcal{S}}\left(\int_{u:K(u)>0,\,x+uh\notin\mathcal{S}}K^{l}(u)u^q D^q f(x) du\right)^{t}dx = v_{t}(h), \label{eq:assume_boundary-1}
\end{equation}
 where $v_{t}(h)$ admits the expansion 
\[
v_{t}(h)=\sum_{i=1}^{r-|q|}e_{i,q,t,l}h^{i}+o\left(h^{r-|q|}\right),
\]
 for some constants $e_{i,q,t,l}$.

 
\end{itemize}

These assumptions can largely be summarized as follows: 1) $f_{X}$,
$f_{Y}$, $f_{XY}$, and $g$ are smooth ($\mathcal{A}.2$-$\mathcal{A}.4$)
; 2) $f_{X}$ and $f_{Y}$ have bounded support sets $\mathcal{S}_{X}$
and $\mathcal{S}_{Y}$ with respective dimensions $d_{X}$ and $d_{Y}$
($\mathcal{A}.1$); 3) $f_{X}$, $f_{Y}$, and $f_{XY}$ are strictly
lower bounded on their support sets ($\mathcal{A}.1$); and 4) the
boundary of the support set is smooth ($\mathcal{A}.5$). More specifically,
assumption $\mathcal{A}.5$ states that the support of the density
is smooth with respect to the kernel $K$ in the sense that the expected value of a polynomial with coefficients consisting of the densities and their derivatives near the boundary is a smooth function of the bandwidth $h$. The inner integral in (\ref{eq:assume_boundary-1}) captures this
expectation while the outer integral averages this inner integral
over all points near the boundary of the support. The $v_{t}(h)$
term captures the fact that the smoothness of this expectation is
proportional to the smoothness of the function $D^q f(x)$. 



While these assumptions may appear highly technical, they are satisfied
for relatively simple support sets and for common kernels, functions
$g$, and densities and thus are widely applicable~\cite{moon2018ensemble,moon2017knn}.
These assumptions are also comparable to those in similar studies
on asymptotic convergence analysis \cite{moon2016isit,moon2014nips,moon2014isit,singh2014renyi,singh2014exponential,sricharan2013ensemble,krishnamurthy2014divergence,kandasamy2015nonparametric,moon2018ensemble}.
Some studies consider the case where the densities are not strictly lower bounded, which makes the problem different~\cite{singh2016finite,han2020optimal} with different minimax rates (see~\cite{han2020optimal} for the entropy estimation case).

In particular, assumption $\mathcal{A}.5$  is satisfied if the kernel $K$ is smooth, has either circular or rectangular support (which includes product kernels), and the density support set consists of the unit cube. See Appendix~\ref{sec:boundary} for details. The unit cube assumption is common in the nonparametric density functional estimation literature~\cite{gao2018demystifying,krishnamurthy2014divergence,singh2014exponential,singh2014renyi,sricharan2013ensemble,moon2017ensemble} as the results can then be extended to density support sets that are isomorphic to the unit cube. 

To derive the convergence rates of many state-of-the art distributional functional estimators, authors commonly assume that the derivatives of the density $f(x)$ vanish near the boundary~\cite{krishnamurthy2014divergence,kandasamy2015nonparametric,singh2014exponential,singh2014renyi,noshad2017direct}. Note that in this assumption, the density itself is not required to vanish near the boundary. Thus densities such as the uniform distribution satisfy this common assumption. However, this assumption is stronger than $\mathcal{A}.5$ as formalized in Proposition~\ref{prop:boundary} below. Our weaker assumption $\mathcal{A}.5$ comes at a small cost as we require the $f$-divergence shaping function $g$ to be infinitely differentiable. In contrast, the authors in~\cite{krishnamurthy2014divergence,kandasamy2015nonparametric,singh2014exponential,singh2014renyi,noshad2017direct} assume that the shaping function has a finite number of derivatives. In practice, this tradeoff does not have a major practical impact as most shaping functions of interest are either infinitely differentiable everywhere (e.g. Shannon and Renyi information) or not differentiable on a set of measure zero (e.g. the total variation distance and the Bayes error rate in the divergence case).
\begin{prop}
\label{prop:boundary}
\label{prop:A5}Let the density support set $\mathcal{S}$ be the unit cube. Let the derivatives of the density $f$ up to order $r$ vanish at the boundary of the density support set. Assume that $||K||_\infty<\infty$ and the support of $K$ is bounded with either rectangular or circular support. Then assumption $\mathcal{A}.5$ is satisfied.
\end{prop}

The proof is given in Appendix~\ref{sub:prop_proof}. The assumption of vanishing density derivatives at the boundary is strictly weaker than assumption $\mathcal{A}.5$. As an example, consider a standard Gaussian distribution truncated to the support $[-1,1]^d$. Clearly, the derivatives of this density do not vanish at the boundary. However, we show in Appendix~\ref{sub:Gauss_truncated} that this density satisfies $\mathcal{A}.5$.

We note that the boundary assumption $\mathcal{A}.5$ does not directly
result in parametric convergence rates for the plug-in estimator $\gt$,
which is in contrast with the boundary assumptions in \cite{singh2014exponential,singh2014renyi,krishnamurthy2014divergence,kandasamy2015nonparametric}.
The estimators in \cite{singh2014exponential,singh2014renyi,krishnamurthy2014divergence,kandasamy2015nonparametric}
perform boundary correction, which requires knowledge of the density
support boundary and complex calculations at the boundary in addition
to the boundary assumptions, to achieve the parametric convergence
rates. In contrast, we use ensemble methods to improve the resulting
convergence rates of $\gt$ without boundary correction, greatly simplifying
our estimator.

\begin{thm}
[Bias Expansion for Continuous $\mathbf{X},\mathbf{Y}$] \label{thm:bias}Under assumptions $\mathcal{A}.0$-$\mathcal{A}.5$,
the bias of $\gt$ is 
\begin{eqnarray}
\bias\left[\gt\right] & = & \sum_{\substack{j=0\\
i+j\neq0
}
}^{r}\sum_{i=0}^{r}c_{10,i,j}\left(\nu_{1}\nu_{2},\nu_{3}\right)h_{X}^{i}h_{Y}^{j}+\frac{c_{11}}{Nh_{X}^{d_{X}}h_{Y}^{d_{Y}}}\nonumber \\
 &  & +O\left(h_{X}^{s}+h_{Y}^{s}+\frac{1}{Nh_{X}^{d_{X}}h_{Y}^{d_{Y}}}\right),\label{eq:bias1}
\end{eqnarray}
where the constants in (\ref{eq:bias1}) are independent of the bandwidths
$h_{X}$ and $h_{Y}$ and depend on the
densities and their derivatives, the functional $g$ and its derivatives,
and the kernels. They also include polynomial terms of $\nu_{1}\nu_{2}$
and $\nu_{3}$ when $\nu_i\neq 1$.
 
\end{thm}
Expressions for
the constants in  (\ref{eq:bias1}) are not given in this paper due to their complexity. These constants are not needed as the bias rates in Theorem~\ref{thm:bias} are sufficient to implement ensemble bias reduction. The resultant ensemble estimator  achieves the parametric MSE convergence
rate $O(1/N)$ (see Section~\ref{sec:mixed_ensemble} for the mixed case and
Appendix~\ref{subsec:cont_ensemble} for the continuous case). 

We also derive a refined expression for the bias that enables
us to achieve the parametric convergence rate under less restrictive
smoothness assumptions on the densities ($s>(d_{X}+d_{Y})/2$ compared
to $s\geq d_{X}+d_{Y}$ for (\ref{eq:bias1})). However, the resulting
expansion has more terms and the ensemble estimator is more complicated to implement. Thus we have chosen
to present the simpler case here. The more complex expansion and estimator are presented
in Appendix~\ref{subsec:Odin2}.

Having obtained an expression for the bias of $\gt$, we now present
an upper bound on its variance to complete the derivation of its MSE.
\begin{thm}
[Variance Bound for Continuous $\mathbf{X}, \mathbf{Y}$] \label{thm:variance}If the functional $g$
is Lipschitz continuous in both of its arguments with Lipschitz constant
$C_{g}$, then the variance of $\gt$ is 
\[
\var\left[\gt\right]\leq\frac{22C_{g}^{2}||K_{X}\cdot K_{Y}||_{\infty}^{2}}{N}.
\]
\end{thm}
The Lipschitz assumption on $g$ for the variance result is comparable
to assumptions made by others for nonparametric estimation of distributional
functionals~\cite{kandasamy2015nonparametric,singh2014exponential,singh2014renyi,moon2016arxiv,krishnamurthy2014divergence}
and is satisfied for Shannon and Renyi informations when the densities
are bounded above and below. Note that Theorem~\ref{thm:variance}
requires much less strict assumptions than Theorem~\ref{thm:bias}.
The proofs of Theorems~\ref{thm:bias} and \ref{thm:variance} are
given in Appendix~\ref{sec:biasProof} and~\ref{sec:VarProof},
respectively.

Theorems~\ref{thm:bias} and \ref{thm:variance} indicate that for
the MSE to go to zero, we require $h_{X},h_{Y}\rightarrow0$ and $Nh_{X}^{d_{X}}h_{Y}^{d_{Y}}\rightarrow\infty$.
In Section~\ref{sec:mixed}, we will use Theorems \ref{thm:bias}
and \ref{thm:variance} to derive bias and variance expressions for
the MI plug-in estimators under the more general cases where $\X$
and/or $\Y$ may contain a mixture of discrete and continuous components.
We will then use these convergence rate results to derive MI ensemble
estimators for both cases (purely continuous random variables and
mixed random variables) that achieve the parametric MSE convergence
rate regardless of the dimension as long as the densities are sufficiently
smooth. 

\subsection{Minimax Rate for MI estimation}

\label{subsec:Minimax}We wrap up this section with a minimax lower
bound on the MSE rate of convergence for the continuous MI estimation
problem. 
\begin{thm}
[Bound on the Minimax Rate for Continuous $\mathbf{X}, \mathbf{Y}$]\label{thm:minimax} Assume that $g$ is
at least twice differentiable and that given $\epsilon>0,$ $\left|g''(\epsilon)\right|>0.$
Define the set of functions $\Sigma\left(s,H,\epsilon_{0},\epsilon_{\infty}\right)$
to be the set of H\"{o}lder continuous functions $\Sigma(s,H)$ that
are bounded between $\epsilon_{0}$ and $\epsilon_{\infty}$. Then
with $\gamma=\min\left\{ 8s/\left(4s+d_{X}+d_{Y}\right),1\right\} $,
there exists a strictly positive constant $c$ such that 
\[
\liminf_{N\rightarrow\infty}\inf_{\hat{G}_{N}}\sup_{f_{XY}\in\Sigma\left(s,H,\epsilon_{0},\epsilon_{\infty}\right)}\bE\left[\left(\hat{G}_{N}-I(\X;\Y)\right)^{2}\right]\geq cN^{-\gamma}.
\]
\end{thm}
The proof uses Le Cam's method~\cite{le2012asymptotic} and is
given in Appendix~\ref{sec:MinimaxProof}. Theorem~\ref{thm:minimax}
indicates that the minimax rate is the parametric rate $N^{-1}$ as
long as $s\geq\left(d_{X}+d_{Y}\right)/4$. This is consistent with
minimax rates for divergence~\cite{krishnamurthy2014divergence,kandasamy2015nonparametric}
and entropy~\cite{birge1995estimation} functional estimation, thus
expanding the previous theory on minimax estimation of information
theoretic functionals. 

In Section~\ref{sec:mixed_ensemble} and Appendix~\ref{sec:extensions},
we derive MI estimators that achieve the minimax rate when $s\geq d_X+d_Y$ and $s>\left(d_{X}+d_{Y}\right)/2$, respectively.
While estimators have been derived for the divergence estimation problem
that achieve the minimax rate for less smooth densities, they require
numerical integration and are thus computationally slow~\cite{krishnamurthy2014divergence,kandasamy2015nonparametric}.
Deriving estimators of these functionals (e.g. MI and divergence)
that are known to achieve the minimax rate in this less smooth regime
and that are computationally reasonable thus remains an open problem.

\section{Mixed Random Variables}

\label{sec:mixed}In this section, we extend the results of Section~\ref{sec:MI_est}
to general MI estimation when $\mathbf{X}$ and $\mathbf{Y}$ may
have a mixture of discrete and continuous components. We focus on
the most complex case: $\X$ and $\Y$ both have discrete and continuous
components. The MI between $\X$ and $\Y$ is written in (\ref{eq:MI_mixed}).

\subsection{KDE Plug-in Estimator}

We first define the KDE plug-in estimator of (\ref{eq:MI_mixed}).
Let $\mathcal{S}_{Y_{C}}$ and $\mathcal{S}_{X_{C}}$ be the respective
supports of the corresponding densities of $\Y_{C}$ and $\X_{C}$
and let $\mathcal{S}_{Y_{D}}$ and $\mathcal{S}_{X_{D}}$ be the respective
supports of the corresponding probability mass functions of $\Y_{D}$
and $\X_{D}$. Suppose we have $N$ i.i.d. samples of $(\X,\Y)$ drawn
from $f_{XY}$ where the $i$th samples are denoted as $\left(\X_{i},\Y_{i}\right)=\left(\X_{i,C},\X_{i,D},\Y_{i,C},\Y_{i,D}\right)$.
Define the following random variables:
\begin{align}
\mathbf{N}_{y} & =\sum_{i=1}^{N}1_{\left\{ \mathbf{Y}_{i,D}=y\right\} ,}\nonumber \\
\N_{x} & =\sum_{i=1}^{N}1_{\left\{ \X_{i,D}=x\right\} ,}\nonumber \\
\N_{xy} & =\sum_{i=1}^{N}1_{\left\{ \X_{i,D}=x,\Y_{i,D}=y\right\} ,}\label{eq:discrete_counts}
\end{align}
where $x\in\mathcal{S}_{X_{D}}$, $y\in\mathcal{S}_{Y_{D}}$, and
$1_{\{\cdot\}}$ is the indicator function. These will be used to
estimate the pmfs of the discrete components of $\X$ and $\Y$.

For the continuous components, we will condition on the discrete components
and construct KDEs for the conditional probability density functions.
Let $\mathcal{S}_{X_{C}}$ and $\mathcal{S}_{Y_{C}}$ be the respective
supports of the marginal densities $f_{X_{C}}$ and $f_{Y_{C}}$ with
corresponding dimensions of $d_{X}$ and $d_{Y}$. As before, let
$K_{X}(\cdot)$ and $K_{Y}(\cdot)$ be kernel functions with $\int K_{X}(x)dx=\int K_{Y}(y)dy=1$,
$||K_{X}||_{\infty},\,||K_{Y}||_{\infty}<\infty$ where $||K||_{\infty}=\sup_{x}|K(x)|$.
Consider the following sets:
\begin{align*}
\mathcal{X}_{x} & =\left\{ \left.\X_{i,C}\in\left\{ \X_{1,C},\dots,\X_{N,C}\right\} \right|\X_{i,D}=x\right\} ,\\
\mathcal{Y}_{y} & =\left\{ \left.\Y_{i,C}\in\left\{ \Y_{1,C},\dots,\Y_{N,C}\right\} \right|\Y_{i,D}=y\right\} .
\end{align*}
The set $\mathcal{X}_{x}$ is the set of the continuous $\X$ data
points where the corresponding discrete component is equal to $x$.
The set $\mathcal{Y}_{y}$ is defined similarly. The KDEs for $f_{X_{C}|X_{D}}$,
$f_{Y_{C}|Y_{D}}$, and $f_{X_{C}Y_{C}|X_{D}Y_{D}}$at $x\in\mathcal{S}_{X_{D}}$
and $y\in\mathcal{S}_{Y_{D}}$ are, respectively, 
\begin{align}
\ft{X_{C}|x}\left(x\right) & =\frac{1}{\N_{x}h_{X_{C}|x}^{d_{X}}}\sum_{\begin{array}{c}
\X_{j,C}\in\mathcal{X}_{x}\\
i\neq j
\end{array}}K_{X}\left(\frac{x-\X_{j,C}}{h_{X_{C}|x}}\right),\nonumber \\
\ft{Y_{C}|y}\left(y\right) & =\frac{1}{\N_{y}h_{Y_{C}|y}^{d_{Y}}}\sum_{\begin{array}{c}
\Y_{j,C}\in\mathcal{Y}_{y}\\
i\neq j
\end{array}}K_{Y}\left(\frac{y-\Y_{j,C}}{h_{Y_{C}|y}}\right),\nonumber \\
\ft{Z_{C}|z}\left(x,y\right) & =\frac{1}{\N_{xy}h_{X_{C}|x}^{d_{X}}h_{Y_{C}|y}^{d_{Y}}}\sum_{\begin{array}{c}
\Y_{j,C}\in\mathcal{Y}_{y}\text{ AND }\X_{j,C}\in\mathcal{X}_{x}\\
i\neq j
\end{array}}K_{X}\left(\frac{x-\X_{j,C}}{h_{X_{C}|x}}\right)K_{Y}\left(\frac{y-\Y_{j,C}}{h_{Y_{C}|y}}\right),\label{eq:cond_KDEs}
\end{align}
where $\Z_{C}=\left(\X_{C},\Y_{C}\right)$ and $h_{Z_{C}|z}=\left(h_{X_{C}|x},h_{Y_{C}|y}\right)$.
Note that we allow the bandwidths to depend on the discrete components
of $\mathbf{X}$ and $\mathbf{Y}$. The reason for this is that the
bandwidth is generally chosen as a function of the number of data
points, which will differ for these conditional distributions as the
discrete components of $\mathbf{X}$ and $\mathbf{Y}$ differ.

The MI $I(\X;\Y)$ can then be estimated by plugging in the conditional
KDEs. Note that the MI in eq.~(\ref{eq:MI_mixed}) is written as
a weighted sum of integral functionals. We therefore first define
an intermediate estimator of the integral functionals:
\[
\g{h_{X_{C}|x},h_{Y_{C}|y}}=\frac{1}{\N_{xy}}\sum_{\X_{C}\in\mathcal{X}_{x}\text{AND}\Y_{C}\in\mathcal{Y}_{y}}g\left(\frac{\ft{X_{C}|x}\left(\X_{C}\right)\ft{Y_{C}|y}\left(\Y_{C}\right)}{\ft{Z_{C}|z}\left(\X_{C},\Y_{C}\right)}\times\frac{\N_{x}\N_{y}}{N\N_{xy}}\right).
\]
Again in practice, we evaluate the KDEs at each of the data points
using a leave-one-out KDE, enabling us to avoid evaluating a high-dimensional
integral. We then define a plug-in KDE estimator of $I(\X;\Y)$:
\begin{equation}
\g{h_{X_{C}|X_{D}},h_{Y_{C}|Y_{D}}}=\sum_{x\in\mathcal{S}_{X_{D}},y\in\mathcal{S}_{Y_{D}}}\frac{\N_{xy}}{N}\g{h_{X_{C}|x},h_{Y_{C}|y}}.\label{eq:mixed_est}
\end{equation}

The quality of the conditional density estimates in terms of bias
and variance depends on the choice of bandwidths $h_{X_{C}|x}$ and
$h_{Y_{C}|y}$. That is, for the KDE $\ft{X_{C}|x}$ to converge in
MSE, it is necessary that $h_{X_{C}|x}\rightarrow0$ and $\N_{x}h_{X_{C}|x}^{d_{X}}\rightarrow0$
as $\N_{x}\rightarrow\infty$ (a similar result holds for $h_{Y_{C}|y}$)~\cite{hansen2009lecture}.
Furthermore, we will see when we derive the bias and variance of $\g{h_{X_{C}|X_{D}},h_{Y_{C}|Y_{D}}}$
that these conditions are also necessary for $\g{h_{X_{C}|X_{D}},h_{Y_{C}|Y_{D}}}$
to converge in MSE. Thus, when deriving the MSE convergence rate of
$\g{h_{X_{C}|X_{D}},h_{Y_{C}|Y_{D}}}$, we will assume that $h_{X_{C}|x}$
is a function of $\N_{x}$ and $h_{Y_{C}|y}$ is a function of $\N_{y}$.

\subsection{MSE Convergence Rates of the Mixed Plug-in Estimator}

\label{subsec:mixed_conv}Here we derive the MSE convergence rate
of a plug-in estimator of MI when the random variables have a mixture
of discrete and continuous components. We will need the following:
\begin{lem}
\label{lem:binom_fractional}Let $\N_{y}$, $\N_{x}$, and $\N_{xy}$
be defined as in (\ref{eq:discrete_counts}). Assume that their corresponding
probability mass functions are bounded away from zero. If $\alpha\in\mathbb{R}\backslash\{0,1\}$
and $\lambda+\beta+\gamma\in\mathbb{R}\backslash\{0,1\}$, then 
\begin{align}
\bE\left[\N_{xy}^{\alpha}\right] & =\left(Nf_{X_{D}Y_{D}}\left(x,y\right)\right)^{\alpha}+O\left(N^{\alpha-1}\right)\label{eq:frac_moment}\\
\bE\left[\N_{xy}^{\lambda}\N_{x}^{\beta}\N_{y}^{\gamma}\right] & =N^{\lambda+\beta+\gamma}\left(f_{X_{D}Y_{D}}\left(x,y\right)\right)^{\lambda}\left(f_{X_{D}}(x)\right)^{\beta}\left(f_{Y_{D}}(y)\right)^{\gamma}+O\left(N^{\lambda+\beta+\gamma-1}\right).\label{eq:frac_combined}
\end{align}
\end{lem}
The proof is in Appendix~\ref{subsec:fracMomentProof} and uses the
generalized binomial theorem, Taylor series expansions, and known
results about the central moments of binomial random variables~\cite{riordan1937moment}.
Lemma~\ref{lem:binom_fractional} provides key results on moments
of products of the binomial random variables $\N_{xy}$, $\N_{x}$,
and $\N_{y}$. These results can be used to derive the bias and variance
of a plug-in estimator of MI with mixed components in (\ref{eq:MI_mixed})
as long as the bias and variance of the corresponding plug-in estimator
for the continuous weighted case in (\ref{eq:MI_weight}) is known.
This is demonstrated in the following theorems for the KDE plug-in
estimator $\g{h_{X_{C}|X_{D}},h_{Y_{C}|Y_{D}}}$.
\begin{thm}
[Bias Expansion for Mixed $\mathbf{X},\mathbf{Y}$]\label{thm:bias_mixed} Assume that assumptions $\mathcal{A}.0$-$\mathcal{A}.5$
hold with respect to the functional $g$, the kernels $K_{X}$ and
$K_{Y}$, and the densities $f_{X_{C}|X_{D}}$, $f_{Y_{C}|Y_{D}}$
and $f_{X_{C}Y_{C}|X_{D}Y_{D}}$. Assume that $\left|\mathcal{S}_{X_{D}}\right|,\left|\mathcal{S}_{Y_{D}}\right|<\infty$.
Assume that $\hx=l_{X}\mathbf{N}_{x}^{-\beta}$ and $\hy=l_{Y}\N_{y}^{-\alpha}$
with $0<\beta<\frac{1}{d_{X}}$, $0<\alpha<\frac{1}{d_{Y}}$, and
$l_{X},l_{Y}>0$. Then the bias of $\g{h_{X_{C}|X_{D}},h_{Y_{C}|Y_{D}}}$
is 
\begin{align}
\bias\left[\g{h_{X_{C}|X_{D}},h_{Y_{C}|Y_{D}}}\right] & =\sum_{\substack{i,j=0\\
i+j\neq0
}
}^{r}c_{13,i,j}l_{X}^{i}l_{Y}^{j}N^{-i\beta-j\alpha}+O\left(N^{-s\alpha}+N^{-s\beta}+N^{\beta d_{X}+\alpha d_{Y}-1}\right).\label{eq:bias_mixed1}
\end{align}
\end{thm}
The constants depend on the underlying densities, the chosen kernels,
the functional $g$, and the probability mass functions and are independent
of $l_{X}$, $l_{Y}$, and $N$. Furthermore, these rates are asymptotically
tight.
\begin{thm}
[Variance Bound for Mixed $\mathbf{X},\mathbf{Y}$]\label{thm:var_mixed} Assume that $\hx=l_{X}\mathbf{N}_{x}^{-\beta}$
and $\hy=l_{Y}\N_{y}^{-\alpha}$ with $0<\beta<\frac{1}{d_{X}}$,
$0<\alpha<\frac{1}{d_{Y}}$, $\beta d_{X}+\alpha d_{Y}\leq1$, and
$l_{X},l_{Y}>0$. Assume that $\left|\mathcal{S}_{X_{D}}\right|,\left|\mathcal{S}_{Y_{D}}\right|<\infty$.
If the shaping function $g$ is Lipschitz continuous in both of its arguments,
then the variance of $\g{h_{X_{C}|X_{D}},h_{Y_{C}|Y_{D}}}$ is $O(1/N)$.
\end{thm}
These theorems provide the necessary information for applying the
theory of optimally weighted ensemble estimation to obtain MI estimators
with improved rates (see Section \ref{sec:mixed_ensemble}). 

\subsection{Proof Sketches of Theorems \ref{thm:bias_mixed} and \ref{thm:var_mixed}}

For Theorem~\ref{thm:bias_mixed}, the proof splits the bias term
into two terms by adding and subtracting $g\left(\mathcal{T}(\X,\Y)\frac{\N_{x}\N_{y}}{N\N_{xy}}\right)$
for each pair $(x,y)$ where $\mathcal{T}(\X,\Y)$ is independent
of the data samples and is defined in Eq. (\ref{eq:ratio}). It can
be shown that the newly added term has bias $O(1/N)$. The other term
is handled by conditioning on the discrete components of the data
samples to obtain the conditional bias terms $\bias\left[\left.\g{h_{X_{C}|x},h_{Y_{C}|y}}\right|\X_{1,D},\dots,\X_{N,D},\Y_{1,D},\dots,\Y_{N,D}\right]$
for each pair $(x,y)$. Theorem~\ref{thm:bias} can then be applied
to each of these terms to obtain expressions of the random variables
$\N_{x}$, $\N_{y}$, and $\N_{xy}$ with terms of the form given
in Lemma~\ref{lem:binom_fractional}. Lemma~\ref{lem:binom_fractional}
can be applied to these terms to obtain the final result, where care
is taken to ensure that all relevant terms have been handled properly.
The full proof is given in Appendix~\ref{subsec:bias_mixed_proof}.

To prove Theorem~\ref{thm:var_mixed}, we use the law of total variance
to split the variance into two terms: the expected value of the variance
conditioned on the discrete components of the data samples and the
variance of the conditional expectation. Theorem~\ref{thm:variance}
is applied to the conditional variance term. For the conditional expectation
term, we use results obtained in the proof of Theorem~\ref{thm:bias_mixed}
combined with the Efron-Stein inequality~\cite{efron1981jackknife}
to obtain expressions of the random variables $\N_{x}$, $\N_{y}$,
and $\N_{xy}$. Lemma~\ref{lem:binom_fractional} can be applied
again to these terms to obtain the final result. The full proof is
given in Appendix~\ref{subsec:varMixedProof}.

\section{Ensemble Estimation of Generalized MI}

\label{sec:mixed_ensemble} 

If no bias correction is performed, then Theorems~\ref{thm:bias}
and \ref{thm:bias_mixed} show that the optimal bias rate of the KDE
plug-in estimators $\gt$ and $\g{h_{X_{C}|X_{D}},h_{Y_{C}|Y_{D}}}$
is $O\left(1/N^{1/(d_{X}+d_{Y}+1)}\right)$, which converges very
slowly to zero when either $d_{X}$ or $d_{Y}$ are not small. Thus
the standard KDE plug-in estimators will perform poorly in  higher-dimensional
settings. We use the theory of optimally weighted ensemble estimation
developed in~\cite{moon2018ensemble} to improve this rate. For brevity,
we focus on the case where $\X$ and $\Y$ both contain a mixture
of discrete and continuous components. The purely continuous case
is described in Appendix~\ref{subsec:cont_ensemble}.

An ensemble of estimators is first formed by choosing different bandwidth
values for the plug-in estimators as follows. Let $\mathcal{L}$ be
a set of real positive numbers with $|\mathcal{L}|=L<\infty$. This
set will parameterize the bandwidths $\hx$ and $\hy$ for $\ft{X_{C}|x}$
and $\ft{Y_{C}|y}$, respectively, resulting in $L$ estimators in
the ensemble. In other words, we set $\hx(l)=l\N_{x}^{-\beta}$ and
$\hy(l)=l\N_{y}^{-\alpha}$ with $l\in\mathcal{L}$. While different
parameter sets for $\hx$ and $\hy$ can be chosen, we only use one
set here for simplicity of exposition. To achieve the parametric rate,
we need to ensure that the final terms in (\ref{eq:bias_mixed1})
are $O(1/\sqrt{N})$. Thus we require the following conditions to
be met: 
\begin{align*}
s\alpha & \geq\frac{1}{2},\\
s\beta & \geq\frac{1}{2},\\
1-\beta d_{X}-\alpha d_{Y} & \geq\frac{1}{2}.
\end{align*}
For all of these conditions to hold, it is necessary that $s\geq d_{X}+d_{Y}$.
Thus for each estimator in the ensemble we choose $\hx(l)=l\N_{x}^{-1/\left(2\left(d_{X}+d_{Y}\right)\right)}$
and $\hy(l)=l\N_{y}^{-1/\left(2\left(d_{X}+d_{Y}\right)\right)}$
where $l\in\mathcal{L}$. Define $w$ to be a weight vector parameterized
by $l\in\mathcal{L}$ with $\sum_{l\in\mathcal{L}}w(l)=1$ and define
\begin{equation}
\g w=\sum_{l\in\mathcal{L}}w(l)\sum_{x\in\mathcal{S}_{X_{D}},y\in\mathcal{S}_{Y_{D}}}\frac{\N_{xy}}{N}\g{h_{X_{C}|x}(l),h_{Y_{C}|y}(l)}.\label{eq:ensemble}
\end{equation}
This is the weighted ensemble estimator. From Theorem \ref{thm:bias_mixed},
the bias of $\g w$ is 
\begin{align}
\bias\left[\g w\right] & =\sum_{l\in\mathcal{L}}\sum_{i=1}^{r}\theta\left(w(l)l^{i}N^{\frac{-i}{2\left(d_{X}+d_{Y}\right)}}\right)\nonumber \\
 & +O\left(\sqrt{L}||w||_{2}\left(N^{\frac{-s}{2\left(d_{X}+d_{Y}\right)}}+N^{\frac{-1}{2}}\right)\right),\label{eq:weight_bias}
\end{align}
where we use $\theta$ notation to omit the constants. 

We use the general theory of optimally weighted ensemble estimation
in \cite{moon2016isit,moon2018ensemble} to improve the MSE convergence
rate of the plug-in estimator by choosing the appropriate weights
to cancel the lower order terms in (\ref{eq:weight_bias}): 
\begin{thm}
[Ensemble MSE]\label{thm:ensemble} Let $\mathcal{L}$ be a set of
real positive numbers with $|\mathcal{L}|=L<\infty$ and let $J=\{1,2,\dots,d_{X}+d_{Y}\}$.
Assume the same conditions in Theorems~\ref{thm:bias_mixed} and
\ref{thm:var_mixed} hold with $\hx(l)=l\N_{x}^{-1/\left(2\left(d_{X}+d_{Y}\right)\right)}$
and $\hy(l)=l\N_{y}^{-1/\left(2\left(d_{X}+d_{Y}\right)\right)}$.
Assume that $s\geq d_{X}+d_{Y}$ and define $\g w$ as in (\ref{eq:ensemble}).
Then the MSE of $\g{w_{0}}$ attains the parametric rate of convergence
of $O\left(1/N\right)$ where $w_{0}$ is the solution to the following
offline convex optimization problem:\emph{
\begin{equation}
\begin{array}{rl}
\min_{w} & ||w||_{2}\\
\text{subject to} & \sum_{l\in\mathcal{L}}w(l)=1,\\
 & \sum_{l\in\mathcal{L}}w(l)l^{i}=0,\,i\in J.
\end{array}\label{eq:optimize}
\end{equation}
}
\end{thm}
To summarize, if the weights are chosen using eq. (\ref{eq:optimize}),
then the weighted ensemble estimator $\g{w_{0}}$ achieves the parametric
MSE rate. In practice, the optimization problem in (\ref{eq:optimize})
typically results in a very large increase in variance for finite
samples. Thus we use a relaxed version of (\ref{eq:optimize}): 
\begin{equation}
\begin{array}{rl}
\min_{w} & \epsilon\\
\text{subject to} & \sum_{l\in\mathcal{L}}w(l)=1,\\
 & \left|\sum_{l\in\mathcal{L}}w(l)l^{i}N^{\frac{1}{2}-\frac{i}{2\left(d_{X}+d_{Y}\right)}}\right|\leq\epsilon,\,\,i\in J,\\
 & \left\Vert w\right\Vert _{2}^{2}\leq\eta\epsilon.
\end{array}\label{eq:relaxed}
\end{equation}
The parameter $\eta$ is chosen to achieve a trade-off between bias
and variance. As shown in \cite{wisler2018direct,moon2018ensemble},
the ensemble estimator $\g{w_{0}}$ using the resulting weight vector
from the optimization problem in (\ref{eq:relaxed}) still achieves
the parametric MSE convergence rate under the same assumptions as
described previously. We denote this estimator as $\g{GENIE}$. Algorithm~\ref{alg:estimator}
summarizes the estimator $\g{GENIE}$.

\begin{algorithm}
\begin{algorithmic}[1]
\renewcommand{\algorithmicrequire}{\textbf{Input:}} \renewcommand{\algorithmicensure}{\textbf{Output:}}

\REQUIRE $L$ positive real numbers $\mathcal{L}$, samples $\left\{ \mathbf{Z}_{1},\dots,\mathbf{Z}_{N}\right\} $
from $f_{XY}$, dimensions $d_{X}$ and $d_{Y}$, function $g$, kernels
$K_{X}$ and $K_{Y}$

\ENSURE The optimally weighted MI estimator $\g{GENIE}$

\STATE Solve for $w_{0}$ using (\ref{eq:relaxed}) 

\FORALL{$l\in\mathcal{L}$ and $(x,y)\in \mathcal{S}_{X_D}\times\mathcal{S}_{Y_D}$}

\STATE Calculate $\N_{xy}$, $\N_{x}$, and $\N_{y}$ as in (\ref{eq:discrete_counts})

\STATE$\hx(l)\leftarrow l\N_{x}^{-1/\left(2\left(d_{X}+d_{Y}\right)\right)}$,
$\hy(l)\leftarrow l\N_{y}^{-1/\left(2\left(d_{X}+d_{Y}\right)\right)}$ 

\STATE$\mathcal{X}_{x}\leftarrow\left\{ \left.\X_{i,C}\in\left\{ \X_{1,C},\dots,\X_{N,C}\right\} \right|\X_{i,D}=x\right\} ,$
$\mathcal{Y}_{y}\leftarrow\left\{ \left.\Y_{i,C}\in\left\{ \Y_{1,C},\dots,\Y_{N,C}\right\} \right|\Y_{i,D}=x\right\} .$

\FOR{$\mathbf{Z}_{i,C}=\left( \X_{i,C},\Y_{i,C}\right)\in\mathcal{X}_x\times\mathcal{Y}_y$}

\STATE Calculate $\ftl{X_{C}|x}\left(\X_{i,C}\right)$, $\ftl{Y_{C}|y}\left(\Y_{i,C}\right)$,
and $\ftl{Z_{C}|z}\left(\Z_{i,C}\right)$ as described in (\ref{eq:cond_KDEs})

\ENDFOR 

\STATE $\g{h_{X_{C}|x}(l),h_{Y_{C}|y}(l)}\leftarrow\frac{1}{\N_{xy}}\sum_{\X_{C}\in\mathcal{X}_{x}\text{ AND }\Y_{C}\in\mathcal{Y}_{y}}g\left(\frac{\ftl{X_{C}|x}\left(\X_{C}\right)\ftl{Y_{C}|y}\left(\Y_{C}\right)}{\ftl{Z_{C}|z}\left(\X_{C},\Y_{C}\right)}\times\frac{\N_{x}\N_{y}}{N\N_{xy}}\right)$

\ENDFOR 

\STATE $\g{GENIE}=\sum_{l\in\mathcal{L}}w_{0}(l)\sum_{x\in\mathcal{S}_{X_{D}},y\in\mathcal{S}_{Y_{D}}}\frac{\N_{xy}}{N}\g{h_{X_{C}|x}(l),h_{Y_{C}|y}(l)}.$

\end{algorithmic}

\caption{Optimally weighted KDE ensemble MI estimator $\protect\g{GENIE}$
\label{alg:estimator}}
\end{algorithm}
A similar approach can be used to derive an ensemble estimator for
the case when $\X$ and $\Y$ are purely continuous. Furthermore,
we can define ensemble estimators for both the continuous and the
mixed cases that achieve the parametric MSE rate if $s>\left(d_{X}+d_{Y}\right)/2$,
although the optimization problem is more complicated. See Appendix~\ref{sec:extensions}
for details.

The weights obtained in (\ref{eq:relaxed}) are optimal in two senses.
First, they are the optimal solution to the problem in (\ref{eq:relaxed}).
This contrasts with other popular ensemble methods such as random
forests, where the ensemble of learners are equally weighted, and
AdaBoost, where the weights are assigned to different regions of the
feature space based on the training data. The weights are also optimal
in an asymptotic sense. It can be shown that the variance of the ensemble
estimator is bounded by a multiple of $||w||_{2}$ \cite{moon2018ensemble,moon2016isit}.
By minimizing the norm of the weights (or an upper bound on it), we
choose a weight vector that reduces the bias (due to the constraints)
while controlling the variance. Thus the weights are also optimal
in the sense that the bias is reduced to the parametric rate while
the variance is controlled as much as possible given the information
that we have. Since the parametric rate is minimax optimal, this is
also asymptotically optimal for sufficiently smooth densities.

We note that the ensemble estimation approach given here can be compared
to the Jackknife bias correction method~\cite{cameron2005microeconometrics,efron1981jackknife}.
Both approaches use a linear combination of estimators to obtain a
less-biased estimator. However, the standard Jackknife approach uses uniform
weights for the linear combination while the ensemble approach presented
here obtains weights from an optimization problem. This results in
a more computationally efficient procedure as only $L$ estimators
are required for the ensemble approach where $L$ is on the order
of 30-50. The weights can also be computed offline and so solving
the optimization problem contributes little to the total computation
time. In contrast, the standard Jackknife approach requires $N$ different
estimators which is less computationally efficient. 

A more general Jackknife approach such as that in~\cite{jiao2020bias} shares more similarities with our ensemble method. In this particular work, the authors similarly compute the weights based on an asymptotic bias expansion. However, they do not control the variance via the norm of the weights as we do. Additionally, the Jackknife approach uses a linear combination of estimators with different samples sizes while we use estimators with different bandwidths. Finally, this general Jackknife approach is still more computationally intensive than our ensemble method which computes the weights offline.

At first glance, the weighted ensemble approach discussed in this
section appears to be quite similar to the optimal kernel approaches
used in~\cite{krishnamurthy2014divergence,kandasamy2015nonparametric,singh2014exponential,singh2014renyi}.
However, the weighted ensemble estimation theory we use is applied
to an ensemble of MI estimators after plugging in an ensemble of KDEs
with different bandwidths. So in some sense, we are optimizing the
ensemble of kernels (whose shape is determined by the bandwidth and
the fixed kernel) for the MI estimation problem. In contrast, the
optimal KDE approach first optimizes the kernel for the KDE problem,
and then plugs in the optimized KDE for MI estimation. It is possible
that a proper modification of the ensemble estimation theory could
be applied to a KDE to obtain an optimal KDE and unify these approaches.
This extension is left for future work.

\subsection{Parameter Selection}

Asymptotically, the theoretical results of the previous sections hold
for any choice of the bandwidth vectors as determined by $\mathcal{L}$.
In practice, we find that the following rules-of-thumb for tuning
the parameters lead to high-quality estimates in the finite sample
regime. 
\begin{enumerate}
\item Select the minimum and maximum bandwidth parameter to produce density
estimates that satisfy the following: first the minimum bandwidth
should not lead to a zero-valued density estimate at any sample point;
second the maximum bandwidth should be smaller than the diameter of
the support. 
\item Ensure the bandwidths are sufficiently distinct. Similar bandwidth
values lead to a negligible decrease in the bias and many bandwidth
values may increase $||w_{0}||_{2}$ resulting in an increase in variance
\cite{sricharan2013ensemble,moon2018ensemble}.
\item Select $L=|\mathcal{L}|>|J|=I$ to obtain a feasible solution for
the optimization problems in (\ref{eq:optimize}) and (\ref{eq:relaxed}).
We find that choosing a value of $30\leq L\leq60$, and setting $\mathcal{L}$
to be $L$ linearly spaced values between the minimum and maximum
values described above works well in practice. 
\end{enumerate}
The resulting ensemble estimators are robust in the sense that they
are not sensitive to the exact choice of the bandwidths or the number
of estimators as long as the the rough rules-of-thumb given above
are followed. Moon et al. \cite{moon2018ensemble} gives more details
on ensemble estimator parameter selection for continuous divergence
estimation. These details also apply to the continuous parts of the
mixed cases for MI estimation in this paper. In particular, the minimum
and maximum bandwidth parameters can be efficiently selected based
on the $k$ nearest neighbor distances of all data points.

Since the optimal weight $w_{0}$ can be calculated offline, the computational
complexity of the estimators is dominated by the construction of the
KDEs which has a complexity of $O\left(N^{2}\right)$ using the standard
implementation. For very large datasets, more efficient KDE implementations
(e.g. \cite{raykar2010fast}) can be used to reduce the computational
burden. 

\subsection{Central Limit Theorem}

\label{subsec:clt}We finish this section with central limit theorems
for the ensemble estimators. This enables us to perform hypothesis
testing on the MI measure.
\begin{thm}
[CLT for Continuous $\mathbf{X},\mathbf{Y}$]\label{thm:clt} Let $\g w^{cont}$ be a weighted
KDE ensemble estimator of $I_{\nu}(\X;\Y)$ when $\X$ and $\Y$ are
continuous with bandwidths $h_{X}(l)$ and $h_{Y}(l)$ for each estimator
in the ensemble. Assume that the shaping function $g$ is Lipschitz in both
arguments with Lipschitz constant $C_{g}$ and that $h_{X}(l),\,h_{Y}(l)\rightarrow0$,
$N\rightarrow\infty$, and $Nh_{X}^{d_{X}}(l),\,Nh_{Y}^{d_{Y}}(l)\rightarrow\infty$
for each $l\in\mathcal{L}$. Then for fixed $\mathcal{L}$, and if
$\mathbf{S}$ is a standard normal random variable, 
\[
\Pr\left(\left(\g w^{cont}-\bE\left[\g w^{cont}\right]\right)/\sqrt{\var\left[\g w^{cont}\right]}\leq t\right)\rightarrow\Pr\left(\mathbf{S}\leq t\right).
\]
\end{thm}
The proof is based on an application of Slutsky's Theorem preceded
by an application of the Efron-Stein inequality (see Appendix \ref{sec:cltProof}).

For the mixed component case, if $\mathcal{S}_{X}$ and $\mathcal{S}_{Y}$
are finite, then the corresponding ensemble estimators also obey a
central limit theorem. The proof follows by an application of Slutsky's
Theorem combined with Theorem~\ref{thm:clt}.
\begin{cor}
[CLT for mixed $\mathbf{X},\mathbf{Y}$]\label{cor:clt} Let $\g w$ be a weighted KDE ensemble
estimator of $I(\X;\Y)$ when $\X$ and $\Y$ contain both continuous
and discrete components. Let the bandwidths for the conditional estimators
be $\hx(l)$ and $\hy(l)$ for each estimator in the ensemble. Assume
that the shaping function $g$ is Lipschitz in both arguments and that $\hx,\,\hy\rightarrow0$,
$N\rightarrow\infty$, and $Nh_{X}^{d_{X}},\,Nh_{X|y}^{d_{X}}\rightarrow\infty$
for each $l\in\mathcal{L}$ and $\forall(x,y)\in\mathcal{S}_{X_{D}}\times\mathcal{S}_{Y_{D}}$
with $\left|\mathcal{S}_{X_{D}}\right|,\left|\mathcal{S}_{Y_{D}}\right|<\infty$.
Then for fixed $\mathcal{L}$, 
\[
\Pr\left(\left(\g w-\bE\left[\g w\right]\right)/\sqrt{\var\left[\g w\right]}\leq t\right)\rightarrow\Pr\left(\mathbf{S}\leq t\right).
\]
\end{cor}

\section{Applications}

\label{sec:experiments}

\subsection{Simulations}

In this section, we validate our theory by estimating the R\'{e}nyi-$\alpha$
MI integral (i.e. $g(x)=x^{\alpha}$ in (\ref{eq:MI_cont}); see~\cite{principe2010information})
where $\mathbf{X}$ is a mixture of truncated Gaussian random variables
restricted to the unit cube and $\mathbf{Y}$ is a categorical random
variable that indicates the corresponding truncated Gaussian random
variable that $\X$ is drawn from in the mixture. In this setting,
$\Y$ can be viewed as a classification variable and $\X$ contains
the chosen features, which are all continuous in this case. Since
$\X$ is purely continuous and $\Y$ is purely discrete, the MI integral
reduces to the following: 
\[
I\left(\X;\Y\right)=\sum_{y\in S_{Y}}f_{Y_{D}}(y)\int\left(\frac{f_{X_{C}}\left(x_{C}\right)}{f_{X_{C}|Y_{D}}\left(x_{C}|y\right)}\right)^{\alpha}f_{X_{C}|Y_{D}}\left(x_{C}|y\right)dx_{C}.
\]

We illustrate with R\'{e}nyi MI as it has received recent interest and the
estimation problem does not reduce to entropy estimation, in contrast
to Shannon MI. Thus this is a clear case where there are no other
nonparametric estimators that are known to achieve the parametric
MSE rate. In fact, to the best of our knowledge, there are no other
nonparametric estimators of R\'{e}nyi MI that are known to be consistent
in this mixed setting.

We consider two cases. In the first case, $\mathbf{Y}$ has three
possible outcomes (i.e. $|\mathcal{S}_{Y}|=3$) and respective probabilities
$\Pr(\mathbf{Y}=0)=\Pr(\mathbf{Y}=1)=2/5$ and $\Pr(\mathbf{Y}=2)=1/5$.
The conditional covariance matrices are all $0.1\times I_{d}$ and
the conditional means are, respectively, $\bar{\mu}_{0}=0.25\times\bar{1}_{d}$,
$\bar{\mu}_{1}=0.75\times\bar{1}_{d}$, and $\bar{\mu}_{2}=0.5\times\bar{1}_{d}$,
where $I_{d}$ is the $d\times d$ identity matrix and $\bar{1}_{d}$
is a $d$-dimensional vector of ones. This experiment can be viewed
as the problem of estimating MI (e.g. for feature selection or Bayes
error bounds) of a classification problem where each discrete value
corresponds to a distinct class, the distribution of each class overlaps
slightly with others, and the class probabilities are unequal. We
use $\alpha=0.5$. We set $\mathcal{L}$ to be 40 linearly spaced
values between 1.2 and 3. The bandwidth in the KDE plug-in estimator
is also set to $2.1N^{-1/(2d)}$. 

\begin{figure}
\centering

\includegraphics[width=0.5\textwidth]{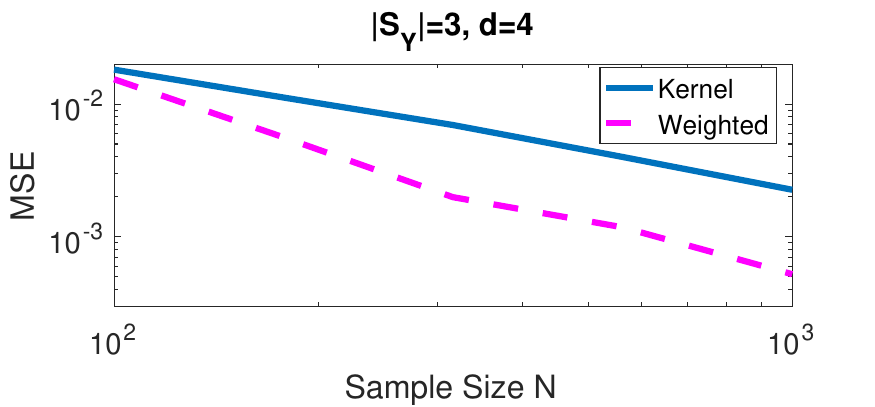}\includegraphics[width=0.5\textwidth]{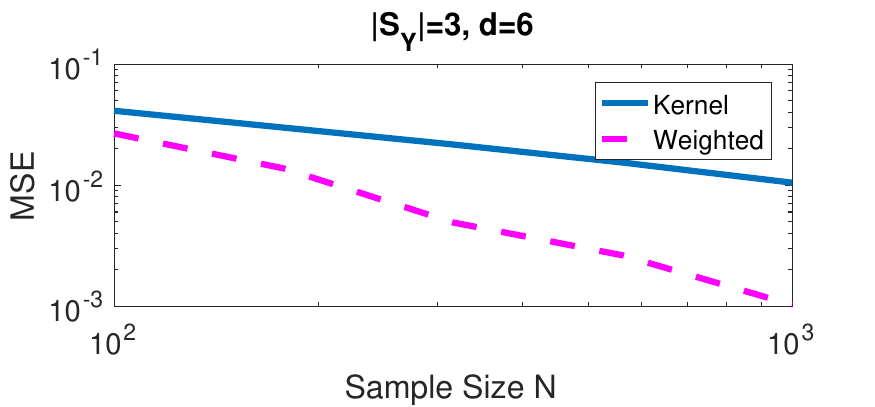}

\includegraphics[width=0.5\textwidth]{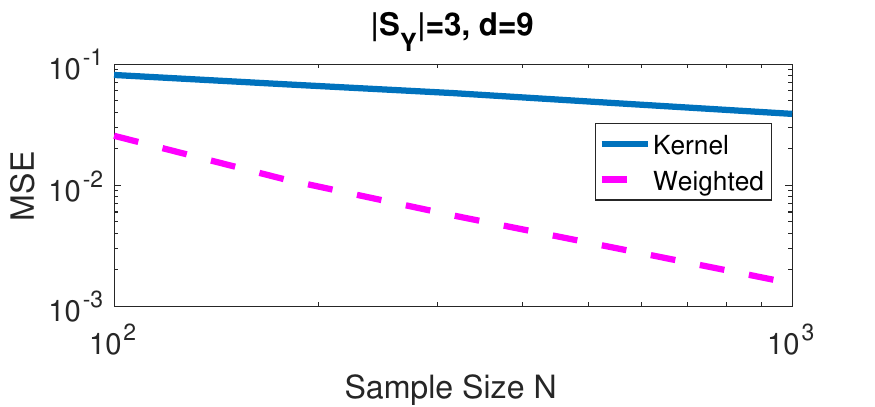}\includegraphics[width=0.5\textwidth]{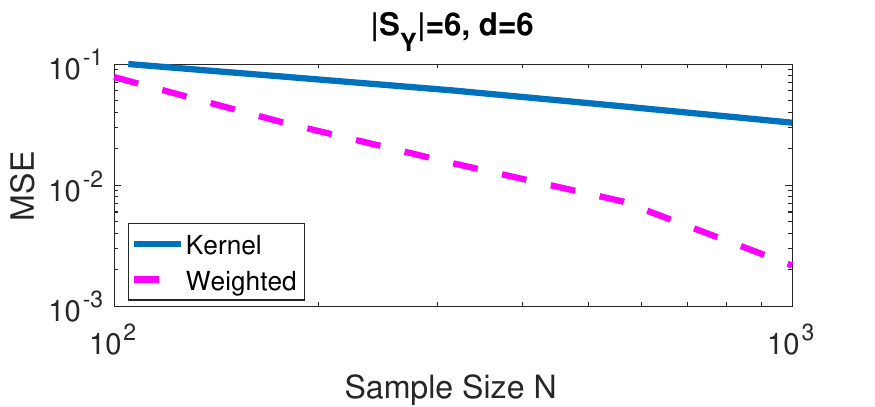}

\caption{MSE log-log plots as a function of sample size for the uniform kernel
plug-in MI estimator (\textquotedbl Kernel\textquotedbl ) and the
proposed optimally weighted ensemble estimator $\protect\g{GENIE}$
(\textquotedbl Weighted\textquotedbl ) for the distributions described
in the text. The ensemble estimator outperforms the kernel plug-in
estimator, especially for larger sample sizes. Note also that as the
dimension increases, the performance gap between the two estimators
increases.\label{fig:mseplot}}
\end{figure}

Figure~\ref{fig:mseplot} shows the MSE (200 trials) of the plug-in
KDE estimator of the MI integral using a uniform kernel and the optimally
weighted ensemble estimator $\g{GENIE}$ for various sample sizes
and for $d=4,\,6,\,9$, respectively. The ensemble estimator GENIE
outperforms the standard plug-in estimator, especially for larger
sample sizes and larger dimensions. This demonstrates that while an
individual kernel estimator performs poorly, an ensemble of estimators
including the individual estimator performs well.

For the second case, $\mathbf{Y}$ has six possible outcomes (i.e.
$|\mathcal{S}_{Y}|=6$) and respective probabilities $\Pr(\Y=0)=0.35$,
$\Pr(\Y=1)=0.2$, $\Pr(\Y=2)=\Pr(\Y=3)=0.15$, $\Pr(\Y=4)=0.1$, and
$\Pr(\Y=5)=0.05$. We chose $\alpha=0.5$ and $d=6$. The conditional
covariance matrices are again $0.1\times I_{d}$ and the conditional
means are, respectively, $\bar{\mu}_{0}=0.25\times\bar{1}_{d}$, $\bar{\mu}_{1}=0.75\times\bar{1}_{d}$,
and $\bar{\mu}_{2}=0.5\times\bar{1}_{d}$, $\bar{\mu}_{3}=\left(0.25\times\bar{1}_{4}^{T},0.5\times\bar{1}_{2}^{T}\right)^{T}$,
$\bar{\mu}_{4}=\left(0.75\times\bar{1}_{2}^{T},0.375\times\bar{1}_{4}^{T}\right)^{T}$,
and $\bar{\mu}_{5}=\left(0.5\times\bar{1}_{4}^{T},0.25\times\bar{1}_{2}^{T}\right)^{T}$.
The results are again given in Figure~\ref{fig:mseplot}. The parameters
for the ensemble estimator and the KDE plug-in estimators are the
same as in the other three plots in Figure \ref{fig:mseplot}. The
ensemble estimator also outperforms the plug-in estimator in this
setting. 

The estimated negative slopes of the log-log plots in Figure~\ref{fig:mseplot}
are given in Table~\ref{tab:slopes}. In all settings, both the plug-in
and ensemble estimators outperform their theoretical rates in this
finite-sample regime. However, the rates are generally approaching
the theoretical rates as the dimension increases. It is also clear
from these slopes that the ensemble estimators greatly outperform
the plug-in estimators. We expect the rates to converge to the theoretical
rates as the sample size increases.

\begin{table}
\centering

\begin{tabular}{lcccc}
\toprule 
Estimator & $\left|S_{Y}\right|=3,d=4$ & $\left|S_{Y}\right|=3,d=6$ & $\left|S_{Y}\right|=3,d=9$ & $\left|S_{Y}\right|=6,d=6$\tabularnewline
\midrule
\midrule 
Kernel Estimator & 0.90 & 0.59 & 0.32 & 0.49\tabularnewline
\midrule 
Kernel Theoretical & 0.50 & 0.33 & 0.22 & 0.33\tabularnewline
\midrule 
Weighted Estimator & 1.45 & 1.42 & 1.21 & 1.51\tabularnewline
\midrule 
Weighted Theoretical & 1 & 1 & 1 & 1\tabularnewline
\bottomrule
\end{tabular}

\caption{Negative log-log slope of the MSE for the plots in Figure~\ref{fig:mseplot}
as a function of the sample size. The estimated values are compared
to the theoretical values. While the estimators outperform the theoretical
rates in this finite sample regime, the estimator rates generally
approach the theoretical rates as the dimension increases.\label{tab:slopes}}

\end{table}

\subsection{Application to Single-Cell RNA-Sequencing Data}

A common application of MI estimation is to measure the strength of
relationships between different variables, especially in a feature
selection setting. Model aggregation, which includes ensemble methods,
for model selection is a classical problem in statistics~\cite{tsybakov2003optimal,bunea2007aggregation,samarov2007aggregation}.
Here we use the GENIE estimator on two different single-cell RNA-sequencing
(scRNA-seq) datasets to demonstrate the estimator's utility for feature
selection. 

Information theory has been used previously in many genomics applications~\cite{krishnaswamy2014conditional,motahari2013information,chee2008improved,hero2012hub,firouzi2016two}.
Single-cell RNA-sequencing data is obtained by measuring the RNA expression
levels in individual (i.e. single) cells~\cite{hwang2018single}.
Thousands of genes are typically measured in thousands of cells. This
allows the data to capture the heterogeneity of cell types within
a sample, in contrast with bulk RNA-sequencing methods which effectively
measure the average RNA expression levels within a sample. To correct
for undersampling that is present in scRNA-seq data, we first performed
imputation on both datasets~\cite{van2018recovering}. 

For these datasets, we estimated two MI measures: the R\'{e}nyi MI
and DREMI~\cite{krishnaswamy2014conditional}. We define the R\'{e}nyi
MI to be equal to the R\'{e}nyi divergence between the joint distribution
of $\X$ and $\Y$ and the product of the marginal distributions.
The DREMI score is a weighted MI developed specifically for analyzing
single-cell data~\cite{krishnaswamy2014conditional}. See Appendix~\ref{sub:DREMI}
for further details. Note that no other estimator has been defined
for $I_{DREMI}$ when the dimension of the continuous component or
components are greater than 1.

\subsubsection{Mouse bone marrow data}

We applied GENIE to scRNA-seq data measured from developing mouse
bone marrow cells~\cite{paul2015transcriptional}. Estimating mutual
information is commonly done in feature selection where features (in
this case the expression levels of genes) are selected based on the
estimated mutual information between the features $\X$ (in this case
the gene expression levels) and the response variable $\Y$ (in this
case the cell type classification). Features with higher MI are chosen
as they provide more information about the response variable. After
preprocessing, the data contained 10,738 genes measured in 2,730 cells.
In \cite{paul2015transcriptional}, the authors assigned each of the
cells to one of 19 different cell types based on its gene expression
profile. Examples of cell types in this data include erythrocytes,
basophils, and monocytes. 

For this data, we estimated the two different MI measures between
the cell type classification (discrete) and selected groups of genes
(continuous). We estimated the MI for different combinations of genes
selected from the Kyoto Encyclopedia of Genes and Genomes (KEGG) pathways
associated with the hematopoietic cell lineage~\cite{kanehisa2000kegg,kanehisa2015kegg,kanehisa2016kegg}.
Each of these collections contained 8-10 genes. Since the number of
cell types is discrete and the gene expression levels are continuous,
the estimation problem corresponds to estimating the MI between $\X$
and $\Y$ for the case where $\Y$ is discrete and $\X$ is continuous.
In this problem, $|\mathcal{S}_{Y}|=19$ and $d_{X}$ is the number
of genes in the chosen collection.

Table~\ref{tab:MI} gives the results. The mean and standard deviation
of the estimated MI (calculated from 1000 bootstrap samples) are reported
for each gene collection including all genes from the four selected
KEGG pathways. Note that the scores for DREMI and R\'{e}nyi MI are
not directly comparable due to different scaling. The estimated R\'{e}nyi
MI for these collections is higher than when selecting 8 genes at
random. This is corroborated by classification accuracies obtained
using either a linear SVM classifier or random forests: the classification
accuracies using the KEGG pathways genes are significantly higher
than those obtained using a random set of genes. This suggests the
genes in KEGG pathways associated with the hematopoietic lineage do
provide some information about cell type in this data. Additionally,
the combined genes from all four pathways have the largest estimated
MI for both measures and classification accuracy, which is expected
as genes from different pathways contain information about different
cell types and are thus necessary for distinguishing between cell
types. 

In general, the estimated DREMI when using the KEGG pathways is higher
than the estimated DREMI obtained using random genes. However, several
of these scores are within a standard deviation of the score obtained
from the random genes. Of the four KEGG pathways collections, the
Erythrocyte pathway genes has the largest estimated R\'{e}nyi MI
and smallest estimated DREMI. Yet, the classification accuracy is
essentially the same as that of the Platelets pathway geneset. These
results highlight the different use cases of these two MI measures.
The Erythrocyte cells are the largest group, containing 1,095 cells.
This suggests that the estimated R\'{e}nyi MI is biased high for
features relevant for overrepresented groups. In contrast, the DREMI
score appears to be biased low in this case. These results indicate
that the DREMI score may be more appropriate than the R\'{e}nyi MI
when analyzing less common populations. On the other hand, when less
common populations are not relevant to the analysis, DREMI may not
be as appropriate as other MI measures. These different use cases
highlight the utility of the GENIE estimator in estimating different
MI measures.

\begin{table}
\centering

\begin{tabular}{lcccccc}
\toprule 
 & Platelets & Erythrocytes & Neutrophils & Macrophages & Combined & Random\tabularnewline
\midrule
\midrule 
Estimated R\'{e}nyi MI & $0.24\pm0.11$ & $0.66\pm0.11$ & $0.27\pm0.10$ & $0.15\pm0.09$ & $1.65\pm0.36$ & $0.007\pm0.07$\tabularnewline
\midrule 
Estimated DREMI & $0.25\pm0.22$ & $0.04\pm0.03$ & $0.20\pm0.12$ & $0.41\pm0.45$ & $0.88\pm0.35$ & $0.03\pm0.08$\tabularnewline
\midrule 
SVM Accuracy & 57.4\% & 57.5\% & 52.9\% & 52.9\% & 65.4\% & 43.2\%\tabularnewline
\midrule 
Random Forests Accuracy & 60.3\% & 60.0\% & 57.8\% & 57.8\% & 65.9\% & 52.3\%\tabularnewline
\bottomrule
\end{tabular}

\caption{Estimated R\'{e}nyi MI and DREMI between collections of genes and
cell type for mouse bone marrow scRNA-seq data~\cite{paul2015transcriptional}
and the corresponding classification accuracies from a linear support
vector machine and random forests using 10-fold cross validation.
Gene collections are selected from the KEGG pathways associated with
the hematopoietic cell lineage. The fifth column (with heading ``Combined'')
gives the result when combining all genes together from the four KEGG
pathways. The last column gives the results when selecting 8 genes
at random averaged over 50 trials. MI results are presented in the
form of mean $\pm$ standard deviation which are calculated from 1000
bootstrapped samples. \label{tab:MI} }
\end{table}

\subsubsection{Human embryoid body data}

\begin{figure}
\centering

\includegraphics[width=1\textwidth]{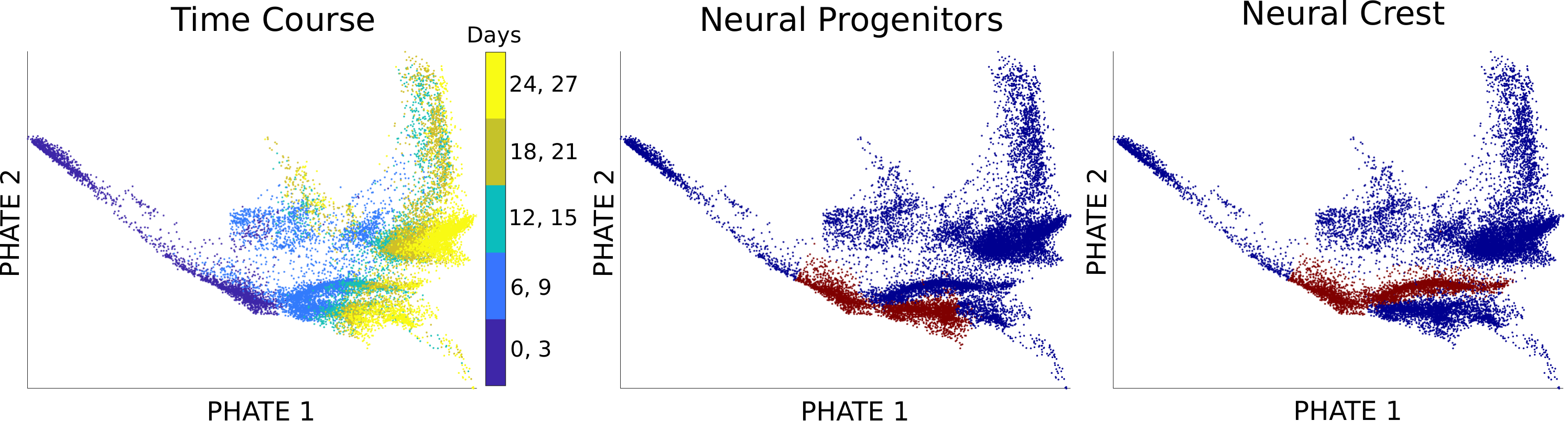}

\caption{PHATE visualizations of the EB scRNA-seq data from~\cite{moon2019phate}
colored by time sample (left), a neural progenitors branch (middle),
and a neural crest branch (right).\label{fig:EB_PHATE}}
\end{figure}

We applied GENIE to scRNA-seq data measured from human embryoid bodies
(EB) collected over a 27-day time course~\cite{moon2019phate}. Cells
were sampled at 3-day intervals and then pooled resulting in 5 different
sample collections over time. Thus sample 1 contains cells from days
0 and 3, sample 2 contains cells from days 6 and 9, etc. After preprocessing,
the data contained 17,580 genes measured in 16,825 cells, with each
of the five time samples containing about 2,400 to 4,100 cells. In~\cite{moon2019phate},
the authors identified and analyzed several branches of cells. We
used GENIE to identify genes associated with a neural progenitor (NP)
branch and a neural crest (NC) branch by estimating the R\'{e}nyi
MI and the DREMI score between the gene expression levels of the cells
in each branch ($\X$) and the timecourse variable ($\Y$). This again
corresponds to the case where $\Y$ is discrete and $\X$ is continuous.
For this problem, $|\mathcal{S}_{Y}|=5$ and $d_{X}$ is allowed to
vary as described below. Figure~\ref{fig:EB_PHATE} shows PHATE visualizations
of the data highlighted by time sample, and the two branches.

We performed three experiments with each of the branches. For all
experiments, we limited ourselves to genes that are on average nondecreasing
in the branch as time goes on. Thus in each branch, we only considered
the genes such that the correlation between the gene expression level
and time is greater than zero. 

For the first experiment, we estimated the MI scores between the time
course variable and a single gene for all genes in the data (i.e.,
$d_{X}=1$). Table~\ref{tab:Exp1} contains the estimated MI scores
of the top 10 genes for each of the measures and branches. Several
of these genes are known to be associated with their respective tissues.
For example, CX3CL1 is often expressed in the brain~\cite{maciejewski1999characterization},
SEPT6 has been found to be important for the developing neural tube
in zebra fish~\cite{zhai2014sept6}, SREBF2 is necessary for normal
brain development in mice~\cite{ferris2017loss}, NR2E1 is predominantly
expressed in the developing brain~\cite{wang2016orphan}, and ZNF804A
may help regulate early brain development~\cite{li2011allelic}.
For the NC branch, multiple HOX genes are listed as having high R\'{e}ny
MI, all of which are known to be important in the NC~\cite{philippidou2013hox}.
Additionally, RBP1 has been found in enteric nerve NC cells~\cite{ishii2012stable},
SHC4 is involved in melanocyte (an NC derivative) development~\cite{colombo2012transcriptomic},
and PRAME is involved in further differentiation of NC cells~\cite{zhang2018msx2}. 

\begin{table}
\centering

\begin{tabular}{lclclclclclc}
\toprule 
\multicolumn{6}{c}{Neural Progenitors Branch} & \multicolumn{6}{c}{Neural Crest Branch}\tabularnewline
\midrule 
\multicolumn{2}{c}{R\'{e}nyi MI} & \multicolumn{2}{c}{DREMI} & \multicolumn{2}{c}{SIS} & \multicolumn{2}{c}{R\'{e}nyi MI} & \multicolumn{2}{c}{DREMI} & \multicolumn{2}{c}{SIS}\tabularnewline
\midrule
\midrule 
LINC00526 & 1.004 & BRWD1-AS2 & 8.768 & FOS & 0.934 & HOXB7 & 0.837 & CRYL1 & 10.324 & RARB & 0.918\tabularnewline
\midrule 
GTF2E2 & 1.003 & NR2E1 & 8.579 & GTF2E2 & 0.934 & SEPT6 & 0.820 & SHC4 & 9.890 & DDIT4 & 0.917\tabularnewline
\midrule 
SEPT6 & 0.966 & ZNF804A & 8.505 & SLC18B1 & 0.932 & HOXA3 & 0.818 & SLITRK2 & 9.692 & HOXB7 & 0.909\tabularnewline
\midrule 
SREBF2 & 0.963 & SYT4 & 8.233 & JAM2 & 0.931 & HOXA7 & 0.818 & GDNF-AS1 & 9.304 & RGCC & 0.908\tabularnewline
\midrule 
EFCAB1 & 0.948 & NTNG1 & 8.146 & EGR1 & 0.930 & RBP1 & 0.806 & PRAME & 9.235 & IGFBP7 & 0.905\tabularnewline
\midrule 
RP11-68606.2 & 0.937 & GPR1 & 8.001 & CX3CL1 & 0.929 & ACADS & 0.804 & PAQR6 & 9.164 & HOXA5 & 0.904\tabularnewline
\midrule 
B2M & 0.936 & POU3F4 & 7.899 & MAGEL2 & 0.927 & HOXB5 & 0.803 & Clorf198 & 9.044 & AEBP1 & 0.903\tabularnewline
\midrule 
CX3CL1 & 0.928 & HSD17B8 & 7.704 & LINC00632 & 0.927 & HOXB6 & 0.794 & HSPB2 & 8.971 & HOXA7 & 0.901\tabularnewline
\midrule 
RP3-525N10.2 & 0.928 & LINC00092 & 7.686 & TPPP3 & 0.927 & HOXB3 & 0.793 & LINC00518 & 8.904 & ACADS & 0.901\tabularnewline
\midrule 
C20orf96 & 0.926 & WNT4 & 7.685 & GSTM3 & 0.927 & RND3 & 0.791 & AZGP1 & 8.890 & PPP1R15A & 0.900\tabularnewline
\bottomrule
\end{tabular}

\caption{Results when computing the R\'{e}nyi MI, DREMI, and the SIS between
the time course variable and a single gene (i.e. $d_{X}=1$) for all
genes in the data with nonnegative correlations with time. The top
10 genes for each branch and score are shown here. The SIS score corresponds
to the correlation coefficient. Many of the genes are known to be
associated with their respective tissues. \label{tab:Exp1}}

\end{table}
For comparison, we also used the sure independence screening (SIS)
approach described in~\cite{fan2008sure}. This approach reduces
to selecting the genes with the largest correlation with $\Y$. Table~\ref{tab:Exp1}
shows the top 10 genes for each of the branches and the corresponding
correlation coefficient. Note that only 1/10 of the SIS-selected genes
match with the R\'{e}nyi MI-selected genes in the NP branch and only
3/10 in the NC branch. None of the DREMI-selected genes match the
SIS-selected genes. Since the SIS approach focuses on linear relationships,
this suggests that our MI estimator is able to effectively detect
strong relationships that are not strictly linear.

\begin{figure}
\centering

\includegraphics[width=0.25\textwidth]{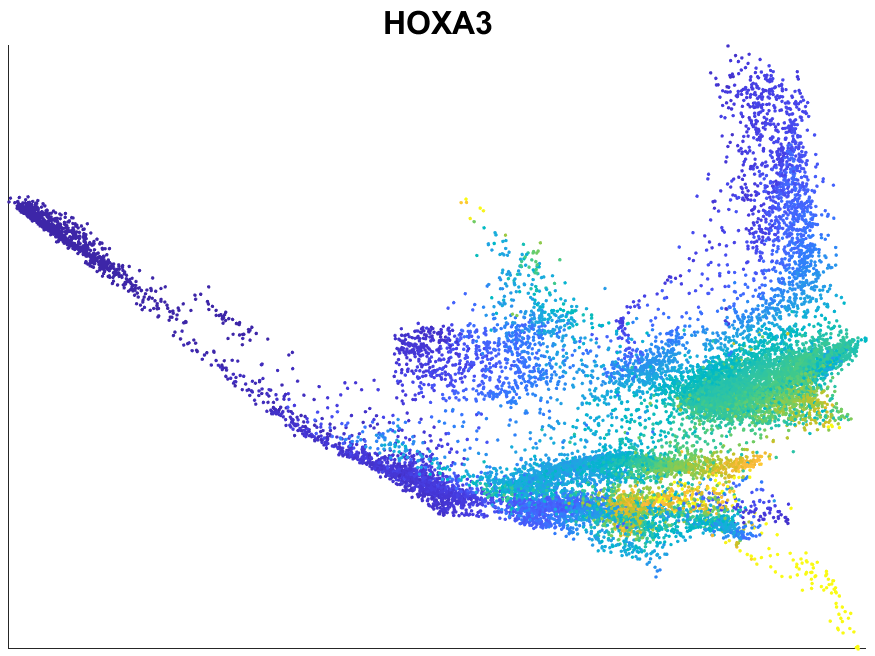}\includegraphics[width=0.25\textwidth]{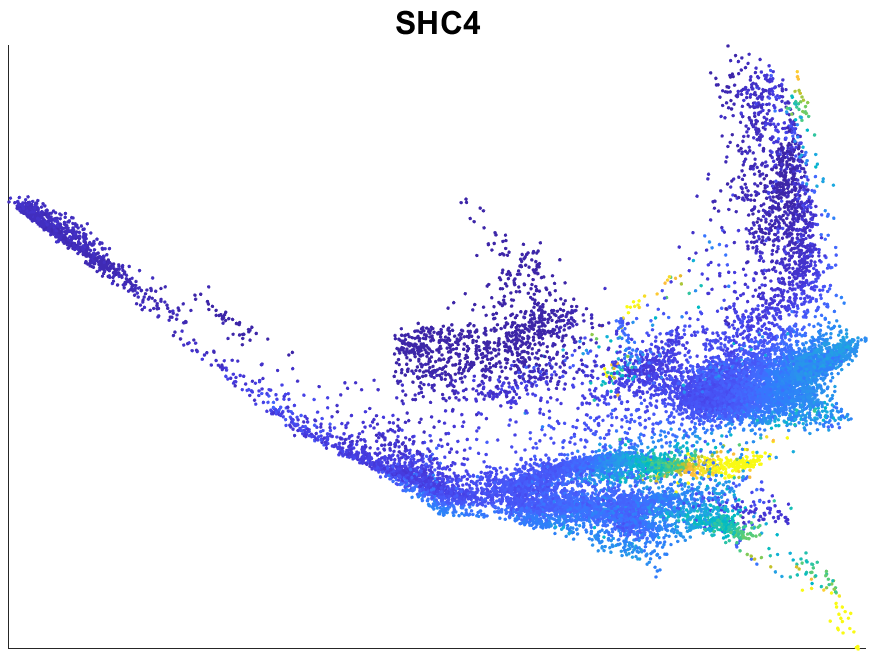}\includegraphics[width=0.25\textwidth]{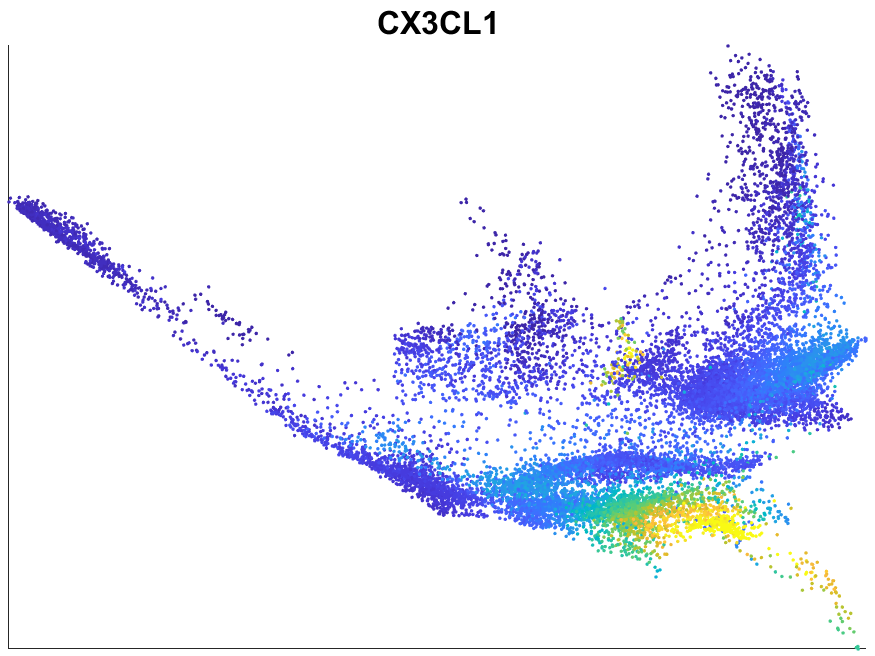}\includegraphics[width=0.25\textwidth]{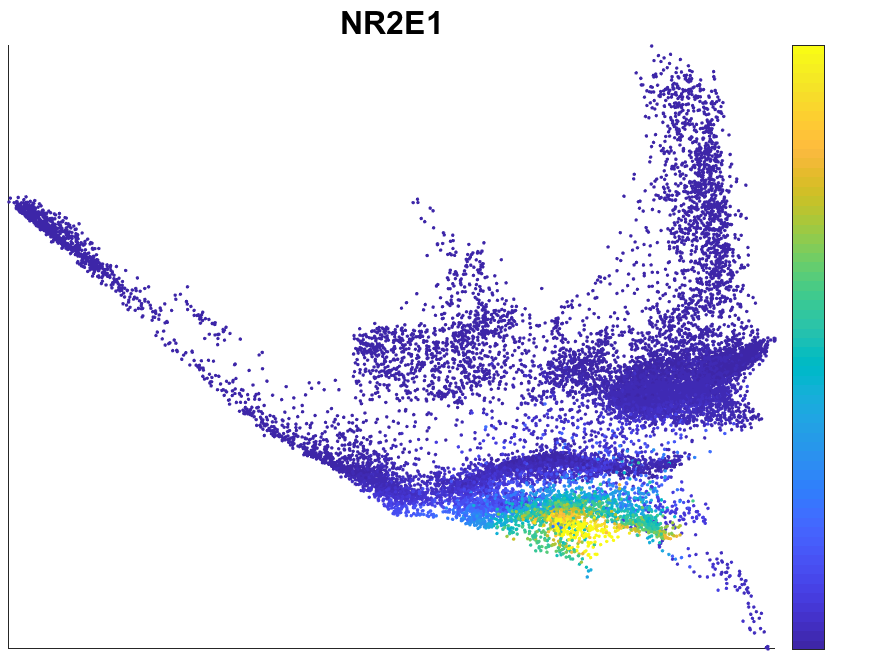}

\caption{PHATE visualization colored by gene expression levels with genes selected
as relevant for a given branch based on the estimated R\'{e}nyi MI
or DREMI (see Table~\ref{tab:Exp1}). These genes showcase differences
in the two MI measures. Genes with high DREMI scores tend to be more
localized to a branch while genes with high R\'{e}nyi MI may be spread
out more.\label{fig:DREMIvsRenyi}}

\end{figure}

None of the DREMI-selected genes match the R\'{e}nyi MI-selected
genes in both branches. Visualizing the gene expression levels of
the selected genes using PHATE indicates that genes with high DREMI
scores tend to be more localized to a branch while genes with high
R\'{e}nyi MI may be spread out more (see Figure~\ref{fig:DREMIvsRenyi}
for some examples). This suggests that the DREMI score may be better
than the R\'{e}nyi MI when the goal is to identify genes that are
uniquely expressed in specific branches. Again, these different use
cases highlight the utility of the GENIE estimator in estimating different
MI measures.

For the second experiment, we used a greedy forward-selection approach
with the GENIE estimator to identify relevant genes for the two branches.
For the third experiment, we used the same forward-selection approach
as in the second experiment except we started by including three or
four relevant genes identified in~\cite{moon2019phate} per branch.
In these experiments, we were able to identify several genes with
known associations with these cell types. See Appendix~\ref{sub:EB_more}
for details.

Our results here indicate that GENIE can be useful in identifying
relevant features under multiple settings, even when the variables
are not purely continuous or purely discrete. In particular, since
GENIE accurately identifies previously known gene relationships, we
propose that GENIE can be used to identify unknown gene relationships
for biological discovery. This use can also be extended to other domains
for scientific discovery. 

\section{Conclusion}

We derived the MSE convergence rates for general plug-in KDE-based
estimators of general MI measures between $\mathbf{X}$ and $\mathbf{Y}$
when they have only continuous components and for the case where $\X$
and/or $\Y$ contain a mixture of discrete and continuous components.
Using these rates, we defined an ensemble estimator GENIE that achieves
an MSE rate of $O(1/N)$ when the densities are sufficiently smooth.
To the best of our knowledge, this is the first nonparametric MI estimator
that achieves the MSE convergence rate of $O(1/N)$ in this setting
of mixed random variables (i.e. $\X$ and $\Y$ are not both purely
discrete or purely continuous). We also derived a minimax lower bound
on the convergence rate for estimating MI in the continuous case,
derived the asymptotic distribution of the estimator, validated the
superior convergence rate of the ensemble estimator via experiments,
and applied the estimator to analyze feature relevance in single cell
data. We show that the ensemble estimators for the continuous case
achieve the minimax rate for sufficiently smooth densities. Future
work includes extending this approach to $k$-nn based estimators
which are generally computationally easier than KDE estimators and
extending the minimax rate for the mixed case considered here. We
conjecture that the minimax rates in the mixed case are at least as
slow as those for the continuous case as the mixed case can be decomposed
into a random sum of continuous MI estimators.



{\small{}\bibliographystyle{ieeetr}
\bibliography{References}

\begin{thebibliography}{100}

\bibitem{moon2017ensemble}
K.~R. Moon, K.~Sricharan, and A.~O. Hero, ``Ensemble estimation of mutual
  information,'' in {\em Information Theory (ISIT), 2017 IEEE International
  Symposium on}, pp.~3030--3034, IEEE, 2017.

\bibitem{gao2017estimating}
W.~Gao, S.~Kannan, S.~Oh, and P.~Viswanath, ``Estimating mutual information for
  discrete-continuous mixtures,'' in {\em Advances in Neural Information
  Processing Systems}, pp.~5986--5997, 2017.

\bibitem{zeng2018jackknife}
X.~Zeng, Y.~Xia, and H.~Tong, ``Jackknife approach to the estimation of mutual
  information,'' {\em Proceedings of the National Academy of Sciences},
  vol.~115, no.~40, pp.~9956--9961, 2018.

\bibitem{jiao2015minimax}
J.~Jiao, K.~Venkat, Y.~Han, and T.~Weissman, ``Minimax estimation of
  functionals of discrete distributions,'' {\em IEEE Transactions on
  Information Theory}, vol.~61, no.~5, pp.~2835--2885, 2015.

\bibitem{jiao2017maximum}
J.~Jiao, K.~Venkat, Y.~Han, and T.~Weissman, ``Maximum likelihood estimation of
  functionals of discrete distributions,'' {\em IEEE Transactions on
  Information Theory}, vol.~63, no.~10, pp.~6774--6798, 2017.

\bibitem{wu2016minimax}
Y.~Wu and P.~Yang, ``Minimax rates of entropy estimation on large alphabets via
  best polynomial approximation,'' {\em IEEE Transactions on Information
  Theory}, vol.~62, no.~6, pp.~3702--3720, 2016.

\bibitem{krishnamurthy2014divergence}
A.~Krishnamurthy, K.~Kandasamy, B.~Poczos, and L.~Wasserman, ``Nonparametric
  estimation of renyi divergence and friends,'' in {\em International
  Conference on Machine Learning}, pp.~919--927, 2014.

\bibitem{kandasamy2015nonparametric}
K.~Kandasamy, A.~Krishnamurthy, B.~Poczos, L.~Wasserman, and J.~Robins,
  ``Nonparametric von mises estimators for entropies, divergences and mutual
  informations,'' in {\em Advances in Neural Information Processing Systems},
  pp.~397--405, 2015.

\bibitem{singh2014exponential}
S.~Singh and B.~P{\'o}czos, ``Exponential concentration of a density functional
  estimator,'' in {\em Advances in Neural Information Processing Systems},
  pp.~3032--3040, 2014.

\bibitem{singh2014renyi}
S.~Singh and B.~P{\'o}czos, ``Generalized exponential concentration inequality
  for r{\'e}nyi divergence estimation,'' in {\em International Conference on
  Machine Learning}, pp.~333--341, 2014.

\bibitem{sricharan2013ensemble}
K.~Sricharan, D.~Wei, and A.~O. Hero, ``Ensemble estimators for multivariate
  entropy estimation,'' {\em Information Theory, IEEE Transactions on},
  vol.~59, no.~7, pp.~4374--4388, 2013.

\bibitem{moon2014nips}
K.~R. Moon and A.~O. Hero, ``Multivariate f-divergence estimation with
  confidence,'' in {\em Adv Neural Inf Process Syst}, pp.~2420--2428, 2014.

\bibitem{moon2014isit}
K.~R. Moon and A.~O. Hero, ``Ensemble estimation of multivariate
  f-divergence,'' in {\em Information Theory (ISIT), 2014 IEEE International
  Symposium on}, pp.~356--360, IEEE, 2014.

\bibitem{moon2018ensemble}
K.~Moon, K.~Sricharan, K.~Greenewald, and A.~Hero, ``Ensemble estimation of
  information divergence,'' {\em Entropy}, vol.~20, no.~8, p.~560, 2018.

\bibitem{noshad2017direct}
M.~Noshad, K.~R. Moon, S.~Y. Sekeh, and A.~O. Hero, ``Direct estimation of
  information divergence using nearest neighbor ratios,'' in {\em Information
  Theory (ISIT), 2017 IEEE International Symposium on}, pp.~903--907, IEEE,
  2017.

\bibitem{wisler2018direct}
A.~Wisler, K.~Moon, and V.~Berisha, ``Direct ensemble estimation of density
  functionals,'' in {\em 2018 IEEE International Conference on Acoustics,
  Speech and Signal Processing (ICASSP)}, pp.~2866--2870, IEEE, 2018.

\bibitem{berrett2019efficient}
T.~B. Berrett, R.~J. Samworth, and M.~Yuan, ``Efficient multivariate entropy
  estimation via $ k $-nearest neighbour distances,'' {\em The Annals of
  Statistics}, vol.~47, no.~1, pp.~288--318, 2019.

\bibitem{cover2012elements}
T.~M. Cover and J.~A. Thomas, {\em Elements of information theory}.
\newblock John Wiley \& Sons, 2012.

\bibitem{avi1996bound}
H.~Avi-Itzhak and T.~Diep, ``Arbitrarily tight upper and lower bounds on the
  {Bayesian} probability of error,'' {\em IEEE Transactions on Pattern Analysis
  and Machine Intelligence}, vol.~18, no.~1, pp.~89--91, 1996.

\bibitem{moon2015Bayes}
K.~Moon, V.~Delouille, and A.~O. Hero, ``Meta learning of bounds on the {Bayes}
  classifier error,'' in {\em IEEE Signal Processing and SP Education
  Workshop}, pp.~13--18, IEEE, 2015.

\bibitem{chernoff1952measure}
H.~Chernoff, ``A measure of asymptotic efficiency for tests of a hypothesis
  based on the sum of observations,'' {\em The Annals of Mathematical
  Statistics}, pp.~493--507, 1952.

\bibitem{berisha2014bound}
V.~Berisha, A.~Wisler, A.~O. Hero~III, and A.~Spanias, ``Empirically estimable
  classification bounds based on a new divergence measure,'' {\em IEEE
  Transactions on Signal Processing}, 2015.

\bibitem{moon2015partI}
K.~R. Moon, J.~J. Li, V.~Delouille, R.~De~Visscher, F.~Watson, and A.~O. Hero,
  ``Image patch analysis of sunspots and active regions. {I. Intrinsic}
  dimension and correlation analysis,'' {\em Journal of Space Weather and Space
  Climate}, vol.~6, no.~A2, 2016.

\bibitem{unnikrishnan2011universal}
J.~Unnikrishnan, D.~Huang, S.~P. Meyn, A.~Surana, and V.~V. Veeravalli,
  ``Universal and composite hypothesis testing via mismatched divergence,''
  {\em IEEE Transactions on Information Theory}, vol.~57, no.~3,
  pp.~1587--1603, 2011.

\bibitem{naghshvar2012extrinsic}
M.~Naghshvar and T.~Javidi, ``Extrinsic jensen-shannon divergence with
  application in active hypothesis testing,'' in {\em 2012 IEEE International
  Symposium on Information Theory Proceedings}, pp.~2191--2195, IEEE, 2012.

\bibitem{pal2010estimation}
D.~P{\'a}l, B.~P{\'o}czos, and C.~Szepesv{\'a}ri, ``Estimation of r{\'e}nyi
  entropy and mutual information based on generalized nearest-neighbor
  graphs,'' in {\em Adv Neural Inf Process Syst}, pp.~1849--1857, 2010.

\bibitem{moon2017information}
K.~R. Moon, M.~Noshad, S.~Y. Sekeh, and A.~O. Hero, ``Information theoretic
  structure learning with confidence,'' in {\em Acoustics, Speech and Signal
  Processing (ICASSP), 2017 IEEE International Conference on}, pp.~6095--6099,
  IEEE, 2017.

\bibitem{chai2009fMRI}
B.~Chai, D.~Walther, D.~Beck, and L.~Fei-Fei, ``Exploring functional
  connectivities of the human brain using multivariate information analysis,''
  in {\em Advances in neural information processing systems}, pp.~270--278,
  2009.

\bibitem{liu2012exponential}
H.~Liu, L.~Wasserman, and J.~D. Lafferty, ``Exponential concentration for
  mutual information estimation with application to forests,'' in {\em Advances
  in Neural Information Processing Systems}, pp.~2537--2545, 2012.

\bibitem{lewi2006real}
J.~Lewi, R.~Butera, and L.~Paninski, ``Real-time adaptive information-theoretic
  optimization of neurophysiology experiments,'' in {\em Advances in Neural
  Information Processing Systems}, pp.~857--864, 2006.

\bibitem{schneidman2003information}
E.~Schneidman, W.~Bialek, and M.~J.~B. II, ``An information theoretic approach
  to the functional classification of neurons,'' {\em Advances in Neural
  Information Processing Systems}, vol.~15, pp.~197--204, 2003.

\bibitem{hild2001blind}
K.~E. Hild, D.~Erdogmus, and J.~C. Principe, ``Blind source separation using
  {Renyi's} mutual information,'' {\em Signal Processing Letters, IEEE},
  vol.~8, no.~6, pp.~174--176, 2001.

\bibitem{mohamed2015variational}
S.~Mohamed and D.~J. Rezende, ``Variational information maximisation for
  intrinsically motivated reinforcement learning,'' in {\em Advances in Neural
  Information Processing Systems}, pp.~2116--2124, 2015.

\bibitem{salge2014changing}
C.~Salge, C.~Glackin, and D.~Polani, ``Changing the environment based on
  empowerment as intrinsic motivation,'' {\em Entropy}, vol.~16, no.~5,
  pp.~2789--2819, 2014.

\bibitem{reshef2011detecting}
D.~N. Reshef, Y.~A. Reshef, H.~K. Finucane, S.~R. Grossman, G.~McVean, P.~J.
  Turnbaugh, E.~S. Lander, M.~Mitzenmacher, and P.~C. Sabeti, ``Detecting novel
  associations in large data sets,'' {\em Science}, vol.~334, no.~6062,
  pp.~1518--1524, 2011.

\bibitem{van2018recovering}
D.~Van~Dijk, R.~Sharma, J.~Nainys, K.~Yim, P.~Kathail, A.~Carr, C.~Burdziak,
  K.~R. Moon, C.~L. Chaffer, D.~Pattabiraman, {\em et~al.}, ``Recovering gene
  interactions from single-cell data using data diffusion,'' {\em Cell},
  vol.~174, no.~3, pp.~716--729, 2018.

\bibitem{krishnaswamy2014conditional}
S.~Krishnaswamy, M.~H. Spitzer, M.~Mingueneau, S.~C. Bendall, O.~Litvin,
  E.~Stone, D.~Pe'er, and G.~P. Nolan, ``Conditional density-based analysis of
  t cell signaling in single-cell data,'' {\em Science}, vol.~346, no.~6213,
  p.~1250689, 2014.

\bibitem{moon2019phate}
K.~R. Moon, D.~van Dijk, Z.~Wang, S.~Gigante, D.~B. Burkhardt, W.~S. Chen,
  K.~Yim, A.~van~den Elzen, M.~J. Hirn, R.~R. Coifman, N.~B. Ivanova, G.~Wolf,
  and S.~Krishnaswamy, ``Visualizing structure and transitions in
  high-dimensional biological data,'' {\em Nature Biotechnology}, vol.~37,
  no.~12, pp.~1482--1492, 2019.

\bibitem{belghazi2018mine}
M.~I. Belghazi, A.~Baratin, S.~Rajeshwar, S.~Ozair, Y.~Bengio, A.~Courville,
  and D.~Hjelm, ``Mutual information neural estimation,'' in {\em Proceedings
  of the 35th International Conference on Machine Learning}, vol.~80,
  pp.~531--540, PMLR, 10--15 Jul 2018.

\bibitem{torkkola2003feature}
K.~Torkkola, ``Feature extraction by non parametric mutual information
  maximization,'' {\em J Mach Learn Res}, vol.~3, pp.~1415--1438, 2003.

\bibitem{vergara2014review}
J.~R. Vergara and P.~A. Est{\'e}vez, ``A review of feature selection methods
  based on mutual information,'' {\em Neural Computing and Applications},
  vol.~24, no.~1, pp.~175--186, 2014.

\bibitem{peng2005feature}
H.~Peng, F.~Long, and C.~Ding, ``Feature selection based on mutual information
  criteria of max-dependency, max-relevance, and min-redundancy,'' {\em Pattern
  Analysis and Machine Intelligence, IEEE Transactions on}, vol.~27, no.~8,
  pp.~1226--1238, 2005.

\bibitem{kwak2002input}
N.~Kwak and C.-H. Choi, ``Input feature selection by mutual information based
  on parzen window,'' {\em Pattern Analysis and Machine Intelligence, IEEE
  Transactions on}, vol.~24, no.~12, pp.~1667--1671, 2002.

\bibitem{zhang2020label}
W.~Zhang, J.~S. Rhodes, A.~Garg, J.~Y. Takemoto, X.~Qi, S.~Harihar, C.-W.~T.
  Chang, K.~R. Moon, and A.~Zhou, ``Label-free discrimination and quantitative
  analysis of oxidative stress induced cytotoxicity and potential protection of
  antioxidants using raman micro-spectroscopy and machine learning,'' {\em
  Analytica Chimica Acta}, vol.~1128, pp.~221--230, 2020.

\bibitem{sugiyama2012machine}
M.~Sugiyama, ``Machine learning with squared-loss mutual information,'' {\em
  Entropy}, vol.~15, no.~1, pp.~80--112, 2012.

\bibitem{costa2004geodesic}
J.~A. Costa and A.~O. Hero, ``Geodesic entropic graphs for dimension and
  entropy estimation in manifold learning,'' {\em IEEE Transactions on Signal
  Processing}, vol.~52, no.~8, pp.~2210--2221, 2004.

\bibitem{csiszar1995generalized}
I.~Csisz{\'a}r, ``Generalized cutoff rates and r{\'e}nyi's information
  measures,'' {\em IEEE Transactions on Information Theory}, vol.~41, no.~1,
  pp.~26--34, 1995.

\bibitem{verdu2015alpha}
S.~Verd{\'u}, ``$\alpha$-mutual information,'' in {\em Information Theory and
  Applications Workshop (ITA), 2015}, pp.~1--6, IEEE, 2015.

\bibitem{principe2010information}
J.~C. Principe, {\em Information theoretic learning: Renyi's entropy and kernel
  perspectives}.
\newblock Springer Science \& Business Media, 2010.

\bibitem{tomamichel2018operational}
M.~Tomamichel and M.~Hayashi, ``Operational interpretation of r{\'e}nyi
  information measures via composite hypothesis testing against product and
  markov distributions,'' {\em IEEE Transactions on Information Theory},
  vol.~64, no.~2, pp.~1064--1082, 2018.

\bibitem{dong2016gravity}
X.~Dong, ``The gravity dual of r{\'e}nyi entropy,'' {\em Nature
  Communications}, vol.~7, p.~12472, 2016.

\bibitem{datta2009min}
N.~Datta, ``Min-and max-relative entropies and a new entanglement monotone,''
  {\em IEEE Transactions on Information Theory}, vol.~55, no.~6,
  pp.~2816--2826, 2009.

\bibitem{hayashi2017equivocations}
M.~Hayashi and V.~Y. Tan, ``Equivocations, exponents, and second-order coding
  rates under various r{\'e}nyi information measures,'' {\em IEEE Transactions
  on Information Theory}, vol.~63, no.~2, pp.~975--1005, 2017.

\bibitem{kraskov2004estimating}
A.~Kraskov, H.~St{\"o}gbauer, and P.~Grassberger, ``Estimating mutual
  information,'' {\em Physical review E}, vol.~69, no.~6, p.~066138, 2004.

\bibitem{kozachenko1987sample}
L.~Kozachenko and N.~N. Leonenko, ``Sample estimate of the entropy of a random
  vector,'' {\em Problemy Peredachi Informatsii}, vol.~23, no.~2, pp.~9--16,
  1987.

\bibitem{gao2018demystifying}
W.~Gao, S.~Oh, and P.~Viswanath, ``Demystifying fixed k-nearest neighbor
  information estimators,'' {\em IEEE Transactions on Information Theory},
  2018.

\bibitem{ganguly2018nearest}
S.~Ganguly, J.~Ryu, Y.-H. Kim, Y.-K. Noh, and D.~D. Lee, ``Nearest neighbor
  density functional estimation based on inverse laplace transform,'' {\em
  arXiv preprint arXiv:1805.08342}, 2018.

\bibitem{berrett2019twosample}
T.~B. Berrett and R.~J. Samworth, ``Efficient two-sample functional estimation
  and the super-oracle phenomenon,'' {\em arXiv preprint arXiv:1904.09347},
  2019.

\bibitem{berrett2019nonparametric}
T.~B. Berrett and R.~J. Samworth, ``Nonparametric independence testing via
  mutual information,'' {\em Biometrika}, vol.~106, no.~3, pp.~547--566, 2019.

\bibitem{han2015minimax}
Y.~Han, J.~Jiao, and T.~Weissman, ``Minimax estimation of discrete
  distributions under $\ell_1$ loss,'' {\em IEEE Transactions on Information
  Theory}, vol.~61, no.~11, pp.~6343--6354, 2015.

\bibitem{darbellay1999MIest}
G.~A. Darbellay, I.~Vajda, {\em et~al.}, ``Estimation of the information by an
  adaptive partitioning of the observation space,'' {\em IEEE Trans.
  Information Theory}, vol.~45, no.~4, pp.~1315--1321, 1999.

\bibitem{singh2016finite}
S.~Singh and B.~P{\'o}czos, ``Finite-sample analysis of fixed-k nearest
  neighbor density functional estimators,'' in {\em Advances in neural
  information processing systems}, pp.~1217--1225, 2016.

\bibitem{han2020optimal}
Y.~Han, J.~Jiao, T.~Weissman, and Y.~Wu, ``Optimal rates of entropy estimation
  over lipschitz balls,'' {\em Annals of Statistics}, vol.~48, no.~6,
  pp.~3228--3250, 2020.

\bibitem{moon2017knn}
K.~R. Moon, K.~Sricharan, and A.~O. Hero~III, ``Ensemble estimation of
  distributional functionals via $ k $-nearest neighbors,'' {\em arXiv preprint
  arXiv:1707.03083}, 2017.

\bibitem{gao2015efficient}
S.~Gao, G.~Ver~Steeg, and A.~Galstyan, ``Efficient estimation of mutual
  information for strongly dependent variables,'' in {\em Proceedings of the
  Eighteenth International Conference on Artificial Intelligence and
  Statistics}, pp.~277--286, 2015.

\bibitem{Dua:2019}
D.~Dua and C.~Graff, ``{UCI} machine learning repository,'' 2020.

\bibitem{olkin1961multivariate}
I.~Olkin, R.~F. Tate, {\em et~al.}, ``Multivariate correlation models with
  mixed discrete and continuous variables,'' {\em The Annals of Mathematical
  Statistics}, vol.~32, no.~2, pp.~448--465, 1961.

\bibitem{cox1999likelihood}
D.~Cox and N.~Wermuth, ``Likelihood factorizations for mixed discrete and
  continuous variables,'' {\em Scandinavian journal of statistics}, vol.~26,
  no.~2, pp.~209--220, 1999.

\bibitem{train2008algorithms}
K.~E. Train, ``Em algorithms for nonparametric estimation of mixing
  distributions,'' {\em Journal of Choice Modelling}, vol.~1, no.~1,
  pp.~40--69, 2008.

\bibitem{little1985maximum}
R.~J. Little and M.~D. Schluchter, ``Maximum likelihood estimation for mixed
  continuous and categorical data with missing values,'' {\em Biometrika},
  vol.~72, no.~3, pp.~497--512, 1985.

\bibitem{karunamuni2005boundary}
R.~J. Karunamuni and T.~Alberts, ``On boundary correction in kernel density
  estimation,'' {\em Stat Methodol}, vol.~2, no.~3, pp.~191--212, 2005.

\bibitem{moon2016arxiv}
K.~R. Moon, K.~Sricharan, K.~Greenewald, and A.~O. Hero, ``Nonparametric
  ensemble estimation of distributional functionals,'' {\em arXiv preprint
  arXiv:1601.06884v2}, 2016.

\bibitem{moon2016isit}
K.~R. Moon, K.~Sricharan, K.~Greenewald, and A.~O. Hero, ``Improving
  convergence of divergence functional ensemble estimators,'' in {\em 2016 IEEE
  International Symposium on Information Theory (ISIT)}, 2016.

\bibitem{birge1995estimation}
L.~Birg{\'e} and P.~Massart, ``Estimation of integral functionals of a
  density,'' {\em The Annals of Statistics}, vol.~23, no.~1, pp.~11--29, 1995.

\bibitem{levit1978asymptotically}
B.~Y. Levit, ``Asymptotically efficient estimation of nonlinear functionals,''
  {\em Problemy Peredachi Informatsii}, vol.~14, no.~3, pp.~65--72, 1978.

\bibitem{goldstein1992optimal}
L.~Goldstein and K.~Messer, ``Optimal plug-in estimators for nonparametric
  functional estimation,'' {\em The annals of statistics}, pp.~1306--1328,
  1992.

\bibitem{ali1966div}
S.~M. Ali and S.~D. Silvey, ``A general class of coefficients of divergence of
  one distribution from another,'' {\em Journal of the Royal Statistical
  Society. Series B (Methodological)}, pp.~131--142, 1966.

\bibitem{csiszar1967div}
I.~Csiszar, ``Information-type measures of difference of probability
  distributions and indirect observations,'' {\em Studia Sci. Math. Hungar.},
  vol.~2, pp.~299--318, 1967.

\bibitem{le2012asymptotic}
L.~Le~Cam, {\em Asymptotic methods in statistical decision theory}.
\newblock Springer Science \& Business Media, 2012.

\bibitem{hansen2009lecture}
B.~E. Hansen, ``Lecture notes on nonparametrics,'' 2009.

\bibitem{riordan1937moment}
J.~Riordan, ``Moment recurrence relations for binomial, poisson and
  hypergeometric frequency distributions,'' {\em The Annals of Mathematical
  Statistics}, vol.~8, no.~2, pp.~103--111, 1937.

\bibitem{efron1981jackknife}
B.~Efron and C.~Stein, ``The jackknife estimate of variance,'' {\em The Annals
  of Statistics}, pp.~586--596, 1981.

\bibitem{cameron2005microeconometrics}
A.~C. Cameron and P.~K. Trivedi, {\em Microeconometrics: methods and
  applications}.
\newblock Cambridge university press, 2005.

\bibitem{jiao2020bias}
J.~Jiao and Y.~Han, ``Bias correction with jackknife, bootstrap, and taylor
  series,'' {\em IEEE Transactions on Information Theory}, vol.~66, no.~7,
  pp.~4392--4418, 2020.

\bibitem{raykar2010fast}
V.~C. Raykar, R.~Duraiswami, and L.~H. Zhao, ``Fast computation of kernel
  estimators,'' {\em Journal of Computational and Graphical Statistics},
  vol.~19, no.~1, pp.~205--220, 2010.

\bibitem{tsybakov2003optimal}
A.~B. Tsybakov, ``Optimal rates of aggregation,'' in {\em Learning theory and
  kernel machines}, pp.~303--313, Springer, 2003.

\bibitem{bunea2007aggregation}
F.~Bunea, A.~B. Tsybakov, M.~H. Wegkamp, {\em et~al.}, ``Aggregation for
  gaussian regression,'' {\em The Annals of Statistics}, vol.~35, no.~4,
  pp.~1674--1697, 2007.

\bibitem{samarov2007aggregation}
A.~Samarov and A.~Tsybakov, ``Aggregation of density estimators and dimension
  reduction,'' in {\em Advances In Statistical Modeling And Inference: Essays
  in Honor of Kjell A Doksum}, pp.~233--251, World Scientific, 2007.

\bibitem{motahari2013information}
A.~S. Motahari, G.~Bresler, and N.~David, ``Information theory of dna shotgun
  sequencing,'' {\em IEEE Transactions on Information Theory}, vol.~59, no.~10,
  pp.~6273--6289, 2013.

\bibitem{chee2008improved}
Y.~M. Chee and S.~Ling, ``Improved lower bounds for constant gc-content dna
  codes,'' {\em IEEE Transactions on Information Theory}, vol.~54, no.~1,
  pp.~391--394, 2008.

\bibitem{hero2012hub}
A.~Hero and B.~Rajaratnam, ``Hub discovery in partial correlation graphs,''
  {\em IEEE Transactions on Information Theory}, vol.~58, no.~9,
  pp.~6064--6078, 2012.

\bibitem{firouzi2016two}
H.~Firouzi, A.~O. Hero, and B.~Rajaratnam, ``Two-stage sampling, prediction and
  adaptive regression via correlation screening,'' {\em IEEE Transactions on
  Information Theory}, vol.~63, no.~1, pp.~698--714, 2016.

\bibitem{hwang2018single}
B.~Hwang, J.~H. Lee, and D.~Bang, ``Single-cell rna sequencing technologies and
  bioinformatics pipelines,'' {\em Experimental \& molecular medicine},
  vol.~50, no.~8, pp.~1--14, 2018.

\bibitem{paul2015transcriptional}
F.~Paul, Y.~Arkin, A.~Giladi, D.~A. Jaitin, E.~Kenigsberg, H.~Keren-Shaul,
  D.~Winter, D.~Lara-Astiaso, M.~Gury, A.~Weiner, {\em et~al.},
  ``Transcriptional heterogeneity and lineage commitment in myeloid
  progenitors,'' {\em Cell}, vol.~163, no.~7, pp.~1663--1677, 2015.

\bibitem{kanehisa2000kegg}
M.~Kanehisa and S.~Goto, ``Kegg: kyoto encyclopedia of genes and genomes,''
  {\em Nucleic acids research}, vol.~28, no.~1, pp.~27--30, 2000.

\bibitem{kanehisa2015kegg}
M.~Kanehisa, Y.~Sato, M.~Kawashima, M.~Furumichi, and M.~Tanabe, ``Kegg as a
  reference resource for gene and protein annotation,'' {\em Nucleic acids
  research}, vol.~44, no.~D1, pp.~D457--D462, 2015.

\bibitem{kanehisa2016kegg}
M.~Kanehisa, M.~Furumichi, M.~Tanabe, Y.~Sato, and K.~Morishima, ``Kegg: new
  perspectives on genomes, pathways, diseases and drugs,'' {\em Nucleic acids
  research}, vol.~45, no.~D1, pp.~D353--D361, 2016.

\bibitem{maciejewski1999characterization}
D.~Maciejewski-Lenoir, S.~Chen, L.~Feng, R.~Maki, and K.~B. Bacon,
  ``Characterization of fractalkine in rat brain cells: migratory and
  activation signals for cx3cr-1-expressing microglia,'' {\em The Journal of
  Immunology}, vol.~163, no.~3, pp.~1628--1635, 1999.

\bibitem{zhai2014sept6}
G.~Zhai, Q.~Gu, J.~He, Q.~Lou, X.~Chen, X.~Jin, E.~Bi, and Z.~Yin, ``Sept6 is
  required for ciliogenesis in kupffer's vesicle, the pronephros, and the
  neural tube during early embryonic development,'' {\em Molecular and cellular
  biology}, vol.~34, no.~7, pp.~1310--1321, 2014.

\bibitem{ferris2017loss}
H.~A. Ferris, R.~J. Perry, G.~V. Moreira, G.~I. Shulman, J.~D. Horton, and
  C.~R. Kahn, ``Loss of astrocyte cholesterol synthesis disrupts neuronal
  function and alters whole-body metabolism,'' {\em Proceedings of the National
  Academy of Sciences}, vol.~114, no.~5, pp.~1189--1194, 2017.

\bibitem{wang2016orphan}
T.~Wang and J.-Q. Xiong, ``The orphan nuclear receptor tlx/nr2e1 in neural stem
  cells and diseases,'' {\em Neuroscience bulletin}, vol.~32, no.~1,
  pp.~108--114, 2016.

\bibitem{li2011allelic}
M.~Li, X.-j. Luo, X.~Xiao, L.~Shi, X.-y. Liu, L.-d. Yin, H.-b. Diao, and B.~Su,
  ``Allelic differences between han chinese and europeans for functional
  variants in znf804a and their association with schizophrenia,'' {\em American
  Journal of Psychiatry}, vol.~168, no.~12, pp.~1318--1325, 2011.

\bibitem{philippidou2013hox}
P.~Philippidou and J.~S. Dasen, ``Hox genes: choreographers in neural
  development, architects of circuit organization,'' {\em Neuron}, vol.~80,
  no.~1, pp.~12--34, 2013.

\bibitem{ishii2012stable}
M.~Ishii, A.~C. Arias, L.~Liu, Y.-B. Chen, M.~E. Bronner, and R.~E. Maxson, ``A
  stable cranial neural crest cell line from mouse,'' {\em Stem cells and
  development}, vol.~21, no.~17, pp.~3069--3080, 2012.

\bibitem{colombo2012transcriptomic}
S.~Colombo, D.~Champeval, F.~Rambow, and L.~Larue, ``Transcriptomic analysis of
  mouse embryonic skin cells reveals previously unreported genes expressed in
  melanoblasts,'' {\em Journal of Investigative Dermatology}, vol.~132, no.~1,
  pp.~170--178, 2012.

\bibitem{zhang2018msx2}
L.~Zhang, H.~Wang, C.~Liu, Q.~Wu, P.~Su, D.~Wu, J.~Guo, W.~Zhou, Y.~Xu, L.~Shi,
  {\em et~al.}, ``Msx2 initiates and accelerates mesenchymal stem/stromal cell
  specification of hpscs by regulating twist1 and prame,'' {\em Stem cell
  reports}, vol.~11, no.~2, pp.~497--513, 2018.

\bibitem{fan2008sure}
J.~Fan and J.~Lv, ``Sure independence screening for ultrahigh dimensional
  feature space,'' {\em Journal of the Royal Statistical Society: Series B
  (Statistical Methodology)}, vol.~70, no.~5, pp.~849--911, 2008.

\bibitem{liu2011mutant}
Y.~Liu, S.~Y. Lee, E.~Neely, W.~Nandar, M.~Moyo, Z.~Simmons, and J.~R. Connor,
  ``Mutant hfe h63d protein is associated with prolonged endoplasmic reticulum
  stress and increased neuronal vulnerability,'' {\em Journal of Biological
  Chemistry}, vol.~286, no.~15, pp.~13161--13170, 2011.

\bibitem{sakaguchi1999doc2alpha}
G.~Sakaguchi, T.~Manabe, K.~Kobayashi, S.~Orita, T.~Sasaki, A.~Naito, M.~Maeda,
  H.~Igarashi, G.~Katsuura, H.~Nishioka, {\em et~al.}, ``Doc2$\alpha$ is an
  activity-dependent modulator of excitatory synaptic transmission,'' {\em
  European Journal of Neuroscience}, vol.~11, no.~12, pp.~4262--4268, 1999.

\bibitem{krupp2012transcriptome}
D.~R. Krupp, P.-T. Xu, S.~Thomas, A.~Dellinger, H.~C. Etchevers, M.~Vekemans,
  J.~R. Gilbert, M.~C. Speer, A.~E. Ashley-Koch, and S.~G. Gregory,
  ``Transcriptome profiling of genes involved in neural tube closure during
  human embryonic development using long serial analysis of gene expression
  (long-sage),'' {\em Birth Defects Research Part A: Clinical and Molecular
  Teratology}, vol.~94, no.~9, pp.~683--692, 2012.

\bibitem{tan2013dihydropyrimidinase}
F.~Tan, R.~Wahdan-Alaswad, S.~Yan, C.~J. Thiele, and Z.~Li,
  ``Dihydropyrimidinase-like protein 3 expression is negatively regulated by
  mycn and associated with clinical outcome in neuroblastoma,'' {\em Cancer
  science}, vol.~104, no.~12, pp.~1586--1592, 2013.

\bibitem{alfert2019baf}
A.~Alfert, N.~Moreno, and K.~Kerl, ``The baf complex in development and
  disease,'' {\em Epigenetics \& chromatin}, vol.~12, no.~1, p.~19, 2019.

\bibitem{meganathan2015neuronal}
K.~Meganathan, S.~Jagtap, S.~P. Srinivasan, V.~Wagh, J.~Hescheler,
  J.~Hengstler, M.~Leist, and A.~Sachinidis, ``Neuronal developmental gene and
  mirna signatures induced by histone deacetylase inhibitors in human embryonic
  stem cells,'' {\em Cell death \& disease}, vol.~6, no.~5, p.~e1756, 2015.

\bibitem{yamada2007identification}
S.~Yamada, E.~Uchimura, T.~Ueda, T.~Nomura, S.~Fujita, K.~Matsumoto, D.~P.
  Funeriu, M.~Miyake, and J.~Miyake, ``Identification of twinfilin-2 as a
  factor involved in neurite outgrowth by rnai-based screen,'' {\em Biochemical
  and biophysical research communications}, vol.~363, no.~4, pp.~926--930,
  2007.

\bibitem{sidorova2018role}
E.~Sidorova-Darmos, R.~Sommer, and J.~H. Eubanks, ``The role of sirt3 in the
  brain under physiological and pathological conditions,'' {\em Frontiers in
  cellular neuroscience}, vol.~12, p.~196, 2018.

\bibitem{sakamoto2003basic}
M.~Sakamoto, H.~Hirata, T.~Ohtsuka, Y.~Bessho, and R.~Kageyama, ``The basic
  helix-loop-helix genes hesr1/hey1 and hesr2/hey2 regulate maintenance of
  neural precursor cells in the brain,'' {\em Journal of Biological Chemistry},
  vol.~278, no.~45, pp.~44808--44815, 2003.

\bibitem{qin2006genetic}
L.~Qin, L.~Wine-Lee, K.~J. Ahn, and E.~B. Crenshaw, ``Genetic analyses
  demonstrate that bone morphogenetic protein signaling is required for
  embryonic cerebellar development,'' {\em Journal of Neuroscience}, vol.~26,
  no.~7, pp.~1896--1905, 2006.

\bibitem{florio2018evolution}
M.~Florio, M.~Heide, A.~Pinson, H.~Brandl, M.~Albert, S.~Winkler, P.~Wimberger,
  W.~B. Huttner, and M.~Hiller, ``Evolution and cell-type specificity of
  human-specific genes preferentially expressed in progenitors of fetal
  neocortex,'' {\em Elife}, vol.~7, p.~e32332, 2018.

\bibitem{low2017identification}
S.-K. Low, A.~Takahashi, Y.~Ebana, K.~Ozaki, I.~E. Christophersen, P.~T.
  Ellinor, S.~Ogishima, M.~Yamamoto, M.~Satoh, M.~Sasaki, {\em et~al.},
  ``Identification of six new genetic loci associated with atrial fibrillation
  in the japanese population,'' {\em Nature genetics}, vol.~49, no.~6, p.~953,
  2017.

\bibitem{hu2019downregulation}
B.~Hu, L.~Cao, X.-y. Wang, and L.~Li, ``Downregulation of micro rna-431-5p
  promotes enteric neural crest cell proliferation via targeting lrsam 1 in
  hirschsprung's disease,'' {\em Development, growth \& differentiation},
  vol.~61, no.~4, pp.~294--302, 2019.

\bibitem{szemes2018wnt}
M.~Szemes, A.~Greenhough, Z.~Melegh, S.~Malik, A.~Yuksel, D.~Catchpoole,
  K.~Gallacher, M.~Kollareddy, J.~H. Park, and K.~Malik, ``Wnt signalling
  drives context-dependent differentiation or proliferation in neuroblastoma,''
  {\em Neoplasia}, vol.~20, no.~4, pp.~335--350, 2018.

\bibitem{chng2013high}
Z.~Chng, G.~S. Peh, W.~B. Herath, T.~Y. Cheng, H.-P. Ang, K.-P. Toh, P.~Robson,
  J.~S. Mehta, and A.~Colman, ``High throughput gene expression analysis
  identifies reliable expression markers of human corneal endothelial cells,''
  {\em PLoS One}, vol.~8, no.~7, p.~e67546, 2013.

\bibitem{li2011spherical_cap}
S.~Li, ``Concise formulas for the area and volume of a hyperspherical cap,''
  {\em Asian Journal of Mathematics and Statistics}, vol.~4, no.~1, pp.~66--70,
  2011.

\bibitem{steele1986efron}
J.~M. Steele {\em et~al.}, ``An efron-stein inequality for nonsymmetric
  statistics,'' {\em The Annals of Statistics}, vol.~14, no.~2, pp.~753--758,
  1986.

\bibitem{tsybakov2008introduction}
A.~B. Tsybakov, {\em Introduction to nonparametric estimation}.
\newblock Springer Science \& Business Media, 2008.

\bibitem{bickel1988estimating}
P.~J. Bickel and Y.~Ritov, ``Estimating integrated squared density derivatives:
  sharp best order of convergence estimates,'' {\em Sankhy{\=a}: The Indian
  Journal of Statistics, Series A}, pp.~381--393, 1988.

\bibitem{durrett2010probability}
R.~Durrett, {\em Probability: Theory and Examples}.
\newblock Cambridge University Press, 2010.

\bibitem{gut2012probability}
A.~Gut, {\em Probability: A Graduate Course}.
\newblock Springer Science \& Business Media, 2012.

\end{thebibliography}
}{\small\par}

\appendices

\section{Extended Genomics Details and Results}

Here we provide further details on the genomics experiments. 

\subsection{Imputation and DREMI}

\label{sub:DREMI}Single-cell RNA-sequencing data typically suffers
from undersampling. Therefore, we perform imputation on both datasets
using MAGIC~\cite{van2018recovering} prior to estimating the MI
measures. 

Typically, MI measures are weighted by the joint probability density
of $\X$ and $\Y$. In DREMI, the measure is instead weighted by the
conditional probability density of $\Y|\X$. This allows DREMI to
measure the strength of the relationship between $\Y$and $\X$ regardless
of differences in population density that often arise in single-cell
data. Since $\X$ is continuous and $\Y$ is discrete for both genomics
applications, DREMI can be defined mathematically as

\begin{align*}
I_{DREMI}(\X;\Y) & =\sum_{y\in S_{Y}}\int f_{Y_{D}|X_{C}}(y|x)\log\left(\frac{f_{X_{D}Y_{C}}(x_{C},y)}{f_{X_{C}}(x_{C})f_{Y_{D}}(y)}\right)dx_{C}\\
 & =\sum_{y\in S_{Y}}f_{Y_{D}}(y)\int\log\left(\frac{f_{X_{C}|Y_{D}}\left(x_{C}|y\right)}{f_{X_{C}}\left(x_{C}\right)}\right)\frac{f_{X_{C}|Y_{D}}\left(x_{C}|y\right)}{f_{X_{C}}\left(x_{C}\right)}dx_{C}.
\end{align*}
This measure differs from standard Shannon MI with the inclusion of
the weight $1/f_{X_{C}}\left(x_{C}\right)$ within the integral. While
this does not fit our standard definition of a generalized MI, our
estimation approach allows us to include the inverse of the KDE of
$f_{X_{C}}$ when estimating the integral. The proof techniques are
unaffected and therefore our theoretical results still hold.

\subsection{EB Data Extended Results}

\label{sub:EB_more}For the second experiment, we used a greedy forward-selection
approach with the GENIE estimator to identify relevant genes. We first
selected the gene with the highest estimated MI in a given branch
($d_{X}=1$). We then identified the gene that gave the largest MI
when included with the first gene ($d_{X}=2$). We then repeated this
to obtain the top 10 genes. The results are shown in Table~\ref{tab:Exp2}.
R\'{e}nyi MI should never decrease as we add more genes, and we indeed
see this in Table~\ref{tab:Exp2}. Thus the relative increase in
estimated R\'{e}nyi MI can be used as a measure of the amount of
information each gene adds. Note that for both branches, the largest
increase in R\'{e}nyi MI occurs within the first four genes and the
inclusion of each subsequent gene adds a decreasing amount of R\'{e}nyi
MI. However, several of these genes have known associations with their
respective branches. Mutations of HFE are associated with neurological
disorders~\cite{liu2011mutant} while DOC2A is mainly expressed in
the brain~\cite{sakaguchi1999doc2alpha}. For the NC branch, RGR
is associated with eye development which comes partially from the
neural crest, ITPKB is associated with neurulation~\cite{krupp2012transcriptome},
and DPYSL4 is associated with the development of the nervous system~\cite{tan2013dihydropyrimidinase}. 

\begin{table}
\centering

\begin{tabular}{lclclclclclc}
\toprule 
\multicolumn{6}{c}{Neural Progenitors (NP) Branch} & \multicolumn{6}{c}{Neural Crest (NC) Branch}\tabularnewline
\midrule 
\multicolumn{2}{c}{R\'{e}nyi MI} & \multicolumn{2}{c}{DREMI} & \multicolumn{2}{c}{SIS} & \multicolumn{2}{c}{R\'{e}nyi MI} & \multicolumn{2}{c}{DREMI} & \multicolumn{2}{c}{SIS }\tabularnewline
\midrule
\midrule 
LINC00526 & 1.004 & BRWD1-AS2 & 9 & FOS & 0.934 & HOXB7 & 0.837 & CRYL1 & 10 & RARB & 0.918\tabularnewline
\midrule 
HFE & 1.382 & FOSL1 & 15 & CDC37L1-AS1 & 0.360 & AF127936.9 & 1.119 & IDH3B & 14 & ID3 & 0.394\tabularnewline
\midrule 
DOC2A & 1.675 & TCP11 & 30 & SH3GL2 & 0.194 & RGR & 1.407 & AC142528.1 & 11 & ZNF564 & 0.282\tabularnewline
\midrule 
P4HA1 & 1.931 & BRD9 & 48 & ATP13A3 & 0.229 & RP11-324E6.10 & 2.151 & CFL1 & 33 & MALRD1 & 0.188\tabularnewline
\midrule 
HIST1H1C & 2.030 & HFE & 103 & DARS & 0.178 & LIMA1 & 2.347 & RP11-676J15.1 & 64 & BRWD1-AS2 & 0.165\tabularnewline
\midrule 
PRDM12 & 2.152 & RP11-225H22.4 & 224 & RP11-390P2.4 & 0.171 & ITGA9 & 2.581 & RP5-1098D14.1 & 87 & RP11-10A14.4 & 0.169\tabularnewline
\midrule 
ACTR3C & 2.189 & ATG9A & 335 & ZNF484 & 0.167 & TTLL9 & 2.699 & BMP8B & 193 & SLC10A5 & 0.126\tabularnewline
\midrule 
TRDC & 2.222 & RP11-35015.1 & 848 & SIDT2 & 0.164 & ITPKB & 2.826 & SLITRK2 & 936 & RP3-402G11.26 & 0.117\tabularnewline
\midrule 
MMEL1 & 2.285 & GAS2L3 & 2385 & NEFH & 0.140 & ACOT1 & 2.857 & TMCC & 1846 & ABCC1 & 0.122\tabularnewline
\midrule 
LINC01229 & 2.327 & PRAC1 & 5845 & ZNF587 & 0.127 & DPYSL4 & 2.869 & MLANA & 6201 & RPSA & 0.124\tabularnewline
\bottomrule
\end{tabular}

\caption{Results when computing the R\'{e}nyi MI, DREMI, and the SIS between
the time course variable and multiple genes using a greedy forward-selection
approach. The top 10 genes for each branch and score are shown here.
The SIS score corresponds to the correlation coefficient of gene expression
with the regression residuals. Many of these genes are known to be
associated with their respective tissues. \label{tab:Exp2}}
\end{table}
While the R\'{e}nyi MI does not decrease with the addition of genes,
DREMI may decrease due to the reweighting caused by using the conditional
distribution instead of the joint. Thus the change in score when adding
genes is less informative for DREMI. For a fixed dimension, however,
the relative DREMI scores are informative and thus can be used to
identify relevant genes using the forward-selection approach. Using
this approach with DREMI, we identified several genes with known associations
such as HFE (also identified with R\'{e}nyi MI) and BRD9~\cite{alfert2019baf}
with the NP branch, and CFL1~\cite{ishii2012stable} and BMP8B~\cite{meganathan2015neuronal}
with the NC branch.

We also performed a forward-selection variant on SIS. We first selected
the gene with the highest SIS score (correlation coefficient in this
case). We then performed regression with this gene and the time course
variable $\Y$. We then calculated the SIS score between all of the
other genes individually and the regression residuals to select the
next gene. This process was repeated to obtain a list of the top ten
genes in Table~\ref{tab:Exp2}. Since the SIS criteria is scale-invariant,
this can sometimes result in an increase in the correlation coefficient
as more genes are included, although generally we expect the correlation
to decrease. Thus it is somewhat difficult to assess using SIS the
amount of information added by including each gene. In this case,
the MI and SIS approaches identified unique genes with no shared overlap
in either branch, again suggesting that our MI approaches are identifying
nonlinear relationships. 

For the third experiment, we used the same forward-selection approach
as in the second experiment except we started by including three or
four relevant genes identified in~\cite{moon2019phate}. These genes
were NKX2-8, EN2, and SOX1 for the NP branch, and PAX3, FOXD3, SOX9,
and SOX10 for the NC branch. The results are presented in Table~\ref{tab:Exp3}.
Interestingly, including these ``preset'' genes results in a larger
overall R\'{e}nyi MI and DREMI in the NP branch than when using a
purely greedy approach (Table~\ref{tab:Exp2}) while the opposite
is true for the NC branch. Additionally, the identified genes are
all different from the purely greedy approach. However, many of them
are known to be associated with their respective tissues. PTK6 affects
neurite extension~\cite{yamada2007identification}, SIRT3 regulates
mitochondria in the brain during development~\cite{sidorova2018role},
HEY1 is expressed in neural precursor cells~\cite{sakamoto2003basic},
BMPR1B is important for brain development~\cite{qin2006genetic},
FAM72C is enriched in cortical neural progenitors~\cite{florio2018evolution},
PPFIA4 is involved in neural development~\cite{low2017identification},
LRSAM1 is related to enteric NC cells~\cite{hu2019downregulation},
LINC00327 is associated with regulating neuroblasts~\cite{szemes2018wnt},
and GRHPR is associated with human eye development~\cite{chng2013high}.

\begin{table}
\centering

\begin{tabular}{lclclclclclc}
\toprule 
\multicolumn{6}{c}{Neural Progenitors (NP) Branch} & \multicolumn{6}{c}{Neural Crest (NC) Branch}\tabularnewline
\midrule 
\multicolumn{2}{c}{R\'{e}nyi MI} & \multicolumn{2}{c}{DREMI} & \multicolumn{2}{c}{SIS} & \multicolumn{2}{c}{R\'{e}nyi MI} & \multicolumn{2}{c}{DREMI} & \multicolumn{2}{c}{SIS}\tabularnewline
\midrule
\midrule 
NKX2-8 & - & NKX2-8 & - & NKX2-8 & - & PAX3 & - & PAX3 & - & PAX3 & -\tabularnewline
\midrule 
EN2 & - & EN2 & - & EN2 & - & FOXD3 & - & FOXD3 & - & FOXD3 & -\tabularnewline
\midrule 
SOX1 & - & SOX1 & - & SOX1 & - & SOX9 & - & SOX9 & - & SOX9 & -\tabularnewline
\midrule 
SLX1B & 2.030 & MSS51 & 15 & FAM174A & 0.567 & SOX10 & - & SOX10 & - & SOX10 & -\tabularnewline
\midrule 
PTK6 & 2.613 & KNSTRN & 247 & RAB27A & 0.329 & RP11-867G23.8 & 1.614 & GRHPR & 7 & ARMC5 & 0.543\tabularnewline
\midrule 
SIRT3 & 3.204 & TEKT3 & 548 & CTD-2568A17.1 & 0.321 & ZBED3 & 1.759 & LINC01389 & 23 & ABL1 & 0.325\tabularnewline
\midrule 
HEY1 & 3.596 & FAM72C & 1998 & CECR5 & 0.209 & RP3-460G2.2 & 1.830 & CTB-25B13.5 & 50 & PSPN & 0.338\tabularnewline
\midrule 
OLIG2 & 4.021 & GZMK & 4904 & LOXL1 & 0.163 & LRSAM1 & 1.855 & CFAP58 & 68 & RP11-354P11.3 & 0.262\tabularnewline
\midrule 
MZF1-AS1 & 4.689 & QRICH2 & 16337 & PROB1 & 0.147 & NOTUM & 1.930 & ZNF774 & 271 & WWTR1-AS1 & 0.185\tabularnewline
\midrule 
BMPR1B & 5.133 & PPFIA4 & 42236 & YY2 & 0.161 & CPLX2 & 1.945 & LINC00327 & 588 & KIF25 & 0.144\tabularnewline
\bottomrule
\end{tabular}

\caption{Results when computing the R\'{e}nyi MI, DREMI, and the SIS between
the time course variable and multiple genes using a greedy forward-selection
approach when starting with three or four relevant genes identified
in~\cite{moon2019phate}. The top 10 genes for each branch and score
are shown here. The SIS score corresponds to the correlation coefficient
of gene expression with the regression residuals. Many of these genes
are known to be associated with their respective tissues. \label{tab:Exp3}}
\end{table}

\section{MI Ensemble Estimation Extensions}

\label{sec:extensions}In this appendix, we present several extensions
of the ensemble estimation approach to MI. First, we show how to apply
the theory to the purely continuous case. We then show how the theory
can be applied to obtain estimators that achieve the parametric rate
for less smooth densities.

\subsection{Continuous Random Varables}

\label{subsec:cont_ensemble}We can apply Theorem 3 in~\cite{moon2018ensemble}
to obtain a version of the GENIE MI estimator that achieves the parametric
rate for the case when $\mathbf{X}$ and $\mathbf{Y}$ are purely
continuous. For completeness, we repeat the theorem and its proof here. For a general
estimation problem, let $N$ be the number of available samples and
let $\mathcal{L}=\left\{ l_{1},\dots,l_{L}\right\} $ be a set of
index values. For an indexed ensemble of estimators $\left\{ \hat{\mathbf{E}}_{l}\right\} _{l\in\mathcal{L}}$
of a parameter $E$, the weighted ensemble estimator with weights
$w=\left\{ w\left(l_{1}\right),\dots,w\left(l_{L}\right)\right\} $
satisfying $\sum_{l\in\mathcal{L}}w(l)=1$ is defined as 
\[
\hat{\mathbf{E}}_{w}=\sum_{l\in\mathcal{L}}w\left(l\right)\hat{\mathbf{E}}_{l}.
\]
 Consider the following conditions on $\left\{ \hat{\mathbf{E}}_{l}\right\} _{l\in\mathcal{L}}$: 
\begin{itemize}
\item $\mathcal{C}.1$ The bias is expressible as 
\[
\bias\left[\hat{\mathbf{E}}_{l}\right]=\sum_{i\in J}c_{i}\psi_{i}(l)\phi_{i,d}(N)+O\left(\frac{1}{\sqrt{N}}\right),
\]
where $c_{i}$ are constants depending on the underlying density and
are independent of $N$ and $l$, $J=\left\{ i_{1},\dots,i_{I}\right\} $
is a finite index set with $I<L$, and $\psi_{i}(l)$ are basis functions
depending only on the parameter $l$ and not on the sample size $N$. 
\item $\mathcal{C}.2$ The variance is expressible as 
\[
\var\left[\hat{\mathbf{E}}_{l}\right]=c_{v}\left(\frac{1}{N}\right)+o\left(\frac{1}{N}\right).
\]
\end{itemize}
\begin{thm}
[Theorem 3 in \cite{moon2018ensemble}]Assume conditions $\mathcal{C}.1$
and $\mathcal{C}.2$ hold for an ensemble of estimators $\left\{ \hat{\mathbf{E}}_{l}\right\} _{l\in\mathcal{L}}$.
Then there exists a weight vector $w_{0}$ such that the MSE of the
weighted ensemble estimator attains the parametric rate of convergence:
\[
\mathbb{E}\left[\left(\hat{\mathbf{E}}_{w_{0}}-E\right)^{2}\right]=O\left(\frac{1}{N}\right).
\]
The weight vector $w_{0}$ is the solution to the following convex
optimization problem:\emph{
\begin{equation}
\begin{array}{rl}
\min_{w} & ||w||_{2}\\
\text{subject to} & \sum_{l\in\mathcal{L}}w(l)=1,\\
 & \gamma_{w}(i)=\sum_{l\in\mathcal{L}}w(l)\psi_{i}(l)=0,\,i\in J.
\end{array}\label{eq:optimize-1}
\end{equation}
}
\end{thm}
\begin{proof}
Due to condition $\mathcal{C}.1$, the bias of the weighted ensemble estimator is
\begin{equation}
    \bias\left[\hat{\mathbf{E}}_{w}\right]=\sum_{i\in J}c_i\gamma_w(i)\phi_{i,d}(N)+O\left(\frac{\sqrt{L}||w||_2}{\sqrt{N}}. \right)
\end{equation}
Denote the covariance matrix of the ensemble of estimators as $\Sigma_L$. By the Cauchy-Schwarz inequality and condition $\mathcal{C}.2$, the entries of $\Sigma_L$ are $O(1/N)$. The variance of the weighted estimator is then bounded above as 
\begin{equation}
    \var \left[\hat{\mathbf{E}}_{w}\right]=w^T\Sigma_L w\leq \frac{Trace(\Sigma_L N)||w||_2^2}{N}=\frac{c_v L||w||_2^2}{N}+o\left(\frac{1}{N}\right).
\end{equation}

The optimization problem in (\ref{eq:optimize-1}) zeroes out the lower-order bias terms and limits the $\ell_2$ norm of the weight vector $w$ to prevent the variance from exploding. This results in an MSE of $O(1/N)$ as long as the dimension $d$ is fixed and $L$ is fixed and independent of the sample size $N$. A solution is guaranteed to (\ref{eq:optimize-1}) as long as $L>I$ and the vectors $a_i=[\psi_i(l_1),\dots,\psi_i(l_L)]$ are linearly independent.

\end{proof}

As before, (\ref{eq:optimize-1}) typically results in an ensemble
estimator with a large variance. We can relax this optimization problem
and obtain an estimator that still obtains the parametric rate: 
\begin{equation}
\begin{array}{rl}
\min_{w} & \epsilon\\
\text{subject to} & \sum_{l\in\mathcal{L}}w(l)=1,\\
 & \left|\gamma_{w}(i)N^{\frac{1}{2}}\phi_{i,d}(N)\right|\leq\epsilon,\,\,i\in J,\\
 & \left\Vert w\right\Vert _{2}^{2}\leq\eta\epsilon.
\end{array}\label{eq:relaxed-1}
\end{equation}

We can use (\ref{eq:relaxed-1}) to obtain a GENIE estimator for the
purely continuous case. Theorem~\ref{thm:bias} indicates that we
need $h_{X}^{d_{X}}h_{Y}^{d_{Y}}\propto N^{-1/2}$ for the $O(1/(Nh_{X}^{d_{X}}h_{Y}^{d_{Y}}))$
terms to be $O(1/\sqrt{N})$. We consider the more general case where
the parameters may differ for $h_{X}$ and $h_{Y}$. Let $\mathcal{L}_{X}$
and $\mathcal{L}_{Y}$ be sets of real, positive numbers with $|\mathcal{L}_{X}|=L_{X}$
and $|\mathcal{L}_{Y}|=L_{Y}$. For each estimator in the ensemble,
choose $l_{X}\in\mathcal{L_{X}}$ and $l_{Y}\in\mathcal{L}_{Y}$ and
set $h_{X}(l_{X})=l_{X}N^{-1/(2(d_{X}+d_{Y}))}$ and $h_{Y}(l_{Y})=l_{Y}N^{-1/(2(d_{X}+d_{Y}))}$.
Define the matrix $w$ s.t. $\sum_{l_{X}\in\mathcal{L}_{X},l_{Y}\in\mathcal{L}_{Y}}w(l_{X},l_{Y})=1$.
From Theorems~\ref{thm:bias} and~\ref{thm:variance}, conditions
$\mathcal{C}.1$ and $\mathcal{C}.2$ are satisfied if $s\geq d_{X}+d_{Y}$
with $\psi_{i,j}(l_{X},l_{Y})=l_{X}^{i}l_{Y}^{j}$ and $\phi_{i,j}(N)=N^{-(i+j)/(2(d_{X}+d_{Y}))}$
for $0\leq i,j\leq d_{X}+d_{Y}$ s.t. $0<i+j\leq d_{X}+d_{Y}$. The
optimal weight $w_{0}$ is calculated using (\ref{eq:relaxed-1}).
The resulting estimator 
\[
\g{w_{0}}^{cont}=\sum_{l_{X}\in\mathcal{L}_{X},l_{Y}\in\mathcal{L}_{Y}}w_{0}(l_{X},l_{Y})\g{h_{X}(l_{X}),h_{Y}(l_{Y})}
\]
achieves the parametric MSE rate when $s\geq d_{X}+d_{Y}$. We denote
this estimator as $\g{GENIE}^{cont}$.

\subsection{Less Smooth Densities}

\label{subsec:Odin2}The GENIE estimators $\g{GENIE}$ and $\g{GENIE}^{cont}$
are guaranteed to achieve the parametric convergence rate as long
as $s\geq d_{X}+d_{Y}$. Here we derive ensemble estimators of MI
that achieve the parametric rate under less strict smoothness assumptions
on the densities. To derive the ensemble estimators for less smooth
densities, we need a different expansion of the bias that includes
some higher order terms. We present those results here for the continuous
and mixed cases and show how to apply the ensemble estimation theory
to obtain the parametric MSE convergence rate when the densities are
less smooth ($s>(d_{X}+d_{Y})/2$).

\subsubsection{Continuous Random Variables}

We first consider the case where $\X$ and $\Y$ are both purely continuous.
Consider the following result on the bias of the plug-in estimator: 
\begin{thm}
[Continuous Bias Expanded]\label{thm:bias_cont_mod}Assume that assumptions
$\mathcal{A}.0$-$\mathcal{A}.5$ hold. Then for $\lambda\geq2$ a
positive integer, the bias of $\gt$ is 
\begin{eqnarray}
\bias\left[\gt\right] & = & \sum_{\substack{m,n=0\\
i+j+m+n\neq0
}
}^{\lambda}\sum_{i,j=0}^{r}c_{11,i,j,m,n}\frac{h_{X}^{i}h_{Y}^{j}}{\left(Nh_{X}^{d_{X}}\right)^{m}\left(Nh_{Y}^{d_{Y}}\right)^{n}}\nonumber \\
 &  & +\sum_{m=1}^{\lambda}\sum_{i=0}^{r}\sum_{j=0}^{r}c_{13,i,j,m}h_{X}^{i}h_{Y}^{j}/\left(Nh_{X}^{d_{X}}h_{Y}^{d_{Y}}\right)^{m}\nonumber \\
 &  & +O\left(h_{X}^{s}+h_{Y}^{s}+1/\left(Nh_{X}^{d_{X}}h_{Y}^{d_{Y}}\right)^{\lambda}\right).\label{eq:bias2-1}
\end{eqnarray}
\end{thm}
The proof is given in Appendix~\ref{sec:biasProof}. Note that no
extra assumptions are required to achieve this result compared to
Theorem~\ref{thm:bias}. However, we elected to retain these results
for the appendix to simplify the presentation in the main paper.

We now use these results to define a new ensemble estimator. Set $\delta>0$
and let $\mathcal{L}_{X}$ and $\mathcal{L}_{Y}$ be sets of real,
positive numbers with $|\mathcal{L}_{X}|=L_{X}$ and $|\mathcal{L}_{Y}|=L_{Y}$.
For each estimator in the ensemble, choose $l_{X}\in\mathcal{L_{X}}$
and $l_{Y}\in\mathcal{L}_{Y}$ and set $h_{X}(l_{X})=l_{X}N^{-1/(d_{X}+d_{Y}+\delta)}$
and $h_{Y}(l_{Y})=l_{Y}N^{-1/(d_{X}+d_{Y}+\delta)}$. Then conditions
$\mathcal{C}.1$ and $\mathcal{C}.2$ are satisfied if $s\geq(d_{X}+d_{Y}+\delta)/2$
and $\lambda\geq(d_{X}+d_{Y}+\delta)/\delta$ with $\psi_{1,i,j,m,n}(l_{X},l_{Y})=l_{X}^{i-md_{X}}l_{Y}^{j-nd_{Y}}$
and $\phi_{1,i,j,m,n}(N)=N^{-\frac{i+j+m(d_{Y}+\delta)+n(d_{X}+\delta)}{d_{X}+d_{Y}+\delta}}$
for $0<i+j+m(d_{Y}+\delta)+n(d_{X}+\delta)\leq\frac{d_{X}+d_{Y}+\delta}{2}$
and the terms $\psi_{2,i,j,m}(l_{X},l_{Y})=l_{X}^{i-md_{X}}l_{Y}^{j-md_{Y}}$
and $\phi_{2,i,j,m}(N)=N^{-\frac{i+j+m\delta}{d_{X}+d_{Y}+\delta}}$
for $m\geq1$ and $i+j+m\delta\leq\frac{d_{X}+d_{Y}+\delta}{2}.$
These all correspond to terms that converge to zero slower than $N^{-1/2}$
when left uncorrected. The optimal weight $w_{0}$ is again calculated
using (\ref{eq:relaxed-1}) and the resulting ensemble estimator achieves
the parametric MSE convergence rate when $s\geq(d_{X}+d_{Y}+\delta)/2$.
Since $\delta$ can be chosen arbitrarily close to zero, the parametric
rate can be achieved theoretically as long as $s>(d_{X}+d_{Y})/2$. 

\subsubsection{Mixed Random Variables}

We now consider the case where $\mathbf{X}$ and $\mathbf{Y}$ may
have any mixture of continuous and discrete components. We have a
similar result on the bias as in Theorem~\ref{thm:bias_cont_mod}.
Here we assume that $\hx=l_{X}\mathbf{N}_{x}^{-\beta}$ and $\hy=l_{Y}\N_{y}^{-\alpha}$
with $0<\beta<\frac{1}{d_{X}}$, $0<\alpha<\frac{1}{d_{Y}}$, and
$l_{X},l_{Y}>0$.
\begin{thm}
[Mixed Bias Expanded]Assume that the same assumptions hold as in
Theorem~\ref{thm:bias_mixed}. Then for $\lambda\geq2$ a positive
integer, the bias of $\g{h_{X_{C}|X_{D}},h_{Y_{C}|Y_{D}}}$ is
\begin{align}
\bias\left[\g{h_{X_{C}|X_{D}},h_{Y_{C}|Y_{D}}}\right] & =\sum_{\substack{m,n=0\\
i+j+m+n\neq0
}
}^{\lambda}\sum_{i,j=0}^{r}c_{14,i,j,m,n}\frac{l_{X}^{i}l_{Y}^{j}N^{-i\beta-j\alpha}}{\left(l_{X}^{d_{X}}N^{1-\beta d_{X}}\right)^{m}\left(l_{Y}^{d_{Y}}N^{1-\alpha d_{Y}}\right)^{n}}\nonumber \\
 & +\sum_{m=1}^{\lambda}\sum_{i=0}^{r}\sum_{j=0}^{r}c_{15,i,j,m}\frac{l_{X}^{i}l_{Y}^{j}N^{-i\beta-j\alpha}}{\left(l_{X}^{d_{X}}l_{Y}^{d_{Y}}N^{1-\beta d_{X}-\alpha d_{Y}}\right)^{m}}\nonumber \\
 & +O\left(N^{-s\beta}+N^{-s\alpha}+\frac{1}{\left(N^{1-\beta d_{X}-\alpha d_{Y}}\right)^{\lambda}}\right).\label{eq:mixed_bias-2}
\end{align}
\end{thm}
The proof is given in Appendix~\ref{subsec:bias_mixed_proof}.

We now use these results to define a new ensemble estimator in the
mixed case. The procedure is similar to the continuous case. Set $\delta>0$
and let $\mathcal{L}_{X}$ and $\mathcal{L}_{Y}$ be sets of real,
positive numbers with $|\mathcal{L}_{X}|=L_{X}$ and $|\mathcal{L}_{Y}|=L_{Y}$.
For each estimator in the ensemble, choose $l_{X}\in\mathcal{L_{X}}$
and $l_{Y}\in\mathcal{L}_{Y}$ and set $\hx(l_{X})=l_{X}\mathbf{N}_{x}^{-1/(d_{X}+d_{Y}+\delta)}$
and $\hy(l_{Y})=l_{Y}\N_{y}^{-1/(d_{X}+d_{Y}+\delta)}$. Conditions
$\mathcal{C}.1$ and $\mathcal{C}.2$ are satisfied if $s\geq(d_{X}+d_{Y}+\delta)/2$
and $\lambda\geq(d_{X}+d_{Y}+\delta)/\delta$. The first set of terms
in the optimization problem are $\psi_{1,i,j,m,n}(l_{X},l_{Y})=l_{X}^{i-md_{X}}l_{Y}^{j-nd_{Y}}$
and $\phi_{1,i,j,m,n}(N)=N^{-\frac{i+j+m(d_{Y}+\delta)+n(d_{X}+\delta)}{d_{X}+d_{Y}+\delta}}$
for $0<i+j+m(d_{Y}+\delta)+n(d_{X}+\delta)\leq\frac{d_{X}+d_{Y}+\delta}{2}$.
The second set of terms are $\psi_{2,i,j,m}(l_{X},l_{Y})=l_{X}^{i-md_{X}}l_{Y}^{j-md_{Y}}$
and $\phi_{2,i,j,m}(N)=N^{-\frac{i+j+m\delta}{d_{X}+d_{Y}+\delta}}$
for $m\geq1$ and $i+j+m\delta\leq\frac{d_{X}+d_{Y}+\delta}{2}.$
The optimal weight $w_{0}$ is again calculated using (\ref{eq:relaxed-1})
and the resulting ensemble estimator achieves the parametric MSE convergence
rate when $s\geq(d_{X}+d_{Y}+\delta)/2$. Since $\delta$ can be chosen
arbitrarily close to zero, the parametric rate can be achieved theoretically
as long as $s>(d_{X}+d_{Y})/2$. 

The modified estimators defined in this section have better statistical
properties than the original GENIE estimators defined in Section~\ref{sec:mixed_ensemble}
and Appendix~\ref{subsec:cont_ensemble} as the parametric rate is
guaranteed under less restrictive smoothness assumptions on the densities.
On the other hand, the number of parameters required for the optimization
problem in (\ref{eq:relaxed-1}) is larger for the modified estimator.
In theory, this could lead to larger variance although this is not
necessarily true in practice according to divergence estimation experiments
in~\cite{moon2016arxiv}.

\section{The Boundary Condition}
\label{sec:boundary}
Here we prove that under certain smoothness assumptions on a kernel
with either rectangular or circular support, the boundary assumption $\mathcal{A}.5$ is
satisfied for densities with the unit cube as its support set. We will
prove the following, more general result:
\begin{thm}
\label{thm:assumption5} Let $K$ be a kernel function  with rectangular or circular support,
i.e., $K(u)=0$ for $\left\Vert u\right\Vert_1 >1$ or $\left\Vert u\right\Vert_2 >1$, respectively. Let $K\in\Sigma(s,H_{K})$ in the interior of its support. Let $p_{x}(u):\mathbb{R}^{d}\rightarrow\mathbb{R}$
be a polynomial in $u$ of order $|q|\leq r=\left\lfloor s\right\rfloor $
whose coefficients are a function of $x$ and are $r-|q|$ times differentiable.
Then for $\mathcal{S}=[0,1]^{d}$ and any positive integers $t$ and
$m$, we have that
\begin{equation}
\int_{x\in\mathcal{S}}\left(\int_{u:K(u)>0,\,x+uh\notin\mathcal{S}}K^{l}(u)p_{x}(u)du\right)^{t}dx=v_{t}(h),\label{eq:assume_boundary-1-1}
\end{equation}
 where $v_{t}(h)$ admits the expansion 
\[
v_{t}(h)=\sum_{i=1}^{r-|q|}e_{i,q,t,l}h^{i}+o\left(h^{r-q}\right),
\]
 for some constants $e_{i,q,t,l}$.
\end{thm}

Before proving this, we will relate this result to assumption $\mathcal{A}.5$.

\begin{cor}
Assumption $\mathcal{A}.5$ is satisfied under the same conditions as in Theorem~\ref{thm:assumption5}, i.e., $K\in\Sigma(s,H_{K})$ is a kernel function with either rectangular or circular support and $\mathcal{S}=[0,1]^{d}$.
\end{cor}
This follows immediately from assumption $\mathcal{A}.2$ that $f$ is in the H\"{o}lder class $\Sigma(s,H)$, which implies that $D^q f(x)$ is $r-|q|$ times differentiable. Thus equation (\ref{eq:assume_boundary-1-1}) implies $\mathcal{A}.5$.

\subsection{Proof of Theorem~\ref{thm:assumption5}: Rectangular Support Kernels}
We will first consider points that are boundary points due to a single coordinate and then extend to the general case where multiple coordinates are close to the boundary.

\paragraph*{Single Coordinate Boundary Point} Since $K\in\Sigma(s,H_K)$, we can take a Taylor series expansion of $K^l(u)$ around zero to obtain
\begin{equation}
    K^l(u)=p_{K,l}(u)+o(||u||_2^r),
\end{equation}
where $p_{K,l}(u)$ is a  polynomial function of $u$ with degree $r$.

Consider points $x$ that are boundary points by virtue
of a single coordinate $x_{i}$ such that $x_{i}+u_{i}h\notin\S$.
Without loss of generality, assume that $x_{i}+u_{i}h>1$. After performing the above Taylor series expansion of $K^l$, the inner
integral in (\ref{eq:assume_boundary-1-1}) can then be evaluated first with respect to
all coordinates other than $i$. Since all of these coordinates lie
within the support, the inner integral over these coordinates will
amount to integration of the polynomial $p_{x}(u)$ over a symmetric
$d-1$ dimensional rectangular region $\left|u_{j}\right|\leq 1$
for all $j\neq i$. This yields a function $\sum_{m=1}^{q+r}\tilde{p}_{m}(x)u_{i}^{m}+o(u_i^{r+1})$
where the coefficients $\tilde{p}_{m}(x)$ are each $r-|q|$ times differentiable
wrt $x$.

With respect to the $u_{i}$ coordinate, the inner integral will have
limits from $\frac{1-x_{i}}{h}$ to $1$ for some $1>x_{i}>1-h$.
Consider the $\tilde{p}_{m}(x)u_{i}^{m}$ monomial term. The inner
integral wrt this term yields (ignoring the $o(\cdot)$ term for now)
\begin{equation}
\sum_{m=1}^{|q|+r}\tilde{p}_{m}(x)\int_{\frac{1-x_{i}}{h}}^{1}u_{i}^{m}du_{i}=\sum_{m=1}^{|q|+r}\tilde{p}_{m}(x)\frac{1}{m+1}\left(1-\left(\frac{1-x_{i}}{h}\right)^{m+1}\right).\label{eq:poly1-1}
\end{equation}
Raising the right hand side of (\ref{eq:poly1-1}) to the power of $t$
results in an expression of the form 
\begin{equation}
\sum_{j=0}^{(|q|+r)t}\check{p}_{j}(x)\left(\frac{1-x_{i}}{h}\right)^{j},\label{eq:poly2}
\end{equation}
where the coefficients $\check{p}_{j}(x)$ are $r-|q|$ times differentiable
wrt $x$. Integrating (\ref{eq:poly2}) over all the coordinates in
$x$ other than $x_{i}$ results in an expression of the form 
\begin{equation}
\sum_{j=0}^{(|q|+r)t}\bar{p}_{j}(x_{i})\left(\frac{1-x_{i}}{h}\right)^{j},\label{eq:poly3}
\end{equation}
 where again the coefficients $\bar{p}_{j}(x_{i})$ are $r-|q|$ times
differentiable wrt $x_{i}$. Note that since the other cooordinates
of $x$ other than $x_{i}$ are far away from the boundary, the coefficients
$\bar{p}_{j}(x_{i})$ are independent of $h$. To evaluate the integral
of (\ref{eq:poly3}), consider the $r-|q|$ term Taylor series expansion
of $\bar{p}_{j}(x_{i})$ around $x_{i}=1$, where we use a smooth extension of the function and its derivatives to the boundary. This will yield terms
of the form 
\begin{eqnarray}
\int_{1-h}^{1}\frac{\left(1-x_{i}\right)^{j+k}}{h^{k}}dx_{i} & = & \left.-\frac{\left(1-x_{i}\right)^{j+k+1}}{h^{k}(j+k+1)}\right|_{x_{i}=1-h}^{x_{i}=1}\\
 & = & \frac{h^{j+1}}{(j+k+1)},
 \label{eq:h_int}
\end{eqnarray}
for $0\leq j\leq r-|q|$, and $0\leq k\leq (|q|+r)t$. By a similar analysis, the $o(\cdot)$ terms from the Taylor expansion of the kernel result in $o(h^r)$. Combining terms results
in the expansion $v_{t}(h)=\sum_{i=1}^{r-|q|}e_{i,q,t}h^{i}+o\left(h^{r-|q|}\right)$.

\paragraph*{Multiple Coordinate Boundary Point} The case where multiple coordinates of the point $x$ are near the
boundary is a straightforward extension of the single boundary point
case so we only sketch the main ideas here. As an example, consider
the case where 2 of the coordinates are near the boundary. Assume
for notational ease that they are $x_{1}$ and $x_{2}$ and that $x_{1}+u_{1}h>1$
and $x_{2}+u_{2}h>1$. The inner integral in (\ref{eq:assume_boundary-1-1})
can again be evaluated first wrt all coordinates other than 1 and
2. This yields a function $\sum_{m,j=1}^{q+r}\tilde{p}_{m,j}(x)u_{1}^{m}u_{2}^{j}$
where the coefficients $\tilde{p}_{m,j}(x)$ are each $r-q$ times
differentiable wrt $x$. Integrating this wrt $x_{1}$ and $x_{2}$
and then raising the result to the power of $t$ yields a double sum
similar to (\ref{eq:poly2}). Integrating this over all the coordinates
in $x$ other than $x_{1}$ and $x_{2}$ gives a double sum similar
to (\ref{eq:poly3}). Then a Taylor series expansion of the coefficients
and integration over $x_{1}$ and $x_{2}$ yields the result.

\subsection{Proof of Theorem~\ref{thm:assumption5}: Circular Support Kernels}
The case where the kernel $K$ has a circular support is more complex than when the support is rectangular. We will again first consider points that are boundary points due to a single coordinate. We will then extend to the general case where multiple coordinates are close to the boundary.

\paragraph*{Single Coordinate Boundary Point} Consider points $x$ that are
boundary points due to a single coordinate $x_{i}$ s.t. $x_{i}+u_{i}h\notin\mathcal{S}$.
Without loss of generality, assume that $x_{i}+u_{i}h>1$. We focus
first on the inner integral in (\ref{eq:assume_boundary-1-1}). We will use the
following lemma: 
\begin{lem}
\label{lem:poly_sphere}Let $D_{d}(\rho)$ be a $d$-sphere with radius
$\rho$ and let $\sum_{i=1}^{d}n_{i}=q$. Then 
\[
\int_{D_{d}(\rho)}u_{1}^{n_{1}}u_{2}^{n_{2}}\dots u_{d}^{n_{d}}du_{1}\dots du_{d}=C\rho^{d+q},
\]
 where $C$ is a constant that depends on the $n_{i}$s and $d$.
\end{lem}
\begin{proof}
We convert to $d$-dimensional spherical coordinates to handle the
integration. Let $r$ be the distance of a point $u$ from the origin.
We nave $d-1$ angular coordinates $\phi_{i}$ where $\phi_{d-1}$
ranges from $0$ to $2\pi$ and all other $\phi_{i}$ range from $0$
to $\pi$. The conversion from the spherical coordinates to Cartesian
coordinates is then 
\begin{eqnarray*}
u_{1} & = & r\cos\left(\phi_{1}\right)\\
u_{2} & = & r\sin\left(\phi_{1}\right)\cos\left(\phi_{2}\right)\\
u_{3} & = & r\sin\left(\phi_{1}\right)\sin\left(\phi_{2}\right)\cos\left(\phi_{3}\right)\\
\vdots\\
u_{d-1} & = & r\sin\left(\phi_{1}\right)\cdots\sin\left(\phi_{d-2}\right)\cos\left(\phi_{d-1}\right)\\
u_{d} & = & r\sin\left(\phi_{1}\right)\cdots\sin\left(\phi_{d-2}\right)\sin\left(\phi_{d-1}\right).
\end{eqnarray*}
 The spherical volume element is then 
\[
r^{d-1}\sin^{d-2}\left(\phi_{1}\right)\sin^{d-3}\left(\phi_{1}\right)\cdots\sin\left(\phi_{d-1}\right)dr\,d\phi_{1}\,d\phi_{2}\cdots d\phi_{d-1}.
\]
 Combining these results gives 
\begin{eqnarray*}
\lefteqn{\int_{D_{d}(\rho)}u_{1}^{n_{1}}u_{2}^{n_{2}}\dots u_{d}^{n_{d}}du_{1}\dots du_{d}}\\
 & = & \int_{0}^{\rho}\int_{o}^{2\pi}\int_{0}^{\pi}\cdots\int_{0}^{\pi}r^{q+d-1}\left[\sin^{q-n_{1}+d-2}\left(\phi_{1}\right)\sin^{q-n_{1}-n_{d}+d-3}\left(\phi_{2}\right)\cdots\right.\\
 &  & \left.\sin^{n_{d}+n_{d-1}+1}\left(\phi_{d-2}\right)\sin^{n_{d}}\left(\phi_{d-1}\right)\right]\left[\cos^{n_{1}}\left(\phi_{1}\right)\cdots\cos^{n_{d}}\left(\phi_{d-1}\right)\right]d\phi_{1}\cdots d\phi_{d-1}dr\\
 & = & C\rho^{q+d}.
\end{eqnarray*}
\end{proof}

Since $K\in\Sigma(s,H_K)$, we can take a Taylor series expansion of $K^l(u)$ around zero to obtain
\begin{equation}
    K^l(u)=p_{K,l}(u)+o(||u||_2^r),
\end{equation}
where $p_{K,l}(u)$ is a  polynomial function of $u$ with degree $r$.

The region of integration for the inner integral in (\ref{eq:assume_boundary-1-1})
corresponds to a hyperspherical cap with radius $1$ and height of
$\frac{1-x_{i}}{h}$. The inner integral can be calculated using an
approach similar to that used in~\cite{li2011spherical_cap} to calculate
the volume of a hyperspherical cap. It is obtained by integrating
the polynomial $p_{x}(u)$ over a $d-1$-sphere with radius $\sin\theta$
and height element $d\cos\theta$. This is done using Lemma~\ref{lem:poly_sphere}.
We then integrate over $\theta$ which has a range of $0$ to $\phi=\cos^{-1}\left(\frac{1-x_{i}}{h}\right).$ 
Thus we have 
\begin{eqnarray}
\int_{u:x+uh\notin\mathcal{S}}K^l(u)p_{x}(u)du & = & \int_{u:||u||_{2}\leq1,x+uh\notin\mathcal{S}}\left(p_{K,l}(u)+o(||u||_2^r)\right)p_{x}(u)du \nonumber \\
& = & \sum_{m=0}^{|q|+r}\tilde{p}_{m,l}(x)\int_{0}^{\phi}\sin^{m+d-1}(\theta)\sin\theta \cos^m\theta d\theta+o\left( \int_{u:||u||_{2}\leq1,x+uh\notin\mathcal{S}}||u||_2^rdu\right)\nonumber \\
 & = & \sum_{m=0}^{|q|+r}\tilde{p}_{m,l}(x)\int_{0}^{\phi}\sin^{m+d}(\theta)\cos^{m}\theta d\theta+o\left( \int_{u:||u||_{2}\leq1,x+uh\notin\mathcal{S}}||u||_2^rdu\right),\label{eq:poly1}
\end{eqnarray}
where $\tilde{p}_{m}(x)$ is the polynomial coefficient corresponding to $u_i^m$ after the Taylor series expansion of $K^l$ and integrating over the $d-1$-sphere. 

We will focus on the first term in Eq.~\ref{eq:poly1} as the $o(\cdot)$ term will follow similarly as a polynomial function of $u$. From standard integral tables, we get that for $n\geq2$ and $m\geq0$
\begin{equation}
\int_{0}^{\phi}\sin^{n}\theta\cos^{m}\theta d\theta=-\frac{\sin^{n-1}\phi\cos^{m+1}\phi}{n+m}+\frac{n-1}{n+m}\int_{0}^{\phi}\sin^{n-2}\theta\cos^{m}\theta d\theta.\label{eq:sincos}
\end{equation}
If $n=1$, then we get 
\[
\int_{0}^{\phi}\sin\theta\cos^{m}\theta d\theta=\frac{1}{m+1}-\frac{\cos^{m+1}\phi}{m+1}.
\]
Since $\phi=\cos^{-1}\left(\frac{1-x_{i}}{h}\right)$, we have 
\begin{eqnarray*}
\cos\phi & = & \frac{1-x_{i}}{h},\\
\sin\phi & = & \sqrt{1-\left(\frac{1-x_{i}}{h}\right)^{2}}.
\end{eqnarray*}
Therefore, if $n$ is odd, we obtain 
\begin{equation}
\int_{0}^{\phi}\sin^{n}\theta\cos^{m}\theta d\theta=\sum_{\ell=0}^{(n-1)/2}c_{\ell}\left(\sqrt{1-\left(\frac{1-x_{i}}{h}\right)^{2}}\right)^{2\ell}\left(\frac{1-x_{i}}{h}\right)^{m+1}+c,\label{eq:nodd}
\end{equation}
where the constants depend on $m$ and $n$.

If $n$ is even and $m>0$, then the final term in the recursion in
(\ref{eq:sincos}) reduces to 
\[
\int_{0}^{\phi}\cos^{m}\theta d\theta=\frac{\cos^{m-1}\phi\sin\phi}{m}+\frac{m-1}{m}\int_{0}^{\phi}\cos^{m-2}\theta d\theta.
\]
If $m=2$, then 
\begin{eqnarray*}
\int_{0}^{\phi}\cos^{2}\theta d\theta & = & \frac{\phi}{2}+\frac{1}{4}\sin(2\phi)\\
 & = & \frac{\phi}{2}+\frac{1}{2}\sin\phi\cos\phi.
\end{eqnarray*}
Therefore, if $n$ and $m$ are both even, then this gives 
\begin{eqnarray}
\int_{0}^{\phi}\sin^{n}\theta\cos^{m}\theta d\theta & = & \sum_{\ell=0}^{(n-2)/2}c_{\ell}^{'}\left(\sqrt{1-\left(\frac{1-x_{i}}{h}\right)^{2}}\right)^{2\ell+1}\left(\frac{1-x_{i}}{h}\right)^{m+1}+c^{'}\cos^{-1}\left(\frac{1-x_{i}}{h}\right)\nonumber \\
 &  & +\sum_{\ell=0}^{(m-2)/2}c_{\ell}^{''}\left(\sqrt{1-\left(\frac{1-x_{i}}{h}\right)^{2}}\right)\left(\frac{1-x_{i}}{h}\right)^{2\ell+1}.\label{eq:nm_even}
\end{eqnarray}
On the other hand, if $n$ is even and $m$ is odd, we get 
\begin{eqnarray}
\int_{0}^{\phi}\sin^{n}\theta\cos^{m}\theta d\theta & = & \sum_{\ell=0}^{(n-2)/2}c_{\ell}^{'''}\left(\sqrt{1-\left(\frac{1-x_{i}}{h}\right)^{2}}\right)^{2\ell+1}\left(\frac{1-x_{i}}{h}\right)^{m+1}\nonumber \\
 &  & +\sum_{\ell=0}^{(m-1)/2}c_{\ell}^{''''}\left(\sqrt{1-\left(\frac{1-x_{i}}{h}\right)^{2}}\right)\left(\frac{1-x_{i}}{h}\right)^{2\ell}.\label{eq:neven_modd}
\end{eqnarray}

If $d$ is odd, then combining (\ref{eq:nodd}) and (\ref{eq:neven_modd})
with (\ref{eq:poly1}) gives 
\begin{equation}
\sum_{m=0}^{|q|+r}\tilde{p}_{m,l}(x)\int_{0}^{\phi}\sin^{m+d}(\theta)\cos^{m}\theta d\theta =  \sum_{m=0}^{|q|+r}\sum_{\ell=0}^{d+|q|}p_{m,\ell,l}(x)\left(\sqrt{1-\left(\frac{1-x_{i}}{h}\right)^{2}}\right)^{\ell}\left(\frac{1-x_{i}}{h}\right)^{m},\label{eq:polyodd}
\end{equation}
where the coefficients $p_{m,\ell,l}(x)$ are $r-|q|$ times differentiable
wrt $x$. Similarly, if $d$ is even, then 
\begin{eqnarray}
\sum_{m=0}^{|q|+r}\tilde{p}_{m,l}(x)\int_{0}^{\phi}\sin^{m+d}(\theta)\cos^{m}\theta d\theta & = & \sum_{m=0}^{|q|+r}\sum_{\ell=0}^{d+|q|}p_{m,\ell,l}^{'}(x)\left(\sqrt{1-\left(\frac{1-x_{i}}{h}\right)^{2}}\right)^{\ell}\left(\frac{1-x_{i}}{h}\right)^{m}\nonumber \\
 &  & +p^{'}(x)\cos^{-1}\left(\frac{1-x_{i}}{h}\right),\label{eq:polyeven}
\end{eqnarray}
where again the coefficients $p_{m,\ell,l}^{'}(x)$ and $p^{'}(x)$
are $r-q$ times differentiable wrt $x$. Raising (\ref{eq:polyodd})
and (\ref{eq:polyeven}) to the power of $t$ gives respective expressions
of the form 
\begin{equation}
\sum_{m=0}^{(|q|+r)t}\sum_{\ell=0}^{(d+|q|)t}\check{p}_{m,\ell,l}(x)\left(\sqrt{1-\left(\frac{1-x_{i}}{h}\right)^{2}}\right)^{\ell}\left(\frac{1-x_{i}}{h}\right)^{m},\label{eq:polyodd2}
\end{equation}
\begin{equation}
\sum_{m=0}^{(|q|+r)t}\sum_{\ell=0}^{(d+|q|)t}\sum_{n=0}^{t}\check{p}_{m,\ell,l,n}(x)\left(\sqrt{1-\left(\frac{1-x_{i}}{h}\right)^{2}}\right)^{\ell}\left(\frac{1-x_{i}}{h}\right)^{m}\left(\cos^{-1}\left(\frac{1-x_{i}}{h}\right)\right)^{n},\label{eq:polyeven2}
\end{equation}
where the coefficients $\check{p}_{m,\ell,l}(x)$ and $\check{p}_{m,\ell,l,n}(x)$
are all $r-q$ times differentiable wrt $x$. Integrating (\ref{eq:polyodd2})
and (\ref{eq:polyeven2}) over all the coordinates in $x$ except
for $x_{i}$ affects only the $\check{p}_{m,\ell,l}(x)$ and $\check{p}_{m,\ell,l,n}(x)$
coefficients, resulting in respective expressions of the form 
\begin{equation}
\sum_{m=0}^{|q|t}\sum_{\ell=0}^{(d+|q|)t}\bar{p}_{m,\ell,l}(x_{i})\left(\sqrt{1-\left(\frac{1-x_{i}}{h}\right)^{2}}\right)^{\ell}\left(\frac{1-x_{i}}{h}\right)^{m},\label{eq:polyodd3}
\end{equation}
\begin{equation}
\sum_{m=0}^{|q|t}\sum_{\ell=0}^{(d+|q|)t}\sum_{n=0}^{t}\bar{p}_{m,\ell,l,n}(x_{i})\left(\sqrt{1-\left(\frac{1-x_{i}}{h}\right)^{2}}\right)^{\ell}\left(\frac{1-x_{i}}{h}\right)^{m}\left(\cos^{-1}\left(\frac{1-x_{i}}{h}\right)\right)^{n}.\label{eq:polyeven3}
\end{equation}

The coefficients $\bar{p}_{m,\ell,l}(x_{i})$ and $\bar{p}_{m,\ell,l,n}(x_{i})$
are $r-q$ times differentiable wrt $x_{i}$. Since the other coordinates
of $x$ other than $x_{i}$ are far away from the boundary, the coefficients
are independent of $h$. For the integral wrt $x_{i}$ of (\ref{eq:polyodd3}),
taking a Taylor series expansion of $\bar{p}_{m,\ell}(x_{i})$ around
$x_{i}=1$ (again using a smooth extension of the function and its derivatives to the boundary) yields terms of the form 
\begin{eqnarray*}
\int_{1-h}^{1}\left(\sqrt{1-\left(\frac{1-x_{i}}{h}\right)^{2}}\right)^{\ell}\left(\frac{1-x_{i}}{h}\right)^{m+j}h^{j}dx_{i} & = & h^{j+1}\int_{0}^{1}\left(1-y_{i}\right)^{\frac{\ell}{2}}y_{i}^{\frac{m+j-1}{2}}dy_{i}\\
 & = & h^{j+1}B\left(\frac{\ell+2}{2},\frac{m+j+1}{2}\right),
\end{eqnarray*}
where $0\leq j\leq r-q$, $0\leq\ell\leq(d+|q|)t$, $0\leq m\leq |q|t$,
and $B(x,y)$ is the beta function. Note that the first step uses
the substitution of $y_{i}=\left(\frac{1-x_{i}}{h}\right)^{2}$.

If $d$ is even (i.e. (\ref{eq:polyeven3})), a simple closed-form
expression is not easy to obtain due to the $\cos^{-1}\left(\frac{1-x_{i}}{h}\right)$
terms. However, by similarly applying a Taylor series expansion to
$\bar{p}_{m,\ell,l,n}(x_{i})$ and substituting $y_{i}=\frac{1-x_{i}}{h}$
gives terms of the form of 
\begin{eqnarray*}
\lefteqn{\int_{1-h}^{1}\left(\sqrt{1-\left(\frac{1-x_{i}}{h}\right)^{2}}\right)^{\ell}\left(\frac{1-x_{i}}{h}\right)^{m+j}\left(\cos^{-1}\left(\frac{1-x_{i}}{h}\right)\right)^{n}h^{j}dx_{i}}\\
 & = & h^{j+1}\int_{0}^{1}\left(1-y_{i}^{2}\right)^{\frac{\ell}{2}}y_{i}^{m+j}\left(\cos^{-1}y_{i}\right)^{n}dy_{i}\\
 & = & h^{j+1}c_{\ell,m,j,n},
\end{eqnarray*}
for $0\leq j\leq r-|q|$, $0\leq\ell\leq(d+|q|)t$, $0\leq m\leq (|q|+r)t$,
and $0\leq n\leq t$. By a similar analysis, the $o(\cdot)$ term in Eq.~\ref{eq:poly1} results in $o(h^r)$. Combining terms results in the expansion $v_{t}(h)=\sum_{i=1}^{r-|q|}e_{i,q,t,l}h^{i}+o(h^{r-|q|})$.

\paragraph*{Multiple Coordinate Boundary Point}

\label{sec:bound_sphere_mult}The case where multiple coordinates
of the point $x$ are near the boundary is a fairly straightforward
extension of the single boundary point case. Consider the case where
2 of the coordinates are near the boundary, e.g., $x_{1}$ and $x_{2}$
with $x_{1}+u_{1}h>1$ and $x_{2}+u_{2}h>1$. The region of integration
for the inner integral can be decomposed into two parts: a hyperspherical
cap wrt $x_{1}$ and the remaining area (denoted, respectively, as
$A_{1}$ and $A_{2}$). The remaining area $A_{2}$ can be decomposed
further into two other areas: a hyperspherical cap wrt $x_{2}$ (denoted
$B_{1}$) and a height chosen s.t. $B_{1}$ just intersects $A_{1}$
on their boundaries. Integrating over the remainder of $A_{2}$ is
achieved by integrating along $x_{2}$ over $d-1$-dimensional hyperspherical
caps from the boundary of $B_{1}$ to the boundary of $A_{2}$. Thus
integrating over these regions yields an expression similar to (\ref{eq:poly1}).
Following a similar procedure will then yield the result. 

\subsection{Proof of Proposition~\ref{prop:boundary}}
\label{sub:prop_proof}
Recall that we assume that the derivatives of the density $f$ vanish near the boundary of the density support set. We will use the following lemma, which is modified from equation (8) in Singh and Poczos~\cite{singh2014renyi}:
\begin{lemma}
\label{lem:singh}
\cite{singh2014renyi} Let the density support set $\mathcal{S}$ be the unit cube with $d\geq2$ and let the derivatives of the density $f$ vanish near the boundary of the support set, which is denoted as $\partial\mathcal{S}$. Assume that $f$ belongs to the H\"{o}lder set of order $s$, i.e. $f\in\Sigma(H,s)$. Assume that $q$ with $0\leq|q|\leq r$ and $x\in\mathcal{B}:=\{x\in\mathcal{S}|\text{dist}(x,\partial\mathcal{S})\leq h \}$. Then if $h\leq 1/2$,
\begin{equation}
    \left| D^q f(x)\right|\leq Hh^{s-|q|}.
\end{equation}
\end{lemma}
\begin{proof}
We include a proof here for completeness. We will use an induction argument on $|q|$ as $|q|$ decreases from $r$ to 0. Let $y\in\partial\mathcal{S}$ such that $||x-y||\leq h$. For $|q|=r$, we have
\begin{align*}
    |D^qf(x)|&=|D^qf(x)-D^qf(y)| \\
    & \leq H||x-y||^{s-|q|} \\
    & \leq Hh^{s-|q|},
\end{align*}
where the first step comes from the fact that $D^qf(y)=0$, the second step uses the H\"{o}lder condition, and the last step follows from the choice of $y$. Suppose we have the desired bound for derivatives of order $|q|+1$. Let $x\in\mathcal{B}$ and let $u=(0,\dots,0,\pm 1,0,\dots,0)\in\mathbb{R}^d$, where $u_j=\pm 1$ for some $j\in[d]$. Then we can write $x=y+hu$ for some $y\in\mathcal{B}$. Furthermore, the point $y+tu\in\mathcal{B}$ for $t\in[0,h]$. To see how this is true, note that if $x\in\mathcal{B}$ then there exists at least one coordinate $x_i$ such that either $0\leq x_i\leq h$ or $1-h\leq x_i\leq 1$. Pick any other coordinate $x_j$. Then choose the sign of $u_j$ such that $y=x-hu\in\mathcal{S}$. This is possible since $h\leq 1/2$.  Since $y$ is constructed by moving from $x\in\mathcal{B}$ in parallel with the boundary, then $y\in\mathcal{B}$ as well as $y+tu\in\mathcal{B}$.

Based on this construction of $y$ and $u$, we have
\begin{align*}
    |D^q f(y+hu)|&\leq \int_0^h \left| \frac{\partial}{\partial x_j}D^q f(y+tu)\right|dt \\
    &\leq \int_0^h Hh^{s-(|q|+1)}dt \\
    &=Hh^{s-|q|}.
\end{align*}
The desired result then follows by induction on $|q|$.

\end{proof}

From Lemma~\ref{lem:singh}, 
\begin{equation}
    \left|\int_{x\in\mathcal{S}}\left(\int_{u:K(u)>0,\,x+uh\notin\mathcal{S}}K^{l}(u)u^q D^q f(x) du\right)^{t}dx\right| \leq \left(Hh^{s-|q|}||K||_\infty\right)^t\left| \int_{x\in\mathcal{S}}\left(\int_{u:K(u)>0,\,x+uh\notin\mathcal{S}}u^q du\right)^tdx\right|.
    \label{eq:prop_bound}
\end{equation}

Consider the case when $K$ has rectangular support from (without loss of generality) $-1$ to 1 in each dimension. We consider the setting when $x$ is close to the 1 boundary in a single coordinate $x_i$. The inner integral in (\ref{eq:prop_bound}) then reduces to 
\begin{align}
    \left(C\int_{\frac{1-x_i}{h}}^1 u_i^m du_i\right)^t & = \left(\frac{C}{m+1}\left(1-\left(\frac{1-x_i}{h}\right)^{m+1}\right)\right)^t \\
    & = \left(\frac{C}{m+1}\right)^t\sum_{j=0}^t (-1)^j\binom{t}{j}\left(\frac{1-x_i}{h}\right)^{(m+1)j},
    \label{eq:prop_inner}
\end{align}
where $C$ is a constant that comes from integrating the other terms of $u^q$ and $m\leq |q|$ is the exponent of $u_i$ in $u^q$. Based on equation (\ref{eq:h_int}), taking the integral of this result with respect to $x$ will then yield terms of the form of 
$h^{j+1}$ for $j\geq 0$. The extension to the case with multiple boundary coordinates is similar to that in the proof of Theorem~\ref{thm:assumption5}.

Combining this with the results in equations (\ref{eq:prop_bound}) and (\ref{eq:prop_inner}) gives that 
\begin{equation}
    \left|\int_{x\in\mathcal{S}}\left(\int_{u:K(u)>0,\,x+uh\notin\mathcal{S}}K^{l}(u)u^q D^q f(x) du\right)^{t}dx\right|=O\left(h^{s-|q|+1}\right)=o\left(h^{r-|q|}\right).
\end{equation}
Therefore, assumption $\mathcal{A}.5$ is satisfied with the coefficients in the expansion being equal to zero.

The proof for the circular support kernel follows similar arguments also adapted from the proof of Theorem~\ref{thm:assumption5}.

\subsection{Assumption $\mathcal{A}.5$ and the Truncated Gaussian Distribution}
\label{sub:Gauss_truncated}
Here we show that the truncated Gaussian distribution satisfies assumption $\mathcal{A}.5$. For simplicity, we will consider the univariate case for $s=2$ and the uniform kernel. We restrict the standard normal distribution to the interval $[-1,1]$. We focus on the $+1$ boundary as the $-1$ boundary follows a similar procedure. This gives 
\begin{align}
    \int_{x\in\mathcal{S}}\left(\int_{u:K(u)>0,\,x+uh\notin\mathcal{S}}K^{l}(u)u^q D^q f(x) du\right)^{t}dx &=\int_{1-h}^1\left(\int_{\frac{1-x}{h}}^1 \frac{d^q}{dx^q}f(x)u^q du\right)^tdx \\
    & = \int_{1-h}^1\left( \left(1-\left(\frac{1-x}{h}\right)^{q+1}\right)\frac{d^q}{dx^q}\frac{f(x)}{q+1}\right)^tdx. 
    \label{eq:Gauss}
\end{align}
We can extend the derivatives of the function $f$ to the boundary by simply using the derivatives prior to truncation. By a Taylor series expansion around $x=1$, we have
\begin{align*}
    f(x)&=f(1)+f'(1)(1-x)+\frac{f''(1)}{2}(1-x)^2+o\left((1-x)^2\right), \\
    \frac{d}{dx}f(x)&=f'(1)+f''(1)(1-x)+o(1-x), \\
    \frac{d^2}{dx^2}f(x)&=f''(1)+o(1).
\end{align*}
Therefore, the integral with respect to $x$ in (\ref{eq:Gauss}) has terms of the form of $\frac{(1-x)^{j+k}}{h^k}$ which integrates to $\frac{h^{j+1}}{j+k+1}$ (see (\ref{eq:h_int})). Combining these results gives the desired expansion. As an example, suppose $t=1$ and $q=1$. Then (\ref{eq:Gauss}) becomes 
\begin{equation}
    \frac{1}{2}\int_{1-h}^1\left(1-\left(\frac{1-x}{h}\right)^{2}\right)(f'(1)+f''(1)(1-x)+o(1-x))dx.
\end{equation}

Distributing the expression for $\frac{d}{dx}f(x)$ through and evaluating the integrals separately gives
\begin{equation}
    \frac{1}{2}\int_{1-h}^1(f'(1)+f''(1)(1-x)+o(1-x))dx=\frac{1}{2}\left(f'(1)h+\frac{f''(1)h^2}{2}\right)+o(h^2),
\end{equation}
\begin{equation}
    \frac{1}{2}\int_{1-h}^1\left(\frac{1-x}{h}\right)^{2}(f'(1)+f''(1)(1-x)+o(1-x))dx=\frac{1}{2}\left( \frac{f'(1)h}{3}+\frac{f''(1)h^2}{4} \right)+o(h^2).
\end{equation}
Taking the difference between these expressions gives the final expansion:
\begin{equation}
    \frac{f'(1)h}{3}+\frac{f''(1)h^2}{8}+o(h^2).
\end{equation}

\section{Proof of Theorem~\ref{thm:bias} (Continuous Bias)}

\label{sec:biasProof}Here we prove the results shown in Theorem~\ref{thm:bias_cont_mod},
which includes the results for Theorem~\ref{thm:bias}. In this section
and throughout all other proofs, let $\ez Z$ denote the conditional
expectation given $\mathbf{Z}$. 

The bias of $\gt$ is
\[
\bias\left[\gt\right]=\bE\left[g\left(\frac{\ft X(\mathbf{X})\ft Y(\mathbf{Y})\nu_{1}\nu_{2}}{\ft Z(\mathbf{X},\mathbf{Y})\nu_{3}}\right)-g\left(\frac{f_{X}(\mathbf{X})f_{Y}(\mathbf{Y})\nu_{1}\nu_{2}}{f_{XY}(\mathbf{X},\mathbf{Y})\nu_{3}}\right)\right],
\]
where $\mathbf{X}$ and $\mathbf{Y}$ are drawn jointly from $f_{XY}$.
We will derive an expression for this in terms of the bandwidths by
applying Taylor series expansions to both the functional $g$ and
the densities.

The Taylor series expansion of $g\left(\frac{\ft X(\mathbf{X})\ft Y(\mathbf{Y})\nu_{1}\nu_{2}}{\ft Z(\mathbf{X},\mathbf{Y})\nu_{3}}\right)$
around $f_{X}(\mathbf{X})f_{Y}(\mathbf{Y})\nu_{1}\nu_{2}$ and $f_{XY}(\mathbf{X},\mathbf{Y})\nu_{3}$
gives an expansion with terms of the form of

Define $g\left(t_{1}/t_{2}\right)=g\left(t_{1},t_{2}\right).$ Define
the following terms:

\begin{align}
\et Z^{q}(\mathbf{Z}) & =\nu_{3}^{q}\left(\ft Z(\mathbf{Z})-f_{XY}(\mathbf{X},\mathbf{Y})\right)^{q},\nonumber \\
\ett XY^{q}(\mathbf{Z}) & =\left(\nu_{1}\nu_{2}\right)^{q}\left(\ft X(\mathbf{X})\ft Y(\mathbf{Y})-f_{X}(\mathbf{X})f_{Y}(\mathbf{Y})\right)^{q}.\label{eq:errTerms}
\end{align}
Then the Taylor series expansion of $g\left(\frac{\ft X(\mathbf{X})\ft Y(\mathbf{Y})\nu_{1}\nu_{2}}{\ft Z(\mathbf{X},\mathbf{Y})\nu_{3}}\right)$
around $f_{X}(\mathbf{X})f_{Y}(\mathbf{Y})\nu_{1}\nu_{2}$ and $f_{XY}(\mathbf{X},\mathbf{Y})\nu_{3}$
gives
\[
g\left(\frac{\ft X(\mathbf{X})\ft Y(\mathbf{Y})\nu_{1}\nu_{2}}{\ft Z(\mathbf{X},\mathbf{Y})\nu_{3}}\right)=\sum_{i=0}^{\infty}\sum_{j=0}^{\infty}\left(\left.\frac{\partial^{i+j}g\left(t_{1},t_{2}\right)}{\partial t_{1}^{i}\partial t_{2}^{j}}\right|_{\begin{array}{c}
t_{1}=f_{X}(\mathbf{X})f_{Y}(\mathbf{Y})\nu_{1}\nu_{2}\\
t_{2}=f_{XY}(\mathbf{X},\mathbf{Y})\nu_{3}
\end{array}}\right)\frac{\et Z^{j}(\mathbf{Z})\ett XY^{i}(\mathbf{Z})}{i!j!}.
\]
To simplify this, we focus on the terms in (\ref{eq:errTerms}). Note
that if $\nu_{i}=1$, then the terms in (\ref{eq:errTerms}) are unaffected.
For other values, $\nu_{i}^{j}$ decreases to zero as $j\rightarrow\infty$
since $0<\nu_{i}<1$. 

By the binomial theorem, we obtain 
\begin{align}
\et Z^{q}(\mathbf{Z}) & =\nu_{3}^{q}\sum_{k=0}^{q}\left(\ft Z(\mathbf{Z})\right)^{k}\left(f_{XY}(\mathbf{X},\mathbf{Y})\right)^{q-k}(-1)^{q-k},\nonumber \\
\ett XY^{q}(\mathbf{Z}) & =\left(\nu_{1}\nu_{2}\right)^{q}\sum_{k=0}^{q}\left(\ft X(\mathbf{X})\ft Y(\mathbf{Y})\right)^{k}\left(f_{X}(\mathbf{X})f_{Y}(\mathbf{Y})\right)^{q-k}(-1)^{q-k}.\label{eq:errBinom}
\end{align}
To derive the bias, we will take the conditional expectation given
$\Z$ of these terms.

Since we are not doing explicit boundary correction, we need to consider
separately the cases when $\mathbf{Z}$ is in the interior of the
support $\mathcal{S}_{X}\times\mathcal{S}_{Y}$ and when $\mathbf{Z}$
is close to the boundary of the support. For precise definitions,
a point $Z=(X,Y)\in\mathcal{S}_{X}\times\mathcal{S}_{Y}$ is in the
interior of $\mathcal{S}_{X}\times\mathcal{S}_{Y}$ if for all $Z^{'}\notin\mathcal{S}_{X}\times\mathcal{S}_{Y}$,
$K_{X}\left(\frac{X-X^{'}}{h_{X}}\right)K_{Y}\left(\frac{Y-Y^{'}}{h_{Y}}\right)=0$,
and a point $Z\in\mathcal{S}_{X}\times\mathcal{S}_{Y}$ is near the
boundary of the support if it is not in the interior. 

\subsection*{Interior Points}

We first consider the case where $\mathbf{Z}=(\mathbf{X},\mathbf{Y})$
is drawn from $f_{XY}$ in the interior of $\mathcal{S}_{X}\times\mathcal{S}_{Y}$.
It can be shown (see~\cite{moon2018ensemble}) by Taylor series expansions
of the probability densities that 

\begin{eqnarray}
\ez X\left[\ft X(\mathbf{X})\right] & = & f_{X}(\mathbf{X})+\sum_{j=1}^{\left\lfloor s/2\right\rfloor }c_{X,j}(\mathbf{X})h_{X}^{2j}+O\left(h_{X}^{s}\right),\label{eq:E_fx-1}\\
\ez Y\left[\ft Y(\mathbf{Y})\right] & = & f_{Y}(\mathbf{Y})+\sum_{j=1}^{\left\lfloor s/2\right\rfloor }c_{Y,j}(\mathbf{Y})h_{Y}^{2j}+O\left(h_{Y}^{s}\right),\nonumber \\
\ez{\Z}\left[\ft Z(\mathbf{Z})\right] & = & f_{XY}(\mathbf{X},\mathbf{Y})+\sum_{\substack{i=0\\
i+j\neq0
}
}^{\left\lfloor s/2\right\rfloor }\sum_{j=0}^{\left\lfloor s/2\right\rfloor }c_{XY,i,j}(\mathbf{X},\mathbf{Y})h_{X}^{2i}h_{Y}^{2j}+O\left(h_{X}^{s}+h_{Y}^{s}\right).\nonumber 
\end{eqnarray}
The constants in the above expressions are independent of the bandwidths
$h_{X}$ and $h_{Y}$ and only depend on the densities and their derivatives.

We will also need an expression for $\ez Z\left[\ft X(\mathbf{X})\ft Y(\mathbf{Y})\right]$:
\begin{align}
\ez Z\left[\ft X(\mathbf{X})\ft Y(\mathbf{Y})\right] & =\ez{\Z}\left[\frac{1}{N^{2}h_{X}^{d_{X}}h_{Y}^{d_{Y}}}\sum_{i=1}^{N}\sum_{j=1}^{N}K_{X}\left(\frac{\X-\X_{i}}{h_{X}}\right)K_{Y}\left(\frac{\Y-\Y_{j}}{h_{Y}}\right)\right]\nonumber \\
 & =\ez Z\left[\frac{1}{N^{2}h_{X}^{d_{X}}h_{Y}^{d_{Y}}}\sum_{i=1}^{N}K_{X}\left(\frac{\X-\X_{i}}{h_{X}}\right)K_{Y}\left(\frac{\Y-\Y_{i}}{h_{Y}}\right)\right]\nonumber \\
 & +\ez Z\left[\frac{1}{N^{2}h_{X}^{d_{X}}h_{Y}^{d_{Y}}}\sum_{\substack{i,j=1\\
i\neq j
}
}^{N}K_{X}\left(\frac{\X-\X_{i}}{h_{X}}\right)K_{Y}\left(\frac{\Y-\Y_{j}}{h_{Y}}\right)\right]\nonumber \\
 & =\frac{1}{N}\ez Z\left[\ft Z(\mathbf{Z})\right]+\frac{N^{2}-N}{N^{2}}\ez X\left[\ft X(\mathbf{X})\right]\ez Y\left[\ft Y(\mathbf{Y})\right]\nonumber \\
 & =\frac{N^{2}-N}{N^{2}}f_{X}(\mathbf{X})f_{Y}(\mathbf{Y})+\sum_{\substack{i=0\\
i+j\neq0
}
}^{\left\lfloor s/2\right\rfloor }\sum_{j=0}^{\left\lfloor s/2\right\rfloor }c_{X,Y,i,j}(\mathbf{X},\mathbf{Y})h_{X}^{2i}h_{Y}^{2j}+O\left(h_{X}^{s}+h_{Y}^{s}+\frac{1}{N}\right),\label{eq:E_marginals}
\end{align}
where we have used the fact that $\Z_{i}$ and $\Z_{j}$ are independent
when $i\neq j$.

We will also need expressions for the conditional expectation of powers
of the KDEs to simplify (\ref{eq:errBinom}). Consider first $\left(\ft Z(\mathbf{Z})\right)^{2}$.
Note that 
\[
\left(\ft Z(\mathbf{Z})\right)^{2}=\frac{1}{N^{2}h_{X}^{2d_{X}}h_{Y}^{2d_{Y}}}\sum_{i=1}^{N}\sum_{j=1}^{N}K_{X}\left(\frac{\X-\X_{i}}{h_{X}}\right)K_{Y}\left(\frac{\Y-\Y_{i}}{h_{Y}}\right)K_{X}\left(\frac{\X-\X_{j}}{h_{X}}\right)K_{Y}\left(\frac{\Y-\Y_{j}}{h_{Y}}\right).
\]
The above double sum can be split into two cases: $i=j$ ($N$ terms)
and $i\neq j$ ($N^{2}-N$ terms). When $i=j$, we have by Taylor
series expansions of the joint density $f_{XY}$:
\begin{align*}
\frac{1}{Nh_{X}^{2d_{X}}h_{Y}^{2d_{Y}}}\ez Z\left[K_{X}^{2}\left(\frac{\X-\X_{i}}{h_{X}}\right)K_{Y}^{2}\left(\frac{\Y-\Y_{i}}{h_{Y}}\right)\right] & =\frac{1}{Nh_{X}^{d_{X}}h_{Y}^{d_{Y}}}\sum_{\substack{i=0}
}^{\left\lfloor s/2\right\rfloor }\sum_{j=0}^{\left\lfloor s/2\right\rfloor }c_{XY,i,j,1}(\mathbf{X},\mathbf{Y})h_{X}^{2i}h_{Y}^{2j}+O\left(h_{X}^{s}+h_{Y}^{s}\right).
\end{align*}

When $i\ne j,$recall that $\Z_{i}$ is independent of $\Z_{j}$.
Therefore, we obtain for these terms
\begin{align*}
\frac{N^{2}-N}{N^{2}h_{X}^{2d_{X}}h_{Y}^{2d_{Y}}}\ez{\Z}\left[K_{X}\left(\frac{\X-\X_{i}}{h_{X}}\right)K_{Y}\left(\frac{\Y-\Y_{i}}{h_{Y}}\right)K_{X}\left(\frac{\X-\X_{j}}{h_{X}}\right)K_{Y}\left(\frac{\Y-\Y_{j}}{h_{Y}}\right)\right]\\
=\frac{N^{2}-N}{N^{2}}\ez Z\left[\ft Z(\mathbf{Z})\right]^{2}.
\end{align*}
Combining these results gives 
\begin{align*}
\ez Z\left[\left(\ft Z(\mathbf{Z})\right)^{2}\right] & =\frac{N^{2}-N}{N^{2}}f_{XY}(\Z)^{2}+\sum_{\substack{i=0\\
i+j\neq0
}
}^{\left\lfloor s/2\right\rfloor }\sum_{j=0}^{\left\lfloor s/2\right\rfloor }c_{XY,i,j,2}(\mathbf{X},\mathbf{Y})h_{X}^{2i}h_{Y}^{2j}\\
 & +\sum_{\substack{i=0}
}^{\left\lfloor s/2\right\rfloor }\sum_{j=0}^{\left\lfloor s/2\right\rfloor }\frac{c_{XY,i,j,3}(\X,\Y)}{Nh_{X}^{d_{X}}h_{Y}^{d_{Y}}}h_{X}^{2i}h_{Y}^{2j}+O\left(h_{X}^{s}+h_{Y}^{s}\right).
\end{align*}

For the cross term ($k=1$) in (\ref{eq:errBinom}) when $q=2$, we
obtain 
\[
\ez Z\left[\ft Z(\mathbf{Z})\right]f_{XY}(\Z)=f_{XY}(\Z)^{2}+f_{XY}(\Z)\sum_{\substack{i=0\\
i+j\neq0
}
}^{\left\lfloor s/2\right\rfloor }\sum_{j=0}^{\left\lfloor s/2\right\rfloor }c_{XY,i,j}(\mathbf{X},\mathbf{Y})h_{X}^{2i}h_{Y}^{2j}+O\left(h_{X}^{s}+h_{Y}^{s}\right).
\]
Combining these results gives 
\begin{align*}
\ez Z\left[\et Z^{2}(\mathbf{Z})\right] & =\sum_{\substack{i=0\\
i+j\neq0
}
}^{\left\lfloor s/2\right\rfloor }\sum_{j=0}^{\left\lfloor s/2\right\rfloor }c_{XY,i,j,4}(\mathbf{X},\mathbf{Y})h_{X}^{2i}h_{Y}^{2j}+\sum_{\substack{i=0}
}^{\left\lfloor s/2\right\rfloor }\sum_{j=0}^{\left\lfloor s/2\right\rfloor }\frac{c_{XY,i,j,5}(\X,\Y)}{Nh_{X}^{d_{X}}h_{Y}^{d_{Y}}}h_{X}^{2i}h_{Y}^{2j}\\
 & -\frac{f_{XY}(\Z)^{2}}{N}+O\left(h_{X}^{s}+h_{Y}^{s}\right).
\end{align*}

By following similar procedures, it can be shown that for $q\geq2$
\begin{align}
\ez Z\left[\et Z^{q}(\mathbf{Z})\right] & =\sum_{\substack{i=0\\
i+j\neq0
}
}^{\left\lfloor s/2\right\rfloor }\sum_{j=0}^{\left\lfloor s/2\right\rfloor }c_{XY,i,j,q,1}(\mathbf{X},\mathbf{Y})h_{X}^{2i}h_{Y}^{2j}+\sum_{\substack{i=0}
}^{\left\lfloor s/2\right\rfloor }\sum_{j=0}^{\left\lfloor s/2\right\rfloor }\frac{c_{XY,i,j,q,2}(\X,\Y)}{\left(Nh_{X}^{d_{X}}h_{Y}^{d_{Y}}\right)^{q-1}}h_{X}^{2i}h_{Y}^{2j}\nonumber \\
 & +O\left(\frac{1}{N^{q-1}}+h_{X}^{s}+h_{Y}^{s}\right).\label{eq:ez}
\end{align}
In particular, the terms related to $f_{XY}(\Z)^{q}$ all combine
to be $O(1/N)$. In the $q=2$ example above, we end up with 
\[
f_{XY}(\Z)^{2}-2f_{XY}(\Z)^{2}+\frac{N^{2}-N}{N^{2}}f_{XY}(\Z)^{2}=-\frac{f_{XY}(\Z)^{2}}{N}.
\]
As another example, for $q=3$, we end up with 
\begin{align*}
\left(\frac{N(N-1)(N-2)}{N^{3}}-3\frac{N^{2}-N}{N^{2}}+3-1\right)f_{XY}(\Z)^{3} & =\left(\frac{N^{3}-2N^{2}-N^{2}+2N-N^{3}}{N^{3}}-3\frac{N^{2}-N-N^{2}}{N^{2}}\right)f_{XY}(\Z)^{3}\\
 & =\left(\frac{-3N+2+3N}{N^{2}}\right)f_{XY}(\Z)^{3}\\
 & =O\left(\frac{1}{N^{2}}\right).
\end{align*}
A similar pattern holds for higher values of $q$. 

Then by following a similar process, we obtain for $q\geq2$
\begin{align*}
\ez Z\left[\ett XY^{q}(\mathbf{Z})\right] & =\sum_{\substack{i=0\\
i+j\neq0
}
}^{\left\lfloor s/2\right\rfloor }\sum_{j=0}^{\left\lfloor s/2\right\rfloor }c_{XY,i,j,q,3}(\mathbf{X},\mathbf{Y})h_{X}^{2i}h_{Y}^{2j}+\sum_{\substack{i=0}
}^{\left\lfloor s/2\right\rfloor }\sum_{j=0}^{\left\lfloor s/2\right\rfloor }\frac{c_{XY,i,j,q,4}(\X,\Y)}{\left(Nh_{X}^{d_{X}}\right)^{q-1}}h_{X}^{2i}h_{Y}^{2j}\\
 & +\sum_{\substack{i=0}
}^{\left\lfloor s/2\right\rfloor }\sum_{j=0}^{\left\lfloor s/2\right\rfloor }\frac{c_{XY,i,j,q,5}(\X,\Y)}{\left(Nh_{Y}^{d_{Y}}\right)^{q-1}}h_{X}^{2i}h_{Y}^{2j}+O\left(\frac{1}{N^{q-1}}+h_{X}^{s}+h_{Y}^{s}\right).
\end{align*}

\subsection*{Points Near the Boundary}

For a point near the boundary of the support, we extend the expectation
beyond the support of the density. As an example if $\mathbf{X}$
is near the boundary of $\mathcal{S}_{X}$, then we get 
\begin{align}
\ez X\left[\ft X(\mathbf{X})\right]-f_{X}(\X) & =\frac{1}{h_{X}^{d_{X}}}\int_{V:V\in\mathcal{S}_{X}}K_{X}\left(\frac{\mathbf{X}-V}{h_{X}}\right)f_{X}(V)dV-f_{X}(\mathbf{X})\nonumber \\
 & =\left[\frac{1}{h_{X}^{d_{X}}}\int_{V:K_{X}\left(\frac{\mathbf{X}-V}{h_{X}}\right)>0}K_{X}\left(\frac{\mathbf{X}-V}{h_{X}}\right)f_{X}(V)dV-f_{X}(\mathbf{X})\right]\nonumber \\
 & -\left[\frac{1}{h_{X}^{d_{X}}}\int_{V:V\notin\mathcal{S}_{X}}K_{X}\left(\frac{\mathbf{X}-V}{h_{X}}\right)f_{X}(V)dV\right]\nonumber \\
 & =T_{1,X}(\mathbf{X})-T_{2,X}(\mathbf{X}).\label{eq:Tdiff-1}
\end{align}

Note that we are technically evaluating the density $f_{X}$ at points
outside of the support in $T_{1,X}(\X)$. However, to obtain an expression
for this integral, we take a Taylor series expansion of $f_{X}$ at
the point $\X$ which is inside the support. Thus the exact manner
in which we define the extension of $f_{X}$ does not matter as long
as the Taylor series remains the same and as long as the extension
is smooth. Thus the expected value of $T_{1,X}(\X)$ gives an expression
similar to that of the interior point case in (\ref{eq:E_fx-1}).

For the $T_{2,X}(\mathbf{X})$ term, we can use multi-index notation
on the expansion of $f_{X}$ to show that 
\begin{align*}
T_{2,X}(\mathbf{X}) & =\left[\frac{1}{h_{X}^{d_{X}}}\int_{V:V\notin\mathcal{S}_{X}}K_{X}\left(\frac{\mathbf{X}-V}{h_{X}}\right)f_{X}(V)dV\right]\\
 & =\int_{u:h_{X}u+\mathbf{X}\notin\mathcal{S}_{X},K_{X}(u)>0}K_{X}(u)f_{X}(\mathbf{X}+h_{X}u)du\\
 & =\sum_{|\alpha|\leq r}\frac{h_{X}^{|\alpha|}}{\alpha!}\int_{u:h_{X}u+\mathbf{X}\notin\mathcal{S}_{X},K_{X}(u)>0}K_{X}(u)D^{\alpha}f_{X}(\mathbf{X})u^{\alpha}du+o(h_{X}^{r}).
\end{align*}
Then since the $|\alpha|$th derivative of $f_{X}$ is $r-|\alpha|$
times differentiable, we apply the condition in assumption $\mathcal{A}.5$
to obtain 
\[
\bE\left[T_{2,X}(\mathbf{X})\right]=\sum_{i=1}^{r}e_{i}h_{X}^{i}+o\left(h_{X}^{r}\right).
\]
Similar expressions can be obtained for $\ft Y$, $\ft Z$, and the
product $\ft Y\ft X$.

The above results considers $\et Z^{q}(\mathbf{Z})$ and $\ett XY^{q}(\mathbf{Z})$
for $q=1$. We now consider when $q\geq2$. We follow a similar procedure
where we extend the density beyond the support, but only evaluate
the densities and their derivatives at points within the support.
Thus by the binomial theorem, we can write
\begin{align*}
\ez Z\left[\et Z^{q}(\mathbf{Z})\right] & =\nu_{3}^{q}\sum_{k=0}^{q}\ez Z\left[\left(\ft Z(\mathbf{Z})\right)^{k}\right]\left(f_{XY}(\mathbf{X},\mathbf{Y})\right)^{q-k}(-1)^{q-k}\\
 & =\nu_{3}^{q}\sum_{k=0}^{q}\ez Z\left[\left(\ft Z(\mathbf{Z})\right)^{k}\right]_{extended}\left(f_{XY}(\mathbf{X},\mathbf{Y})\right)^{q-k}(-1)^{q-k}\\
 & -\nu_{3}^{q}\sum_{k=1}^{q}\ez Z\left[\left(\ft Z(\mathbf{Z})\right)^{k}\right]_{outside}\left(f_{XY}(\mathbf{X},\mathbf{Y})\right)^{q-k}(-1)^{q-k}\\
 & =T_{1,q,Z}(\Z)-T_{2,q,Z}(\Z).
\end{align*}
As before, $T_{1,q,Z}(\Z)$ corresponds to the case where we have
extended the density beyond the support and results in terms of the
form in (\ref{eq:ez}). $T_{2,q,Z}(\Z)$ corresponds to the case where
we integrate outside of the boundary. The additional powers applied
to the KDE simply result in terms with the kernel raised to a power
or $\ez Z\left[\ft Z(\mathbf{Z})\right]$ raised to a power. By applying
assumption $\mathcal{A}.5$, we obtain $\bE\left[T_{2,q,X}(\mathbf{X})\right]=\sum_{i=1}^{r}\sum_{j=1}^{r}e_{q,i}h_{X}^{i}h_{Y}^{j}+o\left(h_{X}^{r}+h_{Y}^{r}\right)$.
Similar results are obtained for $\ett XY^{q}(\mathbf{Z})$.

Combining the results for the interior points and points near the
boundary completes the proof.

\section{Proof of Theorem~\ref{thm:variance} (Continuous Variance)}

\label{sec:VarProof}Here we prove Theorem~\ref{thm:variance}. The
proof uses the Efron-Stein inequality~\cite{steele1986efron,efron1981jackknife}: 
\begin{lem}
[Efron-Stein Inequality] Let $\mathbf{X}_{1},\dots,\mathbf{X}_{n},\mathbf{X}_{1}^{'},\dots,\mathbf{X}_{n}^{'}$
be independent random variables on the space $\mathcal{S}$. Then
if $f:\mathcal{S}\times\dots\times\mathcal{S}\rightarrow\mathbb{R}$,
we have that 
\[
\var\left[f(\mathbf{X}_{1},\dots,\mathbf{X}_{n})\right]\leq\frac{1}{2}\sum_{i=1}^{n}\bE\left[\left(f(\mathbf{X}_{1},\dots,\mathbf{X}_{n})-f(\mathbf{X}_{1},\dots,\mathbf{X}_{i}^{'},\dots,\mathbf{X}_{n})\right)^{2}\right].
\]
\end{lem}
In this case we consider the samples $\left\{ \mathbf{Z}_{1},\dots,\mathbf{Z}_{N}\right\} $
and $\left\{ \mathbf{Z}_{1}^{'},\Z_{2}\dots,\mathbf{Z}_{N}\right\} $
and the respective estimators $\gt$ and $\gt^{'}$. By the triangle
inequality, 
\begin{eqnarray}
\left|\gt-\gt^{'}\right| & \leq & \frac{1}{N}\left|g\left(\frac{\ft X(\mathbf{X}_{1})\ft Y(\mathbf{Y}_{1})\nu_{1}\nu_{2}}{\ft Z(\mathbf{X}_{1},\mathbf{Y}_{1})\nu_{3}}\right)-g\left(\frac{\ft X(\mathbf{X}_{1}^{'})\ft Y(\mathbf{Y}_{1}^{'})\nu_{1}\nu_{2}}{\ft Z(\mathbf{X}_{1}^{'},\mathbf{Y}_{1}^{'})\nu_{3}}\right)\right|\nonumber \\
 &  & +\frac{1}{N}\sum_{j=2}^{N_{2}}\left|g\left(\frac{\ft X(\mathbf{X}_{j})\ft Y(\mathbf{Y}_{j})\nu_{1}\nu_{2}}{\ft Z(\mathbf{X}_{j},\mathbf{Y}_{j})\nu_{3}}\right)-g\left(\frac{\ft X^{'}(\mathbf{X}_{j})\ft Y^{'}(\mathbf{Y}_{1})\nu_{1}\nu_{2}}{\ft Z^{'}(\mathbf{X}_{1},\mathbf{Y}_{1})\nu_{3}}\right)\right|.\label{eq:triangle}
\end{eqnarray}
By the Lipschitz condition on $g$, the first term in (\ref{eq:triangle})
can be decomposed into terms of the form of 
\[
\nu_{3}\left|\ft Z(\mathbf{Z}_{1})-\ft Z(\mathbf{Z}_{1}^{'})\right|,
\]
\[
\nu_{1}\nu_{2}\left|\ft X(\mathbf{X}_{1})\ft Y(\mathbf{Y}_{1})-\ft X(\mathbf{X}_{1}^{'})\ft Y(\mathbf{Y}_{1}^{'})\right|.
\]

For the $\ft Z(\mathbf{Z}_{1})$ term, we first apply Jensen's inequality:

\begin{align*}
\bE\left[\left|\ft Z(\mathbf{Z}_{1})-\ft Z(\mathbf{Z}_{1}^{'})\right|^{2}\right] & =\bE\left[\frac{1}{N^{2}h_{X}^{2d_{X}}h_{Y}^{2d_{Y}}}\left(\sum_{i=1}^{N}\left(K_{X}\left(\frac{\mathbf{X}_{1}-\mathbf{X}_{i}}{h_{X}}\right)K_{Y}\left(\frac{\mathbf{Y}_{1}-\mathbf{Y}_{j}}{h_{Y}}\right)-\left(\frac{\mathbf{X}_{1}^{'}-\mathbf{X}_{i}}{h_{X}}\right)K_{Y}\left(\frac{\mathbf{Y}_{1}^{'}-\mathbf{Y}_{j}}{h_{Y}}\right)\right)\right)^{2}\right]\\
 & \leq\frac{1}{Nh_{X}^{2d_{X}}h_{Y}^{2d_{Y}}}\sum_{i=1}^{N}\bE\left[\left(K_{X}\left(\frac{\mathbf{X}_{1}-\mathbf{X}_{i}}{h_{X}}\right)K_{Y}\left(\frac{\mathbf{Y}_{1}-\mathbf{Y}_{j}}{h_{Y}}\right)-\left(\frac{\mathbf{X}_{1}^{'}-\mathbf{X}_{i}}{h_{X}}\right)K_{Y}\left(\frac{\mathbf{Y}_{1}^{'}-\mathbf{Y}_{j}}{h_{Y}}\right)\right)^{2}\right]
\end{align*}
By making the substitutions $\mathbf{u}_{i}=\frac{\X_{1}-\X_{i}}{h_{X}}$,
$\mathbf{v}_{i}=\frac{\Y_{1}-\Y_{i}}{h_{Y}}$, $\mathbf{u}_{i}^{'}=\frac{\X_{1}-\X_{i}}{h_{X}}$,
and $\mathbf{v}_{i}^{'}=\frac{\Y_{1}^{'}-\Y_{i}}{h_{Y}}$ in the expectation,
we obtain
\[
\bE\left[\frac{1}{h_{X}^{2d_{X}}h_{Y}^{2d_{Y}}}\left(K_{X}\left(\frac{\mathbf{X}_{1}-\mathbf{X}_{i}}{h_{X}}\right)K_{Y}\left(\frac{\mathbf{Y}_{1}-\mathbf{Y}_{j}}{h_{Y}}\right)-\left(\frac{\mathbf{X}_{1}^{'}-\mathbf{X}_{i}}{h_{X}}\right)K_{Y}\left(\frac{\mathbf{Y}_{1}^{'}-\mathbf{Y}_{j}}{h_{Y}}\right)\right)^{2}\right]
\]
\begin{align*}
 & =\frac{1}{h_{X}^{2d_{X}}h_{Y}^{2d_{Y}}}\int\left(K_{X}\left(\frac{\mathbf{X}_{1}-\mathbf{X}_{i}}{h_{X}}\right)K_{Y}\left(\frac{\mathbf{Y}_{1}-\mathbf{Y}_{j}}{h_{Y}}\right)-\left(\frac{\mathbf{X}_{1}^{'}-\mathbf{X}_{i}}{h_{X}}\right)K_{Y}\left(\frac{\mathbf{Y}_{1}^{'}-\mathbf{Y}_{j}}{h_{Y}}\right)\right)^{2}f_{Z}\left(\Z_{i}\right)f_{Z}\left(\Z_{1}^{'}\right)f_{Z}\left(\Z_{1}\right)d\Z_{i}d\Z_{1}d\Z_{1}^{'}\\
 & \leq2||K_{X}\cdot K_{Y}||_{\infty}^{2}.
\end{align*}
This gives 
\[
\bE\left[\nu_{3}^{2}\left|\ft Z(\mathbf{Z}_{1})-\ft Z(\mathbf{Z}_{1}^{'})\right|^{2}\right]\leq2||K_{X}\cdot K_{Y}||_{\infty}^{2},
\]
where we have used the fact that $\nu_{3}\leq1$.

For the product of the marginal KDEs, we have that 
\begin{eqnarray*}
\ft X(\mathbf{X}_{1})\ft Y(\mathbf{Y}_{1}) & = & \frac{1}{N^{2}h_{X}^{d_{X}}h_{Y}^{d_{Y}}}\sum_{i=2}^{N}\sum_{j=2}^{N}K_{X}\left(\frac{\mathbf{X}_{1}-\mathbf{X}_{i}}{h_{X}}\right)K_{Y}\left(\frac{\mathbf{Y}_{1}-\mathbf{Y}_{j}}{h_{Y}}\right)\\
 & = & \frac{1}{N}\ft Z(\mathbf{Z}_{1})+\frac{1}{N^{2}h_{X}^{d_{X}}h_{Y}^{d_{Y}}}\sum_{i\neq j}K_{X}\left(\frac{\mathbf{X}_{1}-\mathbf{X}_{i}}{h_{X}}\right)K_{Y}\left(\frac{\mathbf{Y}_{1}-\mathbf{Y}_{j}}{h_{Y}}\right).
\end{eqnarray*}
By applying the triangle inequality, Jensen's inequality, and similar
substitutions, we get 
\begin{eqnarray*}
\bE\left[\nu_{1}^{2}\nu_{2}^{2}\left|\ft X(\mathbf{X}_{1})\ft Y(\mathbf{Y}_{1})-\ft X(\mathbf{X}_{1}^{'})\ft Y(\mathbf{Y}_{1}^{'})\right|^{2}\right] & \leq & \bE\left[\frac{2}{N^{2}}\left|\ft Z(\mathbf{Z}_{1})-\ft Z(\mathbf{Z}_{1}^{'})\right|^{2}\right]\\
 &  & +\frac{2(M-1)}{N^{3}h_{X}^{2d_{X}}h_{Y}^{2d_{Y}}}\times\\
 &  & \sum_{i\neq j}\bE\left[\left(K_{X}\left(\frac{\mathbf{X}_{1}-\mathbf{X}_{i}}{h_{X}}\right)K_{Y}\left(\frac{\mathbf{Y}_{1}-\mathbf{Y}_{j}}{h_{Y}}\right)\right.\right.\\
 &  & \left.\left.-K_{X}\left(\frac{\mathbf{X}_{1}^{'}-\mathbf{X}_{i}}{h_{X}}\right)K_{Y}\left(\frac{\mathbf{Y}_{1}^{'}-\mathbf{Y}_{j}}{h_{Y}}\right)\right)^{2}\right]\\
 & \leq & \frac{4+2(N-1)^{2}}{N^{2}}||K_{X}\cdot K_{Y}||^{2}.
\end{eqnarray*}

For the second term in (\ref{eq:triangle}), it can be shown that
(see~\cite{moon2018ensemble})
\begin{eqnarray*}
\bE\left[\nu_{3}^{2}\left|\ft Z(\mathbf{Z}_{i})-\ft Z^{'}(\mathbf{Z}_{i})\right|^{2}\right] & = & \frac{\nu_{3}^{2}}{N^{2}h_{X}^{2d_{X}}h_{Y}^{2d_{Y}}}\bE\left[\left(K_{X}\left(\frac{\mathbf{X}_{1}-\mathbf{X}_{i}}{h_{X}}\right)K_{Y}\left(\frac{\mathbf{Y}_{1}-\mathbf{Y}_{j}}{h_{Y}}\right)\right.\right.\\
 &  & \left.\left.-K_{X}\left(\frac{\mathbf{X}_{1}^{'}-\mathbf{X}_{i}}{h_{X}}\right)K_{Y}\left(\frac{\mathbf{Y}_{1}^{'}-\mathbf{Y}_{j}}{h_{Y}}\right)\right)^{2}\right]\\
 & \leq & \frac{2||K_{X}\cdot K_{Y}||_{\infty}^{2}}{N^{2}}.
\end{eqnarray*}
By a similar approach, 
\[
\ft X(\mathbf{X}_{i})\ft Y(\mathbf{Y}_{i})-\ft X^{'}(\mathbf{X}_{i})\ft Y^{'}(\mathbf{Y}_{i})
\]
\begin{eqnarray*}
 & = & \ft Z(\mathbf{Z}_{i})-\ft Z^{'}(\mathbf{Z}_{i})+\frac{1}{M^{2}h_{X}^{d_{X}}h_{Y}^{d_{Y}}}\left(\sum_{\substack{n=2\\
n\neq i
}
}K_{Y}\left(\frac{\mathbf{Y}_{i}-\mathbf{Y}_{n}}{h_{Y}}\right)\left(K_{X}\left(\frac{\mathbf{X}_{i}-\mathbf{X}_{1}}{h_{X}}\right)-K_{X}\left(\frac{\mathbf{X}_{i}-\mathbf{X}_{1}^{'}}{h_{X}}\right)\right)\right.\\
 &  & \left.+\sum_{\substack{n=2\\
n\neq i
}
}K_{X}\left(\frac{\mathbf{X}_{i}-\mathbf{X}_{n}}{h_{X}}\right)\left(K_{Y}\left(\frac{\mathbf{Y}_{i}-\mathbf{Y}_{1}}{h_{Y}}\right)-K_{Y}\left(\frac{\mathbf{Y}_{i}-\mathbf{Y}_{1}^{'}}{h_{Y}}\right)\right)\right),
\end{eqnarray*}
\[
\implies\bE\left[\nu_{1}^{2}\nu_{2}^{2}\left|\ft X(\mathbf{X}_{i})\ft Y(\mathbf{Y}_{i})-\ft X^{'}(\mathbf{X}_{i})\ft Y^{'}(\mathbf{Y}_{i})\right|^{2}\right]\leq6||K_{X}\cdot K_{Y}||_{\infty}^{2}\left(\frac{1}{N^{2}}+\frac{(N-2)^{2}}{N^{4}}\right)
\]

We can then apply the Cauchy Schwarz inequality to bound the square
of the second term in (\ref{eq:triangle}) to get 
\[
\bE\left[\left(\sum_{j=2}^{N_{2}}\left|g\left(\frac{\ft X(\mathbf{X}_{1})\ft Y(\mathbf{Y}_{1})}{\ft Z(\mathbf{X}_{1},\mathbf{Y}_{1})}\right)-g\left(\frac{\ft X^{'}(\mathbf{X}_{1})\ft Y^{'}(\mathbf{Y}_{1})}{\ft Z^{'}(\mathbf{X}_{1},\mathbf{Y}_{1})}\right)\right|\right)^{2}\right]\leq14C_{g}^{2}||K_{X}\cdot K_{Y}||_{\infty}^{2}.
\]
Applying Jensen's inequality in conjunction with these results gives
\[
\bE\left[\left|\gt-\gt^{'}\right|^{2}\right]\leq\frac{44C_{g}^{2}||K_{X}\cdot K_{Y}||_{\infty}^{2}}{N^{2}}.
\]
Applying the Efron-Stein inequality finishes the proof.

\section{Proof of Minimax Rates (Theorem~\ref{thm:minimax})}

\label{sec:MinimaxProof}Here we present a proof of the minimax lower
bound on MI estimation convergence rates given in Theorem~\ref{thm:minimax}.
We follow a similar approach to that given in~\cite{birge1995estimation,krishnamurthy2014divergence,kandasamy2015nonparametric},
which uses the standard approach of Le Cam's method~\cite{le2012asymptotic}.
However, the minimax theory previously derived in these references
are not directly applicable to the MI estimation setting. On the one
hand, MI estimation could be considered to be similar to the entropy
estimation problem as MI is technically a functional of just the joint
distribution as the marginals are derived from the joint. However,
this viewpoint results in a very complicated functional of the joint
distribution, and it is not at all obvious if the theory derived in
\cite{birge1995estimation} is applicable. On the other hand, MI estimation
could be viewed as a divergence estimation problem between the joint
and the product of marginals. However, perturbing the joint density,
as is done in Le Cam's method, should result in a perturbation of
the marginals as well. This is not accounted for in existing approaches
for the divergence estimation problem~\cite{krishnamurthy2014divergence,kandasamy2015nonparametric}.
Therefore, we tailor Le Cam's method directly to the MI estimation
case to avoid these problems.

Le Cam's method reduces the estimation problem to a testing problem
which can then be used to characterize the minimax rate. We will need
the squared Hellinger distance between two densities $p$ and $q$,
which is defined as 
\[
H^{2}(P,Q)=\int\left(\sqrt{p(x,y)}-\sqrt{q(x,y)}\right)^{2}dxdy.
\]

For this proof, we present a shift in notation for clarity. We can
write 
\[
I(\X;\Y)=I\left(f_{XY},f_{X}f_{Y}\right).
\]
We now present Le Cam's method adapted to the continuous MI estimation
setting:
\begin{lem}
\label{lem:LeCam}Let $I$ be a functional defined on some subset
of a parameter space $\Theta\times\Theta$ which contains $\left(f_{XY},f_{X}f_{Y}\right)$
and $\left(f_{XY,\lambda},f_{X}f_{Y}\right)$ for all $\lambda$ in
some index set $\Lambda$. Denote the distributions of $f_{XY},f_{X}f_{Y},$
and $f_{XY,\lambda}$ as $F_{XY}$, $F_{X}F_{Y}$, and $F_{XY,\lambda}$,
respectively. Define $\bar{F}_{XY}^{N}=\frac{1}{|\Lambda|}\sum_{\lambda\in\Lambda}F_{XY,\lambda}^{N}$.
Consider the following two conditions for $\gamma<2$ and $\beta>0$:

\[
(i)\,\,H^{2}\left(\bar{F}_{XY}^{N}\times\left(F_{X}F_{Y}\right)^{N},F_{XY}^{N}\times\left(F_{X}F_{Y}\right)^{N}\right)\leq\gamma<2,
\]
\[
(ii)\,\,I\left(f_{XY},f_{X}f_{Y}\right)\geq2\beta+I\left(f_{XY,\lambda},f_{X}f_{Y}\right)\,\forall\lambda\in\Lambda
\]
\[
\text{OR}\,I\left(f_{XY,\lambda},f_{X}f_{Y}\right)\geq2\beta+I\left(f_{XY},f_{X}f_{Y}\right)\,\forall\lambda\in\Lambda.
\]
 Then 
\[
\inf_{\hat{G}_{N}}\sup_{f_{XY}\in\Theta}\Pr\left[\left|\hat{G}_{N}-I\left(f_{XY},f_{X}f_{Y}\right)\right|>\beta\right]\geq\frac{1}{2}\left(1-\sqrt{\gamma(1-\gamma/4)}\right).
\]
\end{lem}
Lemma~\ref{lem:LeCam} is nearly identical to Lemma 7 in Krishnamurthy
et al. \cite{krishnamurthy2014divergence} which is itself a modification
of Theorem 2.2 in Tsybakov~\cite{tsybakov2008introduction}. The
main change is the addition of the second option to condition (ii).
Thus the only required addition to the proof is to check that the
second option is indeed sufficient. Since we consider the probability
of the absolute value of $\hat{G}_{N}-I\left(f_{XY},f_{X}f_{Y}\right)$
being greater than $\beta$, it is clear that following the original
proof with the second option will give the same result as the first
option in condition (ii).

Note that we are only required to perturb the joint density $f_{XY}$
to derive the minimax rate. In general, perturbing the joint density
will result in perturbed marginal densities as well. However, the
following Lemma shows that it is possible to construct perturbation
functions that do not affect the marginal densities. 
\begin{lem}
\label{lem:functions}Let $R_{1},\dots,R_{\ell}$ be a partition of
$[0,1]^{d_{X}+d_{Y}}$, each being cubes with side length $\ell^{-1/\left(d_{X}+d_{Y}\right)}$.
Then there exists functions $u_{1},\dots,u_{\ell}$ such that, 
\[
\text{supp}(u_{j})\subset\left\{ \left.z\right|B\left(z,\epsilon\right)\subset R_{j}\right\} ,
\]
\[
\int u_{j}^{2}(x,y)dxdy\in\Theta\left(\ell^{-1}\right),
\]
\[
\int u_{j}(x,y)dx=\int u_{j}(x,y)dy=0,
\]
\[
\int g'\left(\frac{f_{X}(x)f_{Y}(y)}{f_{XY}(x,y)}\right)\frac{f_{X}(x)f_{Y}(y)u_{j}(x,y)}{f_{XY}(x,y)}dxdy=0,
\]
\[
\left\Vert D^{r}u_{j}\right\Vert _{\infty}\leq\ell^{r/\left(d_{X}+d_{Y}\right)}\,\forall r\,s.t.\,\sum_{j}r_{j}\leq s+1,
\]
where $B(z,\epsilon)$ denotes an $L_{2}$ ball around $z=\left[\begin{array}{c}
x\\
y
\end{array}\right]$ with radius $\epsilon$$\in(0,1)$.
\end{lem}
\begin{IEEEproof}
Let $\epsilon>0$. We can construct two orthonormal systems of $q>3$
functions. Construct the first system on $[0,1]^{d_{X}}$ such that
$\phi_{X,1}=1$, $\text{supp}\left(\phi_{X,j}\right)\subset[\epsilon,1-\epsilon]^{d_{X}}$,
and $\left\Vert D^{r}\phi_{X,j}\right\Vert _{\infty}\leq J_{X}<\infty$
for all $j$. The second system is constructed on $[0,1]^{d_{Y}}$such
that $\phi_{Y,1}=1$, $\text{supp}\left(\phi_{Y,j}\right)\subset[\epsilon,1-\epsilon]^{d_{Y}}$,
and $\left\Vert D^{r}\phi_{Y,j}\right\Vert _{\infty}\leq J_{Y}<\infty$
for all $j$. We can then construct a combined orthonormal system
$\phi_{i,j}=\phi_{X,i}\phi_{Y,j}$ which has $q^{2}$ functions. It
is clear that this is an orthonormal system since $\int\phi_{i,j}(x,y)\phi_{m,n}(x,y)dxdy=1$
when $i=m$ and $j=n$ and zero otherwise. Also, $\phi_{1,1}=1$,
$\text{supp}\left(\phi_{i,j}\right)\subset[\epsilon,1-\epsilon]^{d_{X}+d_{Y}},$
and there exists some $J<\infty$ such that $\left\Vert D^{r}\phi_{i,j}\right\Vert _{\infty}\leq J$
for all $i,j$.

Now for any given function $f\in L_{2}\left([0,1]^{d_{X}+d_{Y}}\right)$,
we can find a unit-normed function $v\in\text{span}\left(\left\{ \phi_{i,j}\right\} \right)$
such that $v\perp\phi_{1,i}$ for all $i$, $v\perp\phi_{i,1}$ for
all $i$, and $v\perp f$. We can write $v=\sum_{i,j=1}^{q}a_{i}b_{j}\phi_{i,j}$.
Then $D^{r}v=\sum_{i,j=1}^{q}a_{i}b_{j}D^{r}\phi_{i,j}$ which implies
that 
\[
\left\Vert D^{r}v\right\Vert _{\infty}\leq J\sum_{i,j}\left|a_{i}b_{j}\right|\leq Jq,
\]
where the last inequality comes from the fact that $v$ is unit-normed.
Now define $\nu=\frac{1}{Jq}v$. Then clearly $\int\nu^{2}(x,y)dxdy$
is upper and lower bounded and we have that $\left\Vert D^{r}\nu\right\Vert _{\infty}\leq1$.

We can now construct the functions $u_{j}$. First map $R_{j}$ to
$[0,1]^{d_{X}+d_{Y}}$by scaling it appropriately. Then set 
\[
u_{j}(x,y)=\nu\left(\ell^{1/\left(d_{X}+d_{Y}\right)}\left(\left[\begin{array}{c}
x\\
y
\end{array}\right]-\mathbf{j}\right)\right),
\]
where $\mathbf{j}$ is the point in $R_{j}$ that is mapped to $\mathbf{0}$
after scaling. This maps $[0,1]^{d_{X}+d_{Y}}$ back to $R_{j}$ while
inheriting the properties derived from the construction of $\nu$.
Now let $f$ be $g'\left(\frac{f_{X}(x)f_{Y}(y)}{f_{XY}(x,y)}\right)\frac{f_{X}(x)f_{Y}(y)}{f_{XY}(x,y)}$
constrained to $R_{j}$ and scaled to fit $[0,1]^{d_{X}+d_{Y}}$.
Conditions 1, 3, and 4 above are then fulfilled by construction. Also
$\int_{R_{j}}u_{j}^{2}(x,y)dxdy=\frac{1}{\ell}\nu^{2}(x,y)dxdy\in\Theta\left(\ell^{-1}\right)$
which is condition 2 above. It's also clear that $\left\Vert D^{r}u_{j}\right\Vert _{\infty}\leq\ell^{r/\left(d_{X}+d_{Y}\right)}$
which is condition 5, completing the proof.
\end{IEEEproof}
We now can prove Theorem~\ref{thm:minimax}. We will construct the
conditions necessary to apply Lemma~\ref{lem:LeCam}. Apply Lemma~\ref{lem:functions}
to obtain an index set $\tilde{\Lambda}=\{-1,1\}^{\ell}$ and functions
$u_{1},\dots,u_{\ell}$. Define the following set of perturbed functions
around $f_{XY}$:
\[
\Lambda=\left\{ f_{XY,\lambda}=f_{XY}+H_{1}\sum_{j=1}^{\ell}\lambda_{j}u_{j}|\lambda_{j}\in\tilde{\Lambda}\right\} .
\]
This will form our set of alternatives. Due to the fact that $\int u_{j}(x,y)dx=\int u_{j}(x,y)dy=0$,
we have that 
\[
\int f_{XY,\lambda}(x,y)dx=f_{Y}(y),
\]
\[
\int f_{XY,\lambda}(x,y)dy=f_{X}(x).
\]
That is, the perturbations on $f_{XY}$ are chosen so that the resulting
marginal distributions are unperturbed. 

The perturbation functions $u_{j}$ in Lemma~\ref{lem:functions}
are restricted to the small $R_{j}$ bins and thus violate the H\"{o}lder
class assumption. However, by scaling $H_{1}$ appropriately, we can
ensure that $f_{XY,\lambda}\in\Sigma(s,H)$. We show this by following
the same argument as in Krishnamurthy et al.~\cite{krishnamurthy2014divergence},
which we repeat here for completeness. Define $u_{\lambda}=H_{1}\sum_{j=1}^{\ell}\lambda_{j}u_{j}$.
We will first show that $u_{\lambda}$ is H\"{o}lder smooth. Then
$f_{XY,\lambda}$ is H\"{o}lder smooth by the triangle inequality.
For $u_{\lambda}$, fix two points $v,z\in\mathbb{R}^{d_{X}+d_{Y}}$
and fix $r$ with $\sum_{j}r_{j}=s$. Define $z_{1}$ as the boundary
point of the $R_{j}$ bin containing $z$ along the line between $z$
and $v$ and define $v_{1}$ similarly as the boundary point for the
bin containing $v$ along the same line. We then have the following:
\begin{align*}
\left|D^{r}u_{\lambda}(z)-D^{r}u_{\lambda}(v)\right| & \leq\left|D^{r}u_{\lambda}(z)-D^{r}u_{\lambda}(z_{1})\right|+\left|D^{r}u_{\lambda}(z_{1})-D^{r}u_{\lambda}(v_{1})\right|+\left|D^{r}u_{\lambda}(v_{1})-D^{r}u_{\lambda}(v)\right|\\
 & =\left|D^{r}u_{\lambda}(z)-D^{r}u_{\lambda}(z_{1})\right|+\left|D^{r}u_{\lambda}(v_{1})-D^{r}u_{\lambda}(v)\right|\\
 & =\int_{\gamma(z,z_{1})}\nabla D^{r}u_{\lambda}(t)dt+\int_{\gamma(v,v_{1})}\nabla D^{r}u_{\lambda}(t)dt\\
 & \leq H_{1}\left\Vert D^{r+1}u_{j}\right\Vert _{\infty}\left(\left\Vert z-z_{1}\right\Vert _{2}+\left\Vert v-v_{1}\right\Vert _{2}\right)\\
 & \leq H_{1}\ell^{\frac{r+1}{d_{X}+d_{Y}}}\left(\left\Vert z-z_{1}\right\Vert _{2}^{1-(s-r)}\left\Vert z-z_{1}\right\Vert _{2}^{s-r}+\left\Vert v-v_{1}\right\Vert _{2}^{1-(s-r)}\left\Vert v-v_{1}\right\Vert _{2}^{s-r}\right)\\
 & \leq H_{1}\ell^{\frac{r+1}{d_{X}+d_{Y}}}\sqrt{d_{X}+d_{Y}}\ell^{-\frac{1-(s-r)}{d_{X}+d_{Y}}}\left(\left\Vert z-z_{1}\right\Vert _{2}^{s-r}+\left\Vert v-v_{1}\right\Vert _{2}^{s-r}\right)\\
 & \leq H_{1}\ell^{\frac{s}{d_{X}+d_{Y}}}\sqrt{d_{X}+d_{Y}}\left\Vert z-v\right\Vert _{2}^{s-r}.
\end{align*}

The first line is an application of the triangle inequality. The second
line follows from the fact that $u_{\lambda}$ and all of its derivatives
are zero on the boundaries of the cubes $R_{j}$ as $u_{j}$ is not
supported in the band around the border of $R_{j}$. The third line
follows from the fundamental theorem of calculus where $\gamma(z,z_{1})$
is the path between $z$ and $z_{1}$. The fourth line is an application
of H\"{o}lder's inequality where we replace each derivative with
its supremum, leaving just the path integral which simplifies to the
length of the path. The fifth line follows from the assumption that
$\left\Vert D^{r}u_{j}\right\Vert _{\infty}\leq\ell^{r/\left(d_{X}+d_{Y}\right)}$
when $\sum_{j}r_{j}\leq s+1$. For the sixth line, since $z$ and
$z_{1}$ are in the same bin, then $\left\Vert z-z_{1}\right\Vert _{2}\leq\sqrt{d_{X}+d_{Y}}\ell^{-1/\left(d_{X}+d_{Y}\right)}$
as there are $\ell$ boxes with side length $\ell^{-1/\left(d_{X}+d_{Y}\right)}$.
Finally, the last line follows since $z_{1}$ and $v_{1}$ are on
the line segment between $z$ and $v$.

This indicates that $u_{\lambda}$, and therefore $f_{XY,\lambda}$,
is guaranteed to be H\"{o}lder smooth if $H_{1}\ell^{\frac{s}{d_{X}+d_{Y}}}\sqrt{d_{X}+d_{Y}}\leq H$.
Thus we require that $H_{1}=O\left(\ell^{-\frac{s}{d_{X}+d_{Y}}}\right)$.
We will set $\ell$ later on.

Note that for any $f_{XY,\lambda}\in\Lambda$, by a second order Taylor
series approximation in the first argument we have 
\begin{align*}
I\left(f_{XY,\lambda},f_{X}f_{Y}\right) & =I\left(f_{XY},f_{X}f_{Y}\right)-\int g'\left(\frac{f_{X}(x)f_{Y}(y)}{f_{XY}(x,y)}\right)\frac{f_{X}(x)f_{Y}(y)u_{\lambda}(x,y)}{f_{XY}(x,y)}dxdy\\
 & +\frac{1}{2}\int g''\left(\frac{f_{X}(x)f_{Y}(y)}{f_{XY}^{*}(x,y)}\right)\frac{f_{X}^{2}(x)f_{Y}^{2}(y)u_{\lambda}^{2}(x,y)}{\left(f_{XY}^{*}(x,y)\right)^{2}}dxdy,
\end{align*}
where $f_{XY}^{*}$ is the function from Taylor's remainder theorem.
By construction (see Lemma~\ref{lem:functions}), the first order
term vanishes. For the second order term, note that $f_{XY}^{*}$
lies on the line segment between $f_{XY}$ and $f_{XY,\lambda}$ and
is therefore upper and lower bounded. Similarly the density $f_{XY,\lambda}$
will be upper and lower bounded for $N$ sufficiently large as $f_{XY,\lambda}\in\left[f_{XY}-H_{1},f_{XY}+H_{1}\right]$
due to the fact that $\left\Vert D_{0}u_{j}\right\Vert _{\infty}=\left\Vert u_{j}\right\Vert _{\infty}\leq1,$
and $H_{1}$ will be chosen to decrease as $N$ increases. Assume
without loss of generality that given $\epsilon>0$, $g''(\epsilon)>0.$
Thus there exists a constant $c_{0}$ such that 
\[
\frac{1}{2}\int g''\left(\frac{f_{X}(x)f_{Y}(y)}{f_{XY}^{*}(x,y)}\right)\frac{f_{X}^{2}(x)f_{Y}^{2}(y)u_{\lambda}^{2}(x,y)}{\left(f_{XY}^{*}(x,y)\right)^{2}}dxdy\geq c_{0}H_{1}^{2}\sum_{j=1}^{\ell}\left\Vert u_{j}\right\Vert _{2}^{2}\geq c_{1}H_{1}^{2},
\]
where we have used the facts that the $u_{j}$ functions are orthogonal
to each other and $\left\Vert u_{j}\right\Vert _{2}^{2}\in\Theta\left(\ell^{-1}\right)$.
Therefore,
\[
I\left(f_{XY,\lambda},f_{X}f_{Y}\right)-I\left(f_{XY},f_{X}f_{Y}\right)\geq c_{1}H_{1}^{2},
\]
providing us with the necessary separation of $2\beta$ where $\beta=c_{1}H_{1}^{2}/2$.
Note that if $g''(\epsilon)<0$, we can simply consider $I\left(f_{XY},f_{X}f_{Y}\right)-I\left(f_{XY,\lambda},f_{X}f_{Y}\right)$
instead.

We now focus on bounding the squared Hellinger distance $H^{2}\left(\bar{F}_{XY}^{N}\times\left(F_{X}F_{Y}\right)^{N},F_{XY}^{N}\times\left(F_{X}F_{Y}\right)^{N}\right)$
where $\bar{F}_{XY}^{N}=\frac{1}{|\Lambda|}\sum_{\lambda\in\Lambda}F_{XY,\lambda}^{N}$.
The Hellinger distance decomposes across product measures resulting
in:
\begin{align*}
H^{2}\left(\bar{F}_{XY}^{N}\times\left(F_{X}F_{Y}\right)^{N},F_{XY}^{N}\times\left(F_{X}F_{Y}\right)^{N}\right) & =2\left(1-\left(1-H^{2}\left(\bar{F}_{XY}^{N},F_{XY}^{N}\right)/2\right)\left(1-H^{2}\left(\left(F_{X}F_{Y}\right)^{N},\left(F_{X}F_{Y}\right)^{N}\right)\right)\right)\\
 & =H^{2}\left(\bar{F}_{XY}^{N},F_{XY}^{N}\right).
\end{align*}
To bound this, we will use the following result from Birge and Massart~\cite{birge1995estimation}:
\begin{lem}
\label{lem:BirgeBound}Consider a set of densities $p$ and $p_{\lambda}=p\left(1+\sum_{j}\lambda v_{j}\right)$
for $\lambda\in\Lambda=\{-1,1\}^{\ell}$. Suppose that (i) $\left\Vert v_{j}\right\Vert _{\infty}\leq1$,
(ii) $\left\Vert 1_{\{R_{j}^{C}\}}v_{j}\right\Vert _{1}=0,$ (iii)
$\int pv_{j}=0,$ and (iv) $\int pv_{j}^{2}=\alpha_{j}>0$ all hold.
Define $\bar{P}^{N}=\frac{1}{|\Lambda|}\sum_{\lambda\in\Lambda}P_{\lambda}^{N}$.
Then 
\[
H^{2}\left(\bar{P}^{N},P^{N}\right)\leq\frac{N^{2}}{3}\sum_{j=1}^{\ell}\alpha_{j}^{2}.
\]
\end{lem}
To apply Lemma~\ref{lem:BirgeBound}, define $v_{j}(z)=H_{1}u_{j}(z)/f_{XY}(z)$.
Then $f_{XY,\lambda}=f_{XY}\left(1+\sum_{j}\lambda_{j}v_{j}\right)$.
Requirements (i)-(iii) are immediately satisfied based on the properties
of $u_{j}$ (see Lemma~\ref{lem:functions}). Furthermore, 
\[
\alpha_{j}=\int v_{j}^{2}f_{XY}=H_{1}^{2}\int u_{j}^{2}/f_{XY}\leq\frac{H_{1}^{2}C}{\ell},
\]
for some constant $C$. Therefore 
\[
H^{2}\left(\bar{F}_{XY}^{N},F_{XY}^{N}\right)\leq\frac{N^{2}}{3}\sum_{j=1}^{\ell}\alpha_{j}^{2}\leq\frac{N^{2}H_{1}^{4}C^{2}}{\ell}\in\Theta\left(N^{2}\ell^{-\frac{4s+d_{X}+d_{Y}}{d_{X}+d_{Y}}}\right).
\]
Set $\ell=N^{\frac{2\left(d_{X}+d_{Y}\right)}{4s+d_{X}+d_{Y}}}$ resulting
in $H_{1}=N^{-\frac{2s}{4s+d_{X}+d_{Y}}}.$ Then the Hellinger distance
is bounded by a constant. Additionally, the error is larger than $\beta\in\Theta\left(N^{-\frac{4s}{4s+d_{X}+d_{Y}}}\right)$
allowing us to apply Lemma~\ref{lem:LeCam} when $s<\left(d_{X}+d_{Y}\right)/4$.
Markov's inequality then finishes the proof.

For $s>\left(d_{X}+d_{Y}\right)/4,$we get a lower bound of $O\left(N^{-1}\right)$
which is the parametric rate. In general, we cannot do any better
than this~\cite{bickel1988estimating,krishnamurthy2014divergence}
thus establishing the lower bound in this regime. In particular, Krishnamurthy
et al.~\cite{krishnamurthy2014divergence} use a contradiction approach
to establish this for divergence estimation which can be extended
to the MI estimation problem.

\section{Theory for Mixed Random Variables\label{sec:MixedProofs}}

Here we provide proofs of the theory that extends the MI estimators
for the continuous case to the mixed case.

\subsection{Proof of Lemma~\ref{lem:binom_fractional}}

\label{subsec:fracMomentProof}

For (\ref{eq:frac_moment}), note that$\N_{xy}$ is a binomial random
variable with parameter $f_{X_{D}Y_{D}}\left(x,y\right)$, $N$ trials,
and mean $Nf_{X_{D}Y_{D}}\left(x,y\right)$. Thus (\ref{eq:frac_moment})
is the (potentially) fractional moment of a binomial random variable.
By the generalized binomial theorem, we have that
\begin{eqnarray}
\mathbf{N}_{xy}^{\alpha} & = & \left(\mathbf{N}_{xy}-Nf_{X_{D}Y_{D}}\left(x,y\right)+Nf_{X_{D}Y_{D}}\left(x,y\right)\right)^{\alpha}\nonumber \\
 & = & \sum_{i=0}^{\infty}\left(\begin{array}{c}
\alpha\\
i
\end{array}\right)\left(Nf_{X_{D}Y_{D}}\left(x,y\right)\right)^{\alpha-i}\left(\mathbf{N}_{xy}-Nf_{X_{D}Y_{D}}\left(x,y\right)\right)^{i},\nonumber \\
\implies\bE\left[\mathbf{N}_{xy}^{\alpha}\right] & = & \sum_{i=0}^{\infty}\left(\begin{array}{c}
\alpha\\
i
\end{array}\right)\left(Nf_{X_{D}Y_{D}}\left(x,y\right)\right)^{\alpha-i}\bE\left[\left(\mathbf{N}_{xy}-Nf_{X_{D}Y_{D}}\left(x,y\right)\right)^{i}\right].\label{eq:fractional_moment}
\end{eqnarray}
From \cite{riordan1937moment}, the $i$-th central moment of $\mathbf{N}_{xy}$
has the form of 
\[
\bE\left[\left(\mathbf{N}_{xy}-Nf_{X_{D}Y_{D}}\left(x,y\right)\right)^{i}\right]=\sum_{n=0}^{\left\lfloor i/2\right\rfloor }c_{n,i}\left(f_{X_{D}Y_{D}}\left(x,y\right)\right)N^{n}.
\]
Combining this with (\ref{eq:fractional_moment}) gives 
\begin{align*}
\bE\left[\N_{xy}^{\alpha}\right] & =\sum_{i=0}^{\infty}\sum_{n=0}^{\left\lfloor i/2\right\rfloor }\left(\begin{array}{c}
\alpha\\
i
\end{array}\right)\left(f_{X_{D}Y_{D}}\left(x,y\right)\right)^{\alpha-i}c_{n,i}\left(f_{X_{D}Y_{D}}\left(x,y\right)\right)N^{\alpha-i+n}\\
 & =\left(Nf_{X_{D}Y_{D}}\left(x,y\right)\right)^{\alpha}+O\left(N^{\alpha-1}\right).
\end{align*}

For (\ref{eq:frac_combined}), we apply a Taylor series expansion
to obtain
\begin{align*}
\N_{xy}^{\lambda}\N_{x}^{\beta}\N_{y}^{\gamma} & =N^{\lambda+\beta+\gamma}p^{\lambda}p_{x}^{\beta}p_{y}^{\gamma}+\left(\N_{xy}-Np\right)p^{\lambda-1}\left(N^{\lambda+\beta+\gamma-1}p_{x}^{\beta}p_{y}^{\gamma}+N^{\lambda+\beta+\gamma-2}\left(p_{x}^{\beta-1}p_{y}^{\gamma}\left(\N_{x}-Np_{x}\right)+p_{x}^{\beta}p_{y}^{\gamma-1}\left(\N_{y}-Np_{y}\right)\right)\right)\\
 & +N^{\lambda+\beta+\gamma-1}p^{\lambda}\left(p_{x}^{\beta-1}p_{y}^{\gamma}\left(\N_{x}-Np_{x}\right)+p_{x}^{\beta}p_{y}^{\gamma-1}\left(\N_{y}-Np_{y}\right)\right)+O\left(N^{\lambda+\beta+\gamma-2}\left(\left(\N_{x}-Np_{x}\right)\left(\N_{y}-Np_{y}\right)\right)\right),
\end{align*}
where we set $p=f_{X_{D}Y_{D}}\left(x,y\right)$, $p_{x}=f_{X_{D}}(x)$,
and $p_{y}=f_{Y_{D}}(y)$ for notational convenience. By taking the
expected value with respect to $\N_{x}$, $\N_{y}$, and $\N_{xy}$,
we obtain 
\begin{align*}
\bE\left[\N_{xy}^{\lambda}\N_{x}^{\beta}\N_{y}^{\gamma}\right] & =N^{\lambda+\beta+\gamma}p^{\lambda}p_{x}^{\beta}p_{y}^{\gamma}+N^{\lambda+\beta+\gamma-2}p^{\lambda-1}\left(p_{x}^{\beta-1}p_{y}^{\gamma}\cov\left(\N_{xy},\N_{x}\right)+p_{x}^{\beta}p_{y}^{\gamma-1}\cov\left(\N_{xy},\N_{y}\right)\right)\\
 & +O\left(N^{\beta+\gamma-1}\cov\left(\N_{x},\N_{y}\right)\right)\\
 & =N^{\lambda+\beta+\gamma}p^{\lambda}p_{x}^{\beta}p_{y}^{\gamma}+O\left(N^{\lambda+\beta+\gamma-1}\right),
\end{align*}
where the last step follows from the Cauchy-Schwarz inequality and
the variance of a binomial random variable.

\subsection{Proof of Theorem~\ref{thm:bias_mixed} (Bias)}

\label{subsec:bias_mixed_proof}

For notational ease, let 
\begin{equation}
\mathcal{T}(\X,\Y)=\frac{f_{X_{C}|X_{D}}\left(\X_{C}|\X_{D}\right)f_{Y_{C}|Y_{D}}\left(\Y_{C}|\Y_{D}\right)}{f_{X_{C}Y_{C}|X_{D}Y_{D}}\left(\X_{C},\Y_{C}|\X_{D},\Y_{D}\right)}.\label{eq:ratio}
\end{equation}
We have that 
\begin{align}
\bias\left[\g{h_{X_{C}|X_{D}},h_{Y_{C}|Y_{D}}}\right] & =\bE\left[\g{h_{X_{C}|X_{D}},h_{Y_{C}|Y_{D}}}\right]-I(\X;\Y)\nonumber \\
 & =\bE\left[\sum_{x\in\mathcal{S}_{X_{D}},y\in\mathcal{S}_{Y_{D}}}\frac{\N_{xy}}{N}\g{h_{X_{C}|x},h_{Y_{C}|y}}-g\left(\mathcal{T}(\X,\Y)\times\frac{f_{X_{D}}\left(\X_{D}\right)f_{Y_{D}}\left(\Y_{D}\right)}{f_{X_{D}Y_{D}}\left(\X_{D},\Y_{D}\right)}\right)\right]\nonumber \\
 & =\bE\left[\sum_{x\in\mathcal{S}_{X_{D}},y\in\mathcal{S}_{Y_{D}}}\frac{\N_{xy}}{N}\left(\g{h_{X_{C}|x},h_{Y_{C}|y}}-g\left(\mathcal{T}(\X,\Y)\times\frac{\N_{x}\N_{y}}{N\N_{xy}}\right)\right)\right]\nonumber \\
 & +\bE\left[\sum_{x\in\mathcal{S}_{X_{D}},y\in\mathcal{S}_{Y_{D}}}\left(\frac{\N_{xy}}{N}g\left(\mathcal{T}(\X,\Y)\times\frac{\N_{x}\N_{y}}{N\N_{xy}}\right)-f_{X_{D}Y_{D}}(x,y)g\left(\mathcal{T}(\X,\Y)\times\frac{f_{X_{D}}\left(x\right)f_{Y_{D}}\left(y\right)}{f_{X_{D}Y_{D}}\left(x,y\right)}\right)\right)\right].\label{eq:bias_split}
\end{align}
We consider the second term in (\ref{eq:bias_split}) first. A Taylor
series expansion of $g\left(\mathcal{T}(\X,\Y)\times\frac{\N_{x}\N_{y}}{N\N_{xy}}\right)$
evaluated at $\mathcal{T}(\X,\Y)\times\frac{f_{X_{D}}\left(x\right)f_{Y_{D}}\left(y\right)}{f_{X_{D}Y_{D}}\left(x,y\right)}$
gives terms of the form of 
\begin{equation}
\left(f_{X_{C}|X_{D}}\left(\X_{C}|x\right)f_{Y_{C}|Y_{D}}\left(\Y_{C}|y\right)\left(\N_{x}\N_{y}/N^{2}-f_{X_{D}}\left(x\right)f_{Y_{D}}\left(y\right)\right)\right)^{i},\label{eq:disc_bias1}
\end{equation}
\begin{equation}
\left(f_{X_{C}Y_{C}|X_{D}Y_{D}}\left(\X_{C},\Y_{C}|x,y\right)\left(\N_{xy}/N-f_{X_{D}Y_{D}}\left(x,y\right)\right)\right)^{i},\label{eq:disc_bias2}
\end{equation}
where $i$ is a positive integer. For notational ease, set $p=f_{X_{D}Y_{D}}\left(x,y\right)$.
By applying the binomial theorem and (\ref{eq:frac_moment}), we obtain
\begin{align*}
\frac{\N_{xy}}{N}\left(p-\N_{xy}/N\right)^{i} & =\sum_{k=0}^{i}\binom{i}{k}p^{i-k}\left(\frac{\N_{xy}}{N}\right)^{k+1}(-1)^{k}\\
\implies\bE\left[\frac{\N_{xy}}{N}\left(p-\N_{xy}/N\right)^{i}\right] & =p^{i+1}\sum_{k=0}^{i}\binom{i}{k}(-1)^{k}+O\left(\frac{1}{N}\right)\\
 & =O\left(\frac{1}{N}\right).
\end{align*}

Using a similar approach with (\ref{eq:frac_combined}), it can be
shown that 
\[
\bE\left[\frac{\N_{xy}}{N}\left(\frac{\N_{x}\N_{y}}{N^{2}}-f_{X_{D}}\left(x\right)f_{Y_{D}}\left(y\right)\right)^{i}\right]=O\left(\frac{1}{N}\right).
\]
Thus the second term in (\ref{eq:bias_split}) reduces to $O(1/N)$.

By conditioning on $\X_{1,D},\dots,\X_{N,D},\Y_{1,D},\dots,\Y_{N,D}$,
the first term in (\ref{eq:bias_split}) can be written as 
\[
\bE\left[\sum_{x\in\mathcal{S}_{X_{D}},y\in\mathcal{S}_{Y_{D}}}\frac{\N_{xy}}{N}\bias\left[\left.\g{h_{X_{C}|x},h_{Y_{C}|y}}\right|\X_{1,D},\dots,\X_{N,D},\Y_{1,D},\dots,\Y_{N,D}\right]\right].
\]
The conditional bias of $\g{h_{X_{C}|x},h_{Y_{C}|y}}$ given $\X_{1,D},\dots,\X_{N,D},\Y_{1,D},\dots,\Y_{N,D}$
can be obtained from Theorem~\ref{thm:bias} as 
\begin{align}
\bias\left[\left.\g{h_{X_{C}|x},h_{Y_{C}|y}}\right|\X_{1,D},\dots,\X_{N,D},\Y_{1,D},\dots,\Y_{N,D}\right] & =\sum_{\substack{i,j=0\\
i+j\neq0
}
}^{r}c_{10,i,j}\left(\frac{\N_{x}\N_{y}}{N^{2}},\frac{\N_{xy}}{N}\right)\hx^{i}\hy^{j}\nonumber \\
 & +O\left(\hx^{s}+\hy^{s}+\frac{1}{\N_{xy}\hx^{d_{X}}\hy^{d_{Y}}}\right).\label{eq:bias_conditional}
\end{align}
This expression provides the motivation for our choice of $\hx$ and
$\hy$. Since $\hx\propto\N_{x}^{-\beta}$ and $\hy\propto\N_{y}^{-\alpha}$,
then (\ref{eq:mixed_est}) gives terms with the form of $\N_{xy}\N_{x}^{-\beta i}\N_{y}^{-\alpha j}/N$
with $i+j\geq1$. From Lemma~\ref{lem:binom_fractional}, taking
the expected value of these terms gives
\[
\bE\left[\N_{xy}\N_{x}^{-\beta i}\N_{y}^{-\alpha j}/N\right]=N^{-\beta i-\alpha j}f_{X_{D}Y_{D}}\left(x,y\right)\left(f_{X_{D}}(x)\right)^{-\beta i}\left(f_{Y_{D}}(y)\right)^{-\alpha j}+o\left(\frac{1}{N}\right).
\]
Similarly, taking the expectation of $\N_{xy}\N_{x}^{\beta d_{X}}\N_{y}^{\alpha d_{Y}}/N^{2}$
gives $O\left(N^{\beta d_{X}+\alpha d_{Y}-1}\right)$. Note that the
polynomial terms of $\N_{x}\N_{y}/N^{2}$ and $\N_{xy}/N$ in the
constants in (\ref{eq:bias_conditional}) do not contribute to the
bias rate as the $\N_{x}\N_{y}$ and $\N_{xy}$ terms in the numerator
are cancelled by the $N^{2}$ and $N$ terms in the denominator, respectively,
after taking the expectation. Combining all of these results completes
the proof.

\subsection{Proof of Theorem~\ref{thm:var_mixed} (Variance)}

\label{subsec:varMixedProof}By the law of total variance, which can
be derived from the Pythagorean theorem, we have 
\begin{align}
\var\left[\g{h_{X_{C}|X_{D}},h_{Y_{C}|Y_{D}}}\right] & =\bE\left[\var\left[\left.\g{h_{X_{C}|X_{D}},h_{Y_{C}|Y_{D}}}\right|\X_{1,D},\dots,\X_{N,D},\Y_{1,D},\dots,\Y_{N,D}\right]\right]\nonumber \\
 & +\var\left[\bE\left[\left.\g{h_{X_{C}|X_{D}},h_{Y_{C}|Y_{D}}}\right|\X_{1,D},\dots,\X_{N,D},\Y_{1,D},\dots,\Y_{N,D}\right]\right].\label{eq:var_total}
\end{align}
Given all of the $\X_{i,D}$ and $\Y_{i,D}$ random variables, the
estimators $\g{h_{X_{C}|x},h_{Y_{C}|y}}$ are all conditionally independent
since they use different sets of $\mathbf{X}_{i,C}$'s and $\Y_{i,C}$'s
for each pair $(x,y)$. Thus from Theorem~\ref{thm:variance}, we
get 
\begin{align*}
\var\left[\left.\g{h_{X_{C}|X_{D}},h_{Y_{C}|Y_{D}}}\right|\X_{1,D},\dots,\X_{N,D},\Y_{1,D},\dots,\Y_{N,D}\right] & =O\left(\sum_{x\in\mathcal{S}_{X_{D}},y\in\mathcal{S}_{Y_{D}}}\frac{\N_{xy}^{2}}{N^{2}}\frac{1}{\N_{xy}}\right)\\
 & =O\left(\sum_{x\in\mathcal{S}_{X_{D}},y\in\mathcal{S}_{Y_{D}}}\frac{\N_{xy}}{N^{2}}\right).
\end{align*}
 Taking the expectation yields $O(1/N)$.

For the second term in (\ref{eq:var_total}), we know from (\ref{eq:bias_conditional})
that 
\begin{align*}
\bE\left[\left.\g{h_{X_{C}|x},h_{Y_{C}|y}}\right|\X_{1,D},\dots,\X_{N,D},\Y_{1,D},\dots,\Y_{N,D}\right] & =O\left(\sum_{\substack{i,j=0\\
i+j\neq0
}
}^{r}\N_{x}^{-i\beta}\N_{y}^{-j\alpha}+\N_{x}^{-s\beta}+\N_{y}^{-s\alpha}+\frac{\N_{x}^{\beta d_{X}}\N_{y}^{\alpha d_{Y}}}{\N_{xy}}\right)\\
 & =O\left(f\left(\N_{x},\N_{y},\N_{xy}\right)\right).
\end{align*}
Let $\N_{xy}'$, $\N_{x}'$, and $\N_{y}'$ be independent and identically
distributed realizations of $\N_{xy}$, $\N_{x}$, and $\N_{y},$
respectively. Then by the Efron-Stein inequality, 
\begin{align}
\var\left[\sum_{x\in\mathcal{S}_{X_{D}},y\in\mathcal{S}_{Y_{D}}}\frac{\N_{xy}}{N}f\left(\N_{x},\N_{y},\N_{xy}\right)\right] & \leq\frac{1}{2N^{2}}\sum_{x\in\mathcal{S}_{X_{D}},y\in\mathcal{S}_{Y_{D}}}\bE\left[\left(\N_{xy}f\left(\N_{x},\N_{y},\N_{xy}\right)-\N_{xy}'f\left(\N_{x}',\N_{y}',\N_{xy}'\right)\right)^{2}\right],\label{eq:var_efron}
\end{align}
where since $\N_{x}$, $\N_{y}$, and $\N_{xy}$ are not independent,
we consider the effect of resampling all three simultaneously. Note
that 
\begin{align}
\left(\N_{xy}f\left(\N_{x},\N_{y},\N_{xy}\right)-\N_{xy}'f\left(\N_{x},'\N_{y}',\N_{xy}'\right)\right)^{2} & =O\left(\left(\sum_{\substack{i,j=0\\
i+j\neq0
}
}^{r}\left(\N_{xy}\N_{x}^{-i\beta}\N_{y}^{-j\alpha}-\N_{xy}'\left(\N_{x}'\right)^{-i\beta}\left(\N_{y}'\right)^{-j\alpha}\right)\right.\right.\nonumber \\
 & +\left(\N_{xy}\N_{x}^{-s\beta}-\N_{xy}'\left(\N_{x}'\right)^{-s\beta}\right)+\left(\N_{xy}\N_{y}^{-s\alpha}-\N_{xy}'\left(\N_{y}'\right)^{-s\alpha}\right)\nonumber \\
 & \left.\left.\begin{array}{c}
\\
\\
\\
\\
\end{array}+\left(\N_{x}^{\beta d_{X}}\N_{y}^{\alpha d_{Y}}-\left(\N_{x}'\right)^{\beta d_{X}}\left(\N_{y}'\right)^{\alpha d_{Y}}\right)\right)^{2}\right).\label{eq:var_bias}
\end{align}
By Jensen's inequality, we can consider separately each of the squared
differences in (\ref{eq:var_bias}). Then since $\left(\N_{xy},\N_{x},\N_{y}\right)$
is independent of $\left(\N_{xy}',\N_{x}',\N_{y}'\right)$ and they
are identically distributed, then the expected squared difference
is proportional to the variance. For example, applying Lemma \ref{lem:binom_fractional}
gives
\begin{align*}
\bE\left[\left(\N_{xy}\N_{x}^{-s\beta}-\N_{xy}'\left(\N_{x}'\right)^{-s\beta}\right)^{2}\right] & =2\var\left[\N_{xy}\N_{x}^{-s\beta}\right]\\
 & =2N^{2-2s\beta}\left(f_{X_{D}Y_{D}}(x,y)\right)^{2}\left(f_{X_{D}}(x)\right)^{-2s\beta}-2\left(N^{1-s\beta}f_{X_{D}Y_{D}}(x,y)\left(f_{X_{D}}(x)\right)^{-s\beta}\right)^{2}\\
 & +O\left(N^{1-2s\beta}\right)\\
 & =O\left(N^{1-2s\beta}\right).
\end{align*}

By a similar procedure, we obtain 
\begin{align*}
\bE\left[\left(\N_{xy}\N_{y}^{-s\alpha}-\N_{xy}'\left(\N_{y}'\right)^{-s\alpha}\right)^{2}\right] & =O\left(N^{1-2s\alpha}\right),\\
\bE\left[\left(\N_{x}^{\beta d_{X}}\N_{y}^{\alpha d_{Y}}-\left(\N_{x}'\right)^{\beta d_{X}}\left(\N_{y}'\right)^{\alpha d_{Y}}\right)^{2}\right] & =O\left(N^{2\beta d_{X}+2\alpha d_{Y}-1}\right),\\
\bE\left[\left(\sum_{\substack{i,j=0\\
i+j\neq0
}
}^{r}\left(\N_{xy}\N_{x}^{-i\beta}\N_{y}^{-j\alpha}-\N_{xy}'\left(\N_{x}'\right)^{-i\beta}\left(\N_{y}'\right)^{-j\alpha}\right)\right)^{2}\right] & =O\left(N^{1-2\beta}\right)+O\left(N^{1-2\alpha}\right).
\end{align*}
Combining these results with (\ref{eq:var_efron}) and (\ref{eq:var_total})
completes the proof.

\section{Proof of Theorem~\ref{thm:clt} (CLT)}

\label{sec:cltProof} We will first find the asymptotic distribution
of 
\begin{align*}
\sqrt{N}\left(\gt-\bE\left[\gt\right]\right) & =\frac{1}{\sqrt{N}}\sum_{i=1}^{N}\left(g\left(\frac{\ft X(\mathbf{X}_{i})\ft Y(\mathbf{Y}_{i})\nu_{1}\nu_{2}}{\ft Z(\mathbf{X}_{i},\mathbf{Y}_{i})\nu_{3}}\right)-\ez{Z_{i}}\left[g\left(\frac{\ft X(\mathbf{X}_{i})\ft Y(\mathbf{Y}_{i})\nu_{1}\nu_{2}}{\ft Z(\mathbf{X}_{i},\mathbf{Y}_{i})\nu_{3}}\right)\right]\right)\\
 & +\frac{1}{\sqrt{N}}\sum_{i=1}^{N}\left(\ez{Z_{i}}\left[g\left(\frac{\ft X(\mathbf{X}_{i})\ft Y(\mathbf{Y}_{i})\nu_{1}\nu_{2}}{\ft Z(\mathbf{X}_{i},\mathbf{Y}_{i})\nu_{3}}\right)\right]-\bE\left[g\left(\frac{\ft X(\mathbf{X}_{i})\ft Y(\mathbf{Y}_{i})\nu_{1}\nu_{2}}{\ft Z(\mathbf{X}_{i},\mathbf{Y}_{i})\nu_{3}}\right)\right]\right).
\end{align*}
By the standard central limit theorem \cite{durrett2010probability},
the second term converges in distribution to a Gaussian random variable
with variance 
\[
\var\left[\ez Z\left[g\left(\frac{\ft X(\mathbf{X})\ft Y(\mathbf{Y})\nu_{1}\nu_{2}}{\ft Z(\mathbf{X},\mathbf{Y})\nu_{3}}\right)\right]\right].
\]
All that remains is to show that the first term converges in probability
to zero as Slutsky's theorem \cite{gut2012probability} can then be
applied. Denote this first term as $\W_{N}$ and note that $\bE\left[\W_{N}\right]=0$. 

We will use Chebyshev's inequality combined with the Efron-Stein inequality
to bound the variance of $\W_{N}$. Consider the samples $\left\{ \Z_{1},\dots,\Z_{N}\right\} $
and $\left\{ \Z_{1}^{'},\Z_{2},\dots,\Z_{N}\right\} $ and the respective
sequences $\W_{N}$ and $\W_{N}^{'}$. This gives 
\begin{align}
\W_{N}-\W_{N}^{'} & =\frac{1}{\sqrt{N}}\left(g\left(\frac{\ft X(\mathbf{X}_{1})\ft Y(\mathbf{Y}_{1})\nu_{1}\nu_{2}}{\ft Z(\mathbf{X}_{1},\mathbf{Y}_{1})\nu_{3}}\right)-\ez{Z_{1}}\left[g\left(\frac{\ft X(\mathbf{X}_{1})\ft Y(\mathbf{Y}_{1})\nu_{1}\nu_{2}}{\ft Z(\mathbf{X}_{1},\mathbf{Y}_{1})\nu_{3}}\right)\right]\right)\nonumber \\
 & -\frac{1}{\sqrt{N}}\left(g\left(\frac{\ft X(\mathbf{X}_{1}^{'})\ft Y(\mathbf{Y}_{1}^{'})\nu_{1}\nu_{2}}{\ft Z(\mathbf{X}_{1}^{'},\mathbf{Y}_{1}^{'})}\right)-\ez{Z_{1}^{'}}\left[g\left(\frac{\ft X(\mathbf{X}_{1}^{'})\ft Y(\mathbf{Y}_{1}^{'})\nu_{1}\nu_{2}}{\ft Z(\mathbf{X}_{1}^{'},\mathbf{Y}_{1}^{'})}\right)\right]\right)\nonumber \\
 & +\frac{1}{\sqrt{N}}\sum_{i=2}^{N}\left(g\left(\frac{\ft X(\mathbf{X}_{i})\ft Y(\mathbf{Y}_{i})\nu_{1}\nu_{2}}{\ft Z(\mathbf{X}_{i},\mathbf{Y}_{i})\nu_{3}}\right)-g\left(\frac{\ft X^{'}(\mathbf{X}_{i})\ft Y^{'}(\mathbf{Y}_{i})\nu_{1}\nu_{2}}{\ft Z^{'}(\mathbf{X}_{i},\mathbf{Y}_{i})\nu_{3}}\right)\right).\label{eq:Wndiff}
\end{align}

Note that 
\[
\bE\left[\left(g\left(\frac{\ft X(\mathbf{X}_{1})\ft Y(\mathbf{Y}_{1})\nu_{1}\nu_{2}}{\ft Z(\mathbf{X}_{1},\mathbf{Y}_{1})\nu_{3}}\right)-\ez{Z_{1}}\left[g\left(\frac{\ft X(\mathbf{X}_{1})\ft Y(\mathbf{Y}_{1})\nu_{1}\nu_{2}}{\ft Z(\mathbf{X}_{1},\mathbf{Y}_{1})\nu_{3}}\right)\right]\right)^{2}\right]=\bE\left[\var_{\Z_{1}}\left[g\left(\frac{\ft X(\mathbf{X}_{1})\ft Y(\mathbf{Y}_{1})\nu_{1}\nu_{2}}{\ft Z(\mathbf{X}_{1},\mathbf{Y}_{1})\nu_{3}}\right)\right]\right].
\]
We will use the Efron-Stein inequality to bound $\var_{\Z_{1}}\left[g\left(\frac{\ft X(\mathbf{X}_{1})\ft Y(\mathbf{Y}_{1})\nu_{1}\nu_{2}}{\ft Z(\mathbf{X}_{1},\mathbf{Y}_{1})\nu_{3}}\right)\right]$.
We thus need to bound the conditional expectation of the term 
\[
\left|g\left(\frac{\ft X(\mathbf{X}_{1})\ft Y(\mathbf{Y}_{1})\nu_{1}\nu_{2}}{\ft Z(\mathbf{X}_{1},\mathbf{Y}_{1})\nu_{3}}\right)-g\left(\frac{\ft X^{'}(\mathbf{X}_{1})\ft Y^{'}(\mathbf{Y}_{1})\nu_{1}\nu_{2}}{\ft Z^{'}(\mathbf{X}_{1},\mathbf{Y}_{1})\nu_{3}}\right)\right|^{2},
\]
where $\Z_{i}$ is replaced with $\Z_{i}^{'}$ in the KDEs for some
$i\neq1$. Using similar steps as in Appendix~\ref{sec:VarProof},
we have that 
\[
\bE\left[\left|g\left(\frac{\ft X(\mathbf{X}_{1})\ft Y(\mathbf{Y}_{1})\nu_{1}\nu_{2}}{\ft Z(\mathbf{X}_{1},\mathbf{Y}_{1})\nu_{3}}\right)-g\left(\frac{\ft X^{'}(\mathbf{X}_{1})\ft Y^{'}(\mathbf{Y}_{1})\nu_{1}\nu_{2}}{\ft Z^{'}(\mathbf{X}_{1},\mathbf{Y}_{1})\nu_{3}}\right)\right|^{2}\right]=O\left(\frac{1}{N^{2}}\right).
\]
Then by the Efron-Stein inequality, $\var_{\Z_{1}}\left[g\left(\frac{\ft X(\mathbf{X}_{1})\ft Y(\mathbf{Y}_{1})\nu_{1}\nu_{2}}{\ft Z(\mathbf{X}_{1},\mathbf{Y}_{1})\nu_{3}}\right)\right]=O\left(\frac{1}{N}\right)$.
Therefore 
\[
\bE\left[\frac{1}{N}\left(g\left(\frac{\ft X(\mathbf{X}_{1})\ft Y(\mathbf{Y}_{1})\nu_{1}\nu_{2}}{\ft Z(\mathbf{X}_{1},\mathbf{Y}_{1})\nu_{3}}\right)-\ez{Z_{1}}\left[g\left(\frac{\ft X(\mathbf{X}_{1})\ft Y(\mathbf{Y}_{1})\nu_{1}\nu_{2}}{\ft Z(\mathbf{X}_{1},\mathbf{Y}_{1})\nu_{3}}\right)\right]\right)^{2}\right]=O\left(\frac{1}{N^{2}}\right).
\]
A similar result holds for the $g\left(\frac{\ft X(\mathbf{X}_{1}^{'})\ft Y(\mathbf{Y}_{1}^{'})\nu_{1}\nu_{2}}{\ft Z(\mathbf{X}_{1}^{'},\mathbf{Y}_{1}^{'})\nu_{3}}\right)$
term in (\ref{eq:Wndiff}). 

For the third term in (\ref{eq:Wndiff}), 
\begin{align*}
 & \lefteqn{\bE\left[\left(\sum_{i=2}^{N}\left|g\left(\frac{\ft X(\mathbf{X}_{i})\ft Y(\mathbf{Y}_{i})\nu_{1}\nu_{2}}{\ft Z(\mathbf{X}_{i},\mathbf{Y}_{i})\nu_{3}}\right)-g\left(\frac{\ft X^{'}(\mathbf{X}_{i})\ft Y^{'}(\mathbf{Y}_{i})\nu_{1}\nu_{2}}{\ft Z^{'}(\mathbf{X}_{i},\mathbf{Y}_{i})\nu_{3}}\right)\right|\right)^{2}\right]}\\
 & =\sum_{i,j=2}^{N}\bE\left[\left|g\left(\frac{\ft X(\mathbf{X}_{i})\ft Y(\mathbf{Y}_{i})\nu_{1}\nu_{2}}{\ft Z(\mathbf{X}_{i},\mathbf{Y}_{i})\nu_{3}}\right)-g\left(\frac{\ft X^{'}(\mathbf{X}_{i})\ft Y^{'}(\mathbf{Y}_{i})\nu_{1}\nu_{2}}{\ft Z^{'}(\mathbf{X}_{i},\mathbf{Y}_{i})\nu_{3}}\right)\right|\right.\\
 & \times\left.\left|g\left(\frac{\ft X(\mathbf{X}_{j})\ft Y(\mathbf{Y}_{j})\nu_{1}\nu_{2}}{\ft Z(\mathbf{X}_{j},\mathbf{Y}_{j})\nu_{3}}\right)-g\left(\frac{\ft X^{'}(\mathbf{X}_{j})\ft Y^{'}(\mathbf{Y}_{j})\nu_{1}\nu_{2}}{\ft Z^{'}(\mathbf{X}_{j},\mathbf{Y}_{j})\nu_{3}}\right)\right|\right]
\end{align*}
For the $N-1$ terms where $i=j$, we know from Appendix \ref{sec:VarProof}
that 
\[
\bE\left[\left|g\left(\frac{\ft X(\mathbf{X}_{i})\ft Y(\mathbf{Y}_{i})\nu_{1}\nu_{2}}{\ft Z(\mathbf{X}_{i},\mathbf{Y}_{i})\nu_{3}}\right)-g\left(\frac{\ft X^{'}(\mathbf{X}_{i})\ft Y^{'}(\mathbf{Y}_{i})\nu_{1}\nu_{2}}{\ft Z^{'}(\mathbf{X}_{i},\mathbf{Y}_{i})\nu_{3}}\right)\right|^{2}\right]=O\left(\frac{1}{N^{2}}\right).
\]
Thus these terms contribute $O(1/N)$. For the $(N-1)^{2}-(N-1)$
terms where $i\neq j$, we can do multiple substitutions of the form
$\mathbf{u}_{j}=\frac{\X_{j}-\X_{1}}{h_{X}}$ resulting in
\begin{align*}
\bE\left[\left|g\left(\frac{\ft X(\mathbf{X}_{i})\ft Y(\mathbf{Y}_{i})\nu_{1}\nu_{2}}{\ft Z(\mathbf{X}_{i},\mathbf{Y}_{i})}\right)-g\left(\frac{\ft X^{'}(\mathbf{X}_{i})\ft Y^{'}(\mathbf{Y}_{i})\nu_{1}\nu_{2}}{\ft Z^{'}(\mathbf{X}_{i},\mathbf{Y}_{i})}\right)\right|\right.\\
\times\left.\left|g\left(\frac{\ft X(\mathbf{X}_{j})\ft Y(\mathbf{Y}_{j})\nu_{1}\nu_{2}}{\ft Z(\mathbf{X}_{j},\mathbf{Y}_{j})\nu_{3}}\right)-g\left(\frac{\ft X^{'}(\mathbf{X}_{j})\ft Y^{'}(\mathbf{Y}_{j})\nu_{1}\nu_{2}}{\ft Z^{'}(\mathbf{X}_{j},\mathbf{Y}_{j})\nu_{3}}\right)\right|\right] & =O\left(\frac{h_{X}^{2d_{X}}h_{Y}^{2d_{Y}}}{N^{2}}\right).
\end{align*}
Since $h_{X}^{d_{X}}h_{Y}^{d_{Y}}=o(1)$, 
\[
\bE\left[\left(\sum_{i=2}^{N}\left|g\left(\frac{\ft X(\mathbf{X}_{i})\ft Y(\mathbf{Y}_{i})\nu_{1}\nu_{2}}{\ft Z(\mathbf{X}_{i},\mathbf{Y}_{i})\nu_{3}}\right)-g\left(\frac{\ft X^{'}(\mathbf{X}_{i})\ft Y^{'}(\mathbf{Y}_{i})\nu_{1}\nu_{2}}{\ft Z^{'}(\mathbf{X}_{i},\mathbf{Y}_{i})\nu_{3}}\right)\right|\right)^{2}\right]=o(1).
\]
Combining all of these results with Jensen's inequality gives 
\begin{align*}
\bE\left[\left(\W_{N}-\W_{N}^{'}\right)^{2}\right] & \leq\frac{3}{N}\bE\left[\left(g\left(\frac{\ft X(\mathbf{X}_{1})\ft Y(\mathbf{Y}_{1})\nu_{1}\nu_{2}}{\ft Z(\mathbf{X}_{1},\mathbf{Y}_{1})\nu_{3}}\right)-\ez{Z_{1}}\left[g\left(\frac{\ft X(\mathbf{X}_{1})\ft Y(\mathbf{Y}_{1})\nu_{1}\nu_{2}}{\ft Z(\mathbf{X}_{1},\mathbf{Y}_{1})\nu_{3}}\right)\right]\right)^{2}\right]\\
 & +\frac{3}{N}\bE\left[\left(g\left(\frac{\ft X(\mathbf{X}_{1}^{'})\ft Y(\mathbf{Y}_{1}^{'})\nu_{1}\nu_{2}}{\ft Z(\mathbf{X}_{1}^{'},\mathbf{Y}_{1}^{'})\nu_{3}}\right)-\ez{Z_{1}^{'}}\left[g\left(\frac{\ft X(\mathbf{X}_{1}^{'})\ft Y(\mathbf{Y}_{1}^{'})\nu_{1}\nu_{2}}{\ft Z(\mathbf{X}_{1}^{'},\mathbf{Y}_{1}^{'})\nu_{3}}\right)\right]\right)^{2}\right]\\
 & +\frac{3}{N}\bE\left[\left(\sum_{i=2}^{N}\left(g\left(\frac{\ft X(\mathbf{X}_{i})\ft Y(\mathbf{Y}_{i})\nu_{1}\nu_{2}}{\ft Z(\mathbf{X}_{i},\mathbf{Y}_{i})\nu_{3}}\right)-g\left(\frac{\ft X^{'}(\mathbf{X}_{i})\ft Y^{'}(\mathbf{Y}_{i})\nu_{1}\nu_{2}}{\ft Z^{'}(\mathbf{X}_{i},\mathbf{Y}_{i})\nu_{3}}\right)\right)\right)^{2}\right]\\
 & =o\left(\frac{1}{N}\right).
\end{align*}
Applying the Efron-Stein inequality gives that $\var\left[\W_{N}\right]=o(1)$.
Then by ChebyShev's inequality, $\W_{N}$ converges to zero in probability.
This completes the proof for the plug-in estimator.

For the weighted ensemble estimator, we present a more general result
where we have different parameters $l_{X}\in\mathcal{L}_{X}$ and
$l_{Y}\in\mathcal{L}_{Y}$ for $\hx$ and $\hy$, respectively. We
can then write 
\begin{align*}
\sqrt{N}\left(\g w-\bE\left[\g w\right]\right) & =\frac{1}{\sqrt{N}}\sum_{i=1}^{N}\sum_{l_{X}\in\mathcal{L}_{X},l_{Y}\in\mathcal{L}_{Y}}w(l_{X},l_{Y})\left(g\left(\frac{\ftlx XX(\mathbf{X}_{i})\ftlx YY(\mathbf{Y}_{i})\nu_{1}\nu_{2}}{\ftlx ZZ(\mathbf{X}_{i},\mathbf{Y}_{i})\nu_{3}}\right)\right.\\
 & \left.-\ez{Z_{i}}\left[g\left(\frac{\ftlx XX(\mathbf{X}_{i})\ftlx YY(\mathbf{Y}_{i})\nu_{1}\nu_{2}}{\ftlx ZZ(\mathbf{X}_{i},\mathbf{Y}_{i})\nu_{3}}\right)\right]\right)\\
 & +\frac{1}{\sqrt{N}}\sum_{i=1}^{N}\left(\ez{Z_{i}}\left[\sum_{l_{X}\in\mathcal{L}_{X},l_{Y}\in\mathcal{L}_{Y}}w(l_{X},l_{Y})g\left(\frac{\ftlx XX(\mathbf{X}_{i})\ftlx YY(\mathbf{Y}_{i})\nu_{1}\nu_{2}}{\ftlx ZZ(\mathbf{X}_{i},\mathbf{Y}_{i})\nu_{3}}\right)\right]\right.\\
 & \left.-\bE\left[\sum_{l_{X}\in\mathcal{L}_{X},l_{Y}\in\mathcal{L}_{Y}}w(l_{X},l_{Y})g\left(\frac{\ftlx XX(\mathbf{X}_{i})\ftlx YY(\mathbf{Y}_{i})\nu_{1}\nu_{2}}{\ftlx ZZ(\mathbf{X}_{i},\mathbf{Y}_{i})\nu_{3}}\right)\right]\right).
\end{align*}
By the central limit theorem, the second term converges in distribution
to a zero-mean Gaussian random variable with variance 
\[
\var\left[\ez{Z_{i}}\left[\sum_{l_{X}\in\mathcal{L}_{X},l_{Y}\in\mathcal{L}_{Y}}w(l_{X},l_{Y})g\left(\frac{\ftlx XX(\mathbf{X}_{i})\ftlx YY(\mathbf{Y}_{i})\nu_{1}\nu_{2}}{\ftlx ZZ(\mathbf{X}_{i},\mathbf{Y}_{i})\nu_{3}}\right)\right]\right].
\]
From the previous results, the first term converges to zero in probability
as it can be written as 
\begin{align*}
\sum_{l_{X}\in\mathcal{L}_{X},l_{Y}\in\mathcal{L}_{Y}}w(l_{X},l_{Y})\frac{1}{\sqrt{N}}\sum_{i=1}^{N}\left(g\left(\frac{\ftlx XX(\mathbf{X}_{i})\ftlx YY(\mathbf{Y}_{i})\nu_{1}\nu_{2}}{\ftlx ZZ(\mathbf{X}_{i},\mathbf{Y}_{i})\nu_{3}}\right)\right.\\
\left.-\ez{Z_{i}}\left[g\left(\frac{\ftlx XX(\mathbf{X}_{i})\ftlx YY(\mathbf{Y}_{i})\nu_{1}\nu_{2}}{\ftlx ZZ(\mathbf{X}_{i},\mathbf{Y}_{i})\nu_{3}}\right)\right]\right) & =\sum_{l_{X}\in\mathcal{L}_{X},l_{Y}\in\mathcal{L}_{Y}}w(l_{X},l_{Y})o_{P}(1)\\
 & =o_{P}(1),
\end{align*}
where $o_{P}(1)$ denotes convergence to zero in probability and we
use the fact that linear combinations of random variables that converge
in probability individually to constants converge in probability to
the linear combination of the constants. The proof is finished with
Slutsky's theorem.

Note that the proof of Corollary~\ref{cor:clt} follows a similar
procedure as the extension to the ensemble case.
\end{document}